\documentclass[preprintnumbers,superscriptaddress,amsmath,amsfonts,amssymb,nofootinbib,onecolumn]{revtex4}
\usepackage{dcolumn}
\usepackage{bm}
\usepackage{graphicx}
\usepackage[normalem]{ulem}
\usepackage{slashed}
\usepackage{color}
\usepackage{natbib}
\usepackage{twoopt}
\usepackage{mdframed}
\usepackage{framed}
\usepackage{phystex}
\usepackage{enumitem}
\usepackage[pdftex,bookmarks=false,
breaklinks=true,colorlinks=true,linkcolor=red,urlcolor=cyan,citecolor=magenta]{hyperref}
\definecolor{shadecolor}{rgb}{1.0, 0.95, 0.7}
\definecolor{shadecolor_a}{rgb}{0.75,0.55,0.3}
\numberwithin{equation}{section}
\newcommand{\beq}{\begin{equation}}
\newcommand{\eeq}{\end{equation}}

\newcommand{\asv}{\mbox{$\langle \sigma_{\rm ann} {v} \rangle$}}
\newcommand{\asvMW}{\mbox{$\langle \sigma_{\rm ann} {v} \rangle_{\rm MW}$}}

\newcommand{\sigcol}{\mbox{$\sigma_{\rm col}$}}
\DeclareGraphicsExtensions{.png,.pdf}
\graphicspath{{./}{Figs/}{plots/}}
\makeatletter
\def\input@path{{Sections/}{./}}
\makeatother
\begin{document}
%
\title{GeV-scale dark matter with $p$-wave Breit-Wigner enhanced annihilation}
\author{Genevieve B\'elanger}
\email{genevieve.belanger@lapth.cnrs.fr}
\affiliation{LAPTh,
  Universit\'e Savoie Mont Blanc \& CNRS,
  Chemin de Bellevue,
  74941 Annecy Cedex --- France}

\author{Sreemanti Chakraborti}
\email{sreemanti.chakraborti@durham.ac.uk}
\affiliation{Institute for Particle Physics Phenomenology, Department of Physics,\\
            \quad Durham University, Durham, DH1 3LE, United Kingdom}

\author{Yoann G\'enolini}
\email{yoann.genolini@lapth.cnrs.fr}
\affiliation{LAPTh,
  Universit\'e Savoie Mont Blanc \& CNRS,
  Chemin de Bellevue,
  74941 Annecy Cedex --- France}

\author{Pierre Salati}
\email{pierre.salati@lapth.cnrs.fr}
\affiliation{LAPTh,
  Universit\'e Savoie Mont Blanc \& CNRS,
  Chemin de Bellevue,
  74941 Annecy Cedex --- France}

\preprint{LAPTH-002/24, IPPP/23/85}

\begin{abstract}
We consider a light scalar dark matter candidate with mass in the GeV range whose $p$-wave annihilation is enhanced through a Breit-Wigner resonance. The annihilation actually proceeds in the $s$-channel via a dark photon mediator whose mass is nearly equal to the masses of the incoming particles.
We compute the temperature at which kinetic decoupling between dark matter and the primordial plasma occurs and show that  including the effect of  kinetic decoupling  can reduce  the dark matter relic density by orders of magnitude. 
For typical scalar masses ranging from 200~MeV to 5~GeV, we determine the range of values allowed for the dark photon couplings to the scalar and to the standard model particles after requiring the relic density to be in agreement with the value extracted from cosmological observations. We then show that $\mu$ and $y$-distortions of the cosmic microwave background spectrum and X-ray data from XMM-Newton strongly constrain the model and  rule out the region where the dark matter annihilation cross-section is strongly enhanced  at small dispersion velocities.
Constraints from direct detection searches and from accelerator searches for dark photons offer complementary probes of the model.  
\end{abstract}


\maketitle

\section{Introduction}
\label{sec:intro}
Searches for dark matter (DM) have for the last  decades concentrated on a new weakly interacting particle at the electroweak scale.
This was partly motivated by theoretical preferences for new physics at the TeV scale and by the fact that the freeze-out mechanism for a new weakly interacting massive particle (WIMP)  at the electroweak scale naturally leads to the value for the DM relic density extracted from observations~\cite{Planck:2018vyg,Bertone:2004pz}. The lack of signals in (in-)direct detection or at colliders for new particles~\cite{Cirelli:2015gux,Schumann:2019eaa,Buchmueller:2017qhf} aroused interest for a wider range of DM candidates of different scales and/or interaction strengths.
In particular, on the theoretical side a plethora of DM models have emerged that feature the possibility of lighter DM, around the GeV scale or below~\cite{Boehm:2002yz,Boehm:2003hm,DAgnolo:2015ujb,Darme:2017glc,Knapen:2017xzo}.
These candidates largely escape the strongest constraints from direct detection since the minimum recoil energy they can impart to the nucleus falls below the detector threshold~\cite{XENON:2018voc,PandaX-4T:2021bab,LZ:2022lsv}. Efforts to improve the sensitivity for direct detection of light DM are being pursued by taking advantage of scattering on electrons or single phonon excitations in crystals~\cite{Griffin:2019mvc}.
Indirect searches for DM annihilation in the galactic halo or in dwarf spheroidal galaxies (dSPhs) are also mostly sensitive to DM masses above a few GeVs~\cite{Fermi-LAT:2015att,MAGIC:2016xys,Fermi-LAT:2016uux}. However, WIMPs in the MeV-GeV scale can be constrained from cosmic-ray electrons and positrons using Vogager~1 and AMS-02  data~\cite{Boudaud:2016mos,Boudaud:2018oya}. Moreover, light DM annihilating into $e^+e^-$ pairs will leave a signature in X-ray when the electron-positrons scatter on the low-energy photons in the Galaxy. It was shown recently that X-ray data from XMM-Newton can be used to constrain light DM~\cite{Cirelli:2023tnx,Foster:2021ngm}.

Strong constraints on light DM come from cosmology. DM annihilation deposits electromagnetic energy in the primordial plasma and impacts the anisotropies of the cosmic microwave background (CMB)~\cite{Cline:2013fm}. The precise measurements of the CMB by the PLANCK satellite thus put robust constraints on DM annihilation cross-sections into photons or charged particles. These constraints typically exclude the cross-section required to achieve the measured relic density when the DM mass lies below approximately 10~GeV under the assumption of $s$-wave DM annihilation~\cite{Slatyer:2015jla}.
However, such constraints are escaped easily if DM annihilation is $p$-wave, indeed the typical DM velocity during recombination ($v\approx 10^{-8}c$) is much smaller than the typical velocity at freeze-out, leading to a strong suppression of the cross-section at recombination.
The energy injected in the primordial plasma from DM annihilations can also induce deviations from the pure black-body spectrum of the CMB. Depending on the value of the redshift at which the energy injection occurs, the dominant effect is either $\mu$-distortion or, at lower redshifts, $y$-distortion~\cite{Chluba:2018cww}. FIRAS~\cite{Fixsen_1996,Bianchini:2022dqh} has set limits on both types of distortions resulting in important constraints for $p$-wave annihilating DM as we will see in the later sections.

When DM annihilation is dominated by $p$-wave, the cross-section for DM annihilation in galaxies is also suppressed relative to the one at freeze-out, since the DM velocity in the galactic halo ($v\approx 10^{-3}c$) or in dSphs ($v\approx 10^{-4}c$) is also much smaller than at freeze-out.
To have the possibility of a significant signal in indirect detection, one therefore needs to boost the DM annihilation cross-section with respect to its value in the early Universe.
The typical `boost' scenario occurs when DM annihilation proceeds through a $s$-channel exchange of a mediator ($X$) near resonance. The cross-section then strongly depends on the kinetic energy of the DM particles.  When the mass of the mediator is larger than that of a pair of DM particles, $m_{x} > 2 m_{\phi}$, and for a small mass splitting, the maximal enhancement is found at small relative velocities. This gives rise to a Breit-Wigner resonance (BW) enhancement~\cite{Ibe:2008ye,Guo:2009aj}. The boost from BW was exploited to generate large cross-sections for DM annihilation in galaxies for both  sub-GeV~\cite{Binder:2022pmf,Bernreuther:2020koj} or electroweak scale DM in a variety of models featuring $s$-wave~\cite{AlbornozVasquez:2011js,Duch:2017nbe,Feldman:2008xs} or $p$-wave annihilation~\cite{Ding:2021sbj}.

In this paper, we revisit the case of $p$-wave dominated dark matter annihilation for DM at the GeV scale in scenarios with Breit-Wigner enhancement.
For the theoretical framework, we consider a simplified model with a scalar DM coupled to a light vector particle, a `dark photon'~\cite{Holdom:1985ag,Ilten:2018crw,Filippi:2020kii}, (for an analysis of fermionic DM see~\cite{Ding:2021sbj}).
We first calculate the DM relic density corresponding to annihilation near a $s$-channel resonance, and we examine the dependence of DM annihilation on the DM dispersion velocity. This leads to orders of magnitude variations in the relic density~\cite{Bi:2011qm,Binder:2017rgn}.
We define the parameter space allowed by the relic density observations. For a given DM mass nearly half the dark photon mass, the relevant parameters are the couplings of the dark photon to DM and to standard model (SM) particles. We also include constraints on the dark photon coupling from collider and beam dump experiments~\cite{Ilten:2018crw}. We find that these offer complementary probes of the light DM scenario.

This paper is organized as follows. In section~\ref{sec:model}, we describe the model considered and present the DM annihilation cross-section and its dependence on the DM dispersion velocity and on the couplings of the dark photon. The relic abundance calculation is discussed in section~\ref{sec:relic} including the effect of kinetic decoupling. In section~\ref{sec:results} we calculate the allowed parameter space after taking into account the relic density constraint as well as constraints from CMB $\mu$ and $y$-distortions and indirect detection constraints. Section~\ref{sec:conclusion} contains our conclusions. The appendices contain more details on DM annihilation and on DM thermalization that are required to determine when kinetic decoupling occurs.
To help readers find what might be of interest to them, we summarize hereafter the salient results of our work.
\begin{itemize}
\item{We examine the behavior of the annihilation cross-section $\asv$ near a resonance\\
as a function of DM dispersion velocity (section~\ref{sec:model} and appendix~\ref{app:jsigmav}).}
\item{We derive approximate solutions for the evolution of the cosmological DM abundance\\
(section~\ref{subsec:omegah2}).}
\item{We take into account kinetic decoupling assuming that DM is thermally coupled with\\
itself (section~\ref{subsec:withkd} and appendix~\ref{sec:phi_thermalization}).}
\item{We determine the parameter space allowed by the relic density constraint (section~\ref{subsec:lim_gx_epsilon}).}
\item{We impose constraints from DM annihilation in the Milky Way and from DM distorting\\
the CMB (sections~\ref{subsec:lim_xs_mw} and \ref{subsec:lim_gx_epsilon_mudistorsion}).}
\item{We set constraints from DM direct detection and colliders (section~\ref{subsec:lim_gx_epsilon_other}).}
\item{We find that several regions of parameter space evade all the constraints. Summary\\
plots are given in section~\ref{subsubsec:sumplots}.}
\end{itemize}

\section{The generic model}
\label{sec:model}
The Lagrangian which we consider describes a vector boson portal, between DM and the SM sector, in the presence of a Breit-Wigner resonance which may enhance DM annihilations. This Lagrangian contains the couplings of a new photon $A'^{\mu}$, dubbed dark, to ordinary fermions $f$ and to a new complex scalar field $\phi$ which plays the role of the DM candidate. The scalars $\phi$ and their anti-partners $\bar{\phi}$ are charged under the new local gauge group ${\rm U'}(1)$ whose associated gauge boson is the dark photon $A'^{\mu}$. The model to be explored is described by the interaction Lagrangian
\beq
{\cal L}_{\rm int} =
- \left\{ A'_{\mu} J_{\phi}^{\mu} \equiv
i g_{x} A'^{\mu} \left( \phi^{\dagger} \partial_{\mu} \phi - \partial_{\mu} \phi^{\dagger} \phi \right) \right\}
- \epsilon e Q_{\!f} \bar{f} \slashed A' f \,.
\label{eq:Lagrangian_1}
\eeq
The DM species $\phi$ and $\bar{\phi}$ interact weakly on short distances through the exchange of the massive vector boson $A'^{\mu}$. They respectively carry the dark charges $+g_{x}$ and $-g_{x}$. The hidden sector has a broken ${\rm U'}(1)$ symmetry, which facilitates the rotation between true $A^{\mu}$ and dark $A'^{\mu}$ photon states. This explains the presence of the mixing angle $\epsilon$ with which the dark photon couples to ordinary fermions in the second term of the right-hand side of the previous equation. To stabilise the DM, we further impose a discrete $Z_2$ symmetry, where we assume that $\phi$ and $\bar{\phi}$ are odd under $Z_2$, whereas $A_\mu^\prime$ and all other SM particles are even under the symmetry.\footnote{We neglect interaction of the type $\phi\phi^{\dagger} H^2$, where $H$ is the SM Higgs doublet.}

We can readily extract from the Lagrangian~\ref{eq:Lagrangian_1} the cross-section of the annihilation of DM species into ordinary fermions through the reaction
\beq
\phi + \bar{\phi} \to f + \bar{f} \,.
\eeq
The annihilation is mediated by the dark photon $A'^\mu$ exchanged in the $s$-channel. Let us denote by $\vec{v}_{rel}$ the difference $\vec{v}_{\phi} - \vec{v}_{\bar{\phi}}$ between the $\phi$ and $\bar{\phi}$ velocities. A straightforward calculation leads to
\beq
\sigma_{\rm ann} {v} = {\displaystyle \frac{g_{x}^{2} \epsilon^{2} e^{2} \tilde{Q}^{2}}{6 \pi}} \left\{
{\displaystyle \frac{m_{\phi}^{2} {v}^{2}}{(4 m_{\phi}^{2} - m_{x}^{2} + m_{\phi}^{2} {v}^{2})^{2} + m_{x}^{2} \Gamma_{x}^{2}}}
\right\} \,,
\eeq 
where ${v}$ is the norm $|| \vec{v}_{rel} ||$ of the velocity difference $\vec{v}_{rel}$. The masses of the DM scalars $\phi$ and $\bar{\phi}$, and of the dark photon, are respectively denoted by $m_{\phi}$ and $m_{x}$, while $\Gamma_{x}$ is the decay width of the dark photon. The effective dimensionless charge $\tilde{Q}$ is defined by
\beq
\tilde{Q}^{2} = {\displaystyle \sum_{f}} \left\{1 - \frac{4 m_{f}^{2}}{s} \right\}^{\!1/2} \! \left\{ 1 + \frac{2 m_{f}^{2}}{s} \right\}  Q_{\!f}^{2}\,,
\label{eq:definition_tilde_Q_a}
\eeq
where the sum runs over the fermions $f$ produced by the annihilation, i.e. the fermions whose mass $m_{f}$ is less than ${\sqrt{s}}/{2}$. The incoming species $\phi$ and $\bar{\phi}$ are non-relativistic, i.e. ${v}$ is very small compared to the speed of light. As a consequence, the square of the center-of-mass energy $\sqrt{s}$ reduces to
\beq
s = 4\, m_{\phi}^{2} + m_{\phi}^{2} {v}^{2} \,,
\eeq
and, in the non-relativistic limit, the effective charge $\tilde{Q}$ can be defined as
\beq
\tilde{Q}^{2} = {\displaystyle \sum_{f}}
\left\{1 - \frac{m_{f}^{2}}{m_{\phi}^{2}} \right\}^{\!1/2} \! \left\{ 1 + \frac{m_{f}^{2}}{2 m_{\phi}^{2}} \right\}  Q_{\!f}^{2}
\,\,\,\text{with}\,\,\,
m_{f} \leq m_{\phi} \,.
\label{eq:definition_tilde_Q_b}
\eeq

For the relic density computation, we are interested in the product $\sigma_{\rm ann} {v}$ averaged over the velocity distributions of $\phi$ and $\bar{\phi}$ species. There is no asymmetry between these components and the velocity distributions for $\vec{v}_{\phi}$ and $\vec{v}_{\bar{\phi}}$ are taken identical. The thermal average of $\sigma_{\rm ann} {v}$ is defined as
\beq
\asv = {\displaystyle \iint} {\rm d^{3}}\vec{v}_{\phi} \, {\rm d^{3}}\vec{v}_{\bar{\phi}} \, f(\vec{v}_{\phi}) \, f(\vec{v}_{\bar{\phi}}) \; \sigma_{\rm ann} {v} \,.
\label{eq:definition_asv}
\eeq
As discussed in Sec.~\ref{sec:relic}, we assume hereafter that velocities are distributed according to a Maxwellian distribution
\beq
f(\vec{v}) = (2 \pi \Sigma^{2})^{-3/2} \exp (- {\vec{v}^{2}} / {2 \Sigma^{2}}) \,,
\eeq
where $\Sigma$ is the one-dimensional dispersion velocity. 
The integral~\ref{eq:definition_asv} can be carried out analytically by swapping the velocities $\vec{v}_{\phi}$ and $\vec{v}_{\bar{\phi}}$ for the variables
\beq
\vec{v}_{G} = {\displaystyle \frac{\vec{v}_{\phi} + \vec{v}_{\bar{\phi}}}{2}}
\,\,\,\text{and}\,\,\,
\vec{v}_{rel} = \vec{v}_{\phi} - \vec{v}_{\bar{\phi}} \,,
\eeq
and by noticing that ${\rm d^{3}}\vec{v}_{\phi} \, {\rm d^{3}}\vec{v}_{\bar{\phi}} \equiv {\rm d^{3}}\vec{v}_{G} \, {\rm d^{3}}\vec{v}_{rel}$. Integrating out the barycentric velocity $\vec{v}_{G}$ yields the thermally averaged cross-section as follows
\beq
\asv = {\displaystyle \frac{g_{x}^{2} \epsilon^{2} e^{2} \tilde{Q}^{2}}{12 \pi}}
\left\{ {\displaystyle \frac{J(a,b)}{m_{\phi}^{2} \Sigma^{2}}} \right\} \,,
\label{eq:asv_1}
\eeq
The parameters $a$ and $b$ are defined as
\beq
a = - \, {\displaystyle \frac{m_{x}^{2}}{4 m_{\phi}^{2}}} \; {\displaystyle \frac{\Sigma_{0}^{2}}{\Sigma^{2}}}
\;\;\;\text{and}\;\;\;
b = {\displaystyle \frac{m_{x}^{2}}{4 m_{\phi}^{2}}} \; {\displaystyle \frac{\Lambda_{0}^{2}}{\Sigma^{2}}} \,
\label{eq:definition_a_b}
\eeq
where
\beq
\Sigma_{0}^{2} \equiv 1 - {4 m_{\phi}^{2}}/{m_{x}^{2}}
\;\;\;\text{while}\;\;\;
\Lambda_{0}^{2} \equiv \Gamma_{x}/m_{x} \,.
\label{eq:definition_Sigma_02_Lambda_02}
\eeq
They are related to the function $J$ through the integrals
\beq
J(a,b) = \frac{2}{\sqrt{\pi}\,} {\displaystyle \int_{0}^{+ \infty}} \!\!
{\displaystyle \frac{u^{4\,} e^{-u^{2}}}{(u^{2} + a)^{2} + b^{2}}} \; du =
\frac{1}{\sqrt{\pi}\,} {\displaystyle \int_{0}^{+ \infty}} \!\!
{\displaystyle \frac{t^{3/2\,} e^{-t}}{(t + a)^{2} + b^{2}}} \; dt \,.
\label{eq:definition_J_a_b}
\eeq
The dark photon decay width $\Gamma_{x}$ can be expressed as
\beq
\Gamma_{x} = {\displaystyle \frac{m_{x}}{12 \pi}}
\left\{ {\displaystyle \frac{g_{x}^{2}}{4}} \left( 1 - {4 m_{\phi}^{2}}/{m_{x}^{2}} \right)^{3/2} + \epsilon^{2} e^{2} {Q'}^{2} \right\} =
{\displaystyle \frac{m_{x}}{12 \pi}}
\left\{ {\displaystyle \frac{g_{x}^{2}}{4}} \, \Sigma_{0}^{3} + \epsilon^{2} e^{2} {Q'}^{2} \right\}.
\label{eq:xdecay}
\eeq
The terms inside brackets respectively refer to decays into $\phi \bar{\phi}$ pairs and into fermions. The new effective dimensionless charge $Q'$ is defined as
\beq
{Q'}^{2} = {\displaystyle \sum_{f}}
\left\{1 - \frac{4 m_{f}^{2}}{m_{x}^{2}} \right\}^{\!1/2} \! \left\{ 1 + \frac{2 m_{f}^{2}}{m_{x}^{2}} \right\}  Q_{\!f}^{2}
\;\;\;\text{with}\;\;\;
m_{f} \leq {m_{x}}/{2} \,.
\label{eq:definition_Q_prime_a}
\eeq
In our parameter region of interest, $m_{x}$ is slightly larger than $2 m_{\phi}$ and ${Q'}$ boils down to the charge $\tilde{Q}$ as defined in relation~\ref{eq:definition_tilde_Q_b}.

From the definitions of $a$ and $b$, it is obvious that the sign of $a$ determines whether annihilation takes place above or below the resonance. For both cases, however, we can obtain interesting approximations while computing $J(a,b)$ and eventually, $\langle \sigma_{\rm ann} v\rangle$. The details regarding this are given in Appendix~\ref{app:jsigmav}.

\begin{figure}[h!]
\centering
\includegraphics[width=0.65\textwidth]{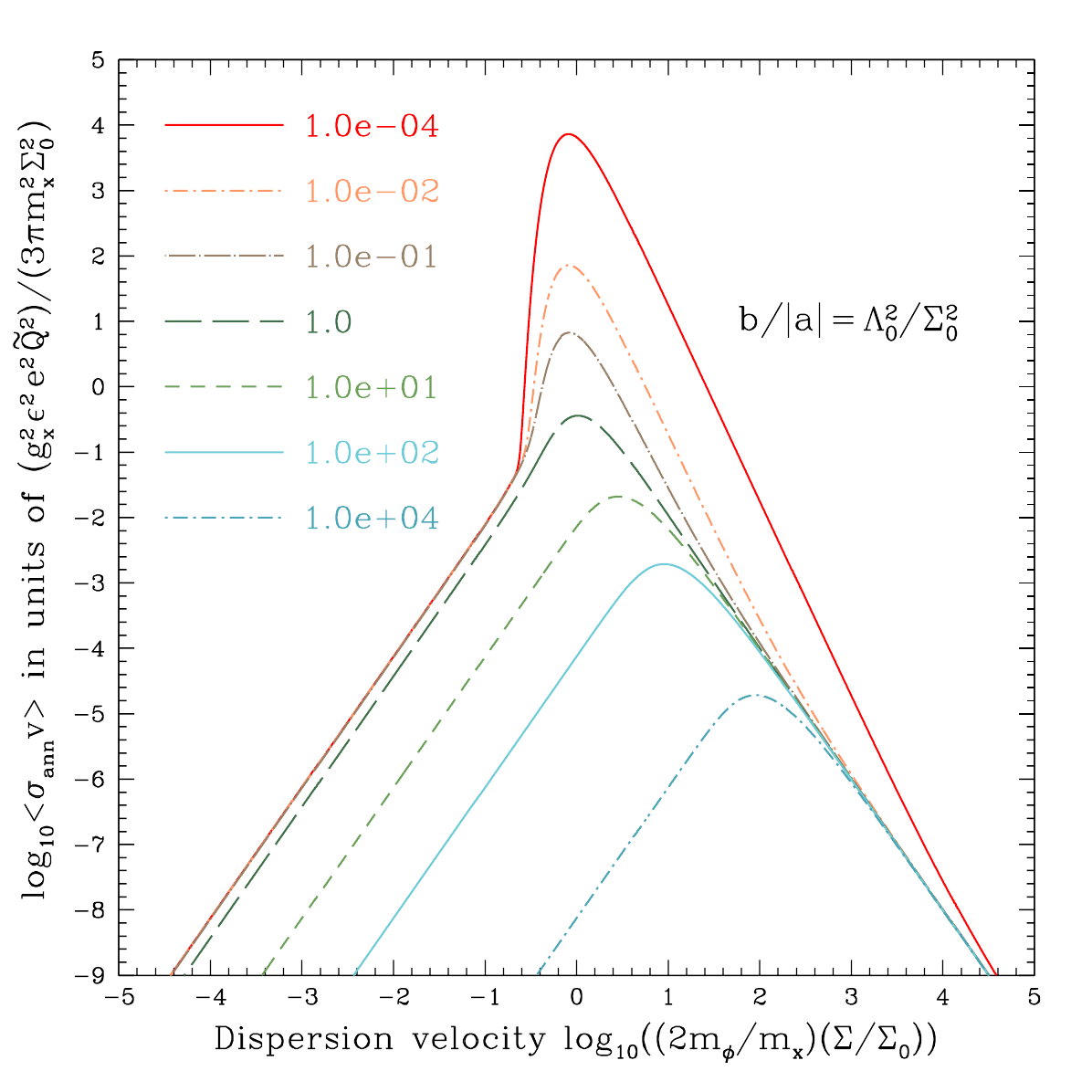}
\caption{The variation of the annihilation cross-section as a function of the dispersion velocity $\Sigma$ for different values of $b/|a|=\Lambda_0^2/\Sigma_0^2$.
On the horizontal axis, the rescaled variable $({2 m_{\phi}}/{m_{x}})({\Sigma}/{\Sigma_{0}})$ has been used.
When $\Lambda_{0}$ is smaller than $\Sigma_{0}$, the cross-section is enhanced by a Breit-Wigner resonance. Above a velocity of order $\Sigma_{0}$, where its peak value is reached, $\asv$ drops like $\Sigma^{-3}$ to reach the asymptotic behavior $\Sigma^{-2}$. Below the peak, the $p$-wave annihilation regime sets in and $\asv$ is proportional to $\Sigma^{2}$. For large values of $\Lambda_{0}$ with respect to $\Sigma_{0}$, the two asymptotic regimes only appear.
}
\label{fig:J_figure_3}
\end{figure}

In this work, we concentrate on the case where the dark photon mass $m_{x}$ is larger than $2 m_{\phi}$.
Fig.~\ref{fig:J_figure_3} shows how the annihilation cross-section $\asv$ depends on the one-dimensional dispersion velocity $\Sigma$ when $a<0$, i.e. when the mass gap $\Delta=m_x-2m_\phi$ is positive. As long as $\Delta$ is small with respect to $m_\phi$, the mass degeneracy parameter $\Sigma_0^2$ is nearly equal to the ratio $\Delta/m_\phi$. Substituting $\Sigma$ from Eq.~\ref{eq:definition_a_b} into relation~\ref{eq:asv_1}, one obtains
\beq
\asv = {\displaystyle \frac{g_{x}^{2} \epsilon^{2} e^{2} \tilde{Q}^{2}}{3 \pi m_{x}^{2} \Sigma_0^{2}}} \times
\left\{ |a| J(a,b) \right\} \,,
\label{eq:asv_2}
\eeq
In Fig.~\ref{fig:J_figure_3}, we plot the term inside brackets as a function of ${1}/{\sqrt{|a|}} \equiv ({2 m_{\phi}}/{m_{x}})({\Sigma}/{\Sigma_{0}})$.
For $\Lambda_{0}$ values that are large with respect to $\Sigma_{0}$, i.e. when $\Gamma_x/2$ is larger than the mass gap $\Delta$, the cross-section is $p$-wave suppressed at low velocities and increases like $\Sigma^{2}$. It reaches its maximum for a velocity of order $\Lambda_{0}$ above which it decreases like $\Sigma^{-2}$. At high velocities, $\asv$ behaves actually as if it was Sommerfeld enhanced. The model has this intriguing and fascinating property to predict together, albeit in different velocity regimes, a $p$-wave suppression as well as a Sommerfeld-like behavior.
For values of $\Lambda_{0}$ small with respect to $\Sigma_{0}$, i.e. when $\Gamma_x/2$ is smaller than the mass gap $\Delta$, a Breit-Wigner resonance comes also into play, which increases dramatically the annihilation cross-section as featured by the bumps of Fig.~\ref{fig:J_figure_3}. The maximum is reached at a velocity of $\Sigma_{\rm M}=\sqrt{2/3}\,(m_x/{2m_\phi})\,\Sigma_{0}$, above which $\asv$ behaves as $\Sigma^{-3}$. At low and high velocities, the annihilation cross-section scales respectively as $\Sigma^{2}$ ($p$-wave) and $\Sigma^{-2}$ (high-velocity) like in the large $\Lambda_{0}$ case.

\section{Scalar dark matter relic abundance}
\label{sec:relic}
From now on, we will assume that scalar DM is in thermal and chemical equilibrium with the primordial plasma when its temperature is of order the DM mass $m_{\phi}$. This is the starting point of our analysis. Going beyond this assumption would require the knowledge of the complete theory in order to determine the thermal behavior of DM at much earlier times than those considered here.
Because there is potentially an infinite number of such theories, the scope of this article is to focus on the cosmological consequences of a vector boson portal with Breit-Wigner resonance as encoded by the Lagrangian~\ref{eq:Lagrangian_1}. The model has only a few parameters, namely the dark charge $g_{x}$, the mixing angle $\epsilon$ and the masses $m_{\phi}$ and $m_{x}$, but its phenomenology is already very rich and subtle as we shall see.


Assuming that scalar DM is in thermodynamical equilibrium with the SM plasma while becoming non-relativistic is quite conceivable. To commence, in the portion of the parameter space where the couplings $g_{x}$ and $\epsilon$ are not too small, scalar DM collides upon, and annihilates into, SM fermions so efficiently that thermodynamical equilibrium ensues at early times.
If this is not so, we can assume that the kinetic mixing $\epsilon$ between the visible and dark photons is triggered by radiative corrections implying loops of heavy dark species $\Psi$ and $\Psi'$ with opposite electric or dark charges. At very early times, these particles are relativistic and interact efficiently with both scalar DM and SM species through collisions and annihilations, allowing for the coupling between these components.
Notice also that the bulk of DM annihilation takes place well after freeze-out and is most efficient at late times, when the Breit-Wigner resonance becomes active. The DM density at freeze-out turns out not to be relevant to compute the relic abundance. This alleviates the problem of determining the actual thermodynamical state of DM before freeze-out.


We will also assume that scalar DM reaches at all times inner thermal equilibrium so that its velocity distribution is well described by a Maxwell-Boltzmann law. Such a condition could be established through collisions of $\phi$ and $\bar{\phi}$ particles with SM fermions, provided that energy is transferred sufficiently rapidly between these components. But this turns out not to be the case in a large portion of the parameter space where DM kinetic decoupling occurs. To simplify an already intricate analysis, we have assumed that the $\phi$ and $\bar{\phi}$ populations reach thermal equilibrium through mutual collisions. This allows us to define a common temperature $T_{\phi}$ for scalar DM which may be different from the plasma temperature $T$ after kinetic decoupling has occurred. Notice that if the dark charge $g_{x}$ is not too small, DM scalars $\phi$ and $\bar{\phi}$ could efficiently collide upon each other through the exchange in the $t$-channel of a virtual dark photon. DM needs to be dense enough though. Alternatively, a self-coupling \`a la $\phi^{4}$ could also lead to such a behavior. The annihilation cross-section of scalar DM into SM species can then be averaged over a Maxwell-Boltzmann distribution of DM velocities, yielding the result of Sec.~\ref{sec:model}.


In this section, we derive the relic abundance $\Omega_{\phi} h^{2}$ of scalar DM as a function of the parameters of the model.
In Sec.~\ref{subsec:omegah2}, we discuss the master equation that drives DM freeze-out and explain how we solve it.
We then present numerical results together with approximate solutions that help understanding how $\Omega_{\phi} h^{2}$ depends on the mass degeneracy parameter $\Sigma_{0}^{2}$. We find three different regimes.
In Sec.~\ref{subsec:nokd}, we assume that scalar DM stays in thermal contact with the SM plasma throughout the entire freeze-out process.
This assumption is abandoned in Sec.~\ref{subsec:withkd} where we determine when scalar DM decouples from kinetic equilibrium and analyse how this strongly affects the relic abundance $\Omega_{\phi} h^{2}$.

\subsection{Calculation of the relic abundance $\Omega_{\phi} h^2$}
\label{subsec:omegah2}
The scalar DM particles $\phi$ and $\bar{\phi}$ can annihilate into, and be produced from, SM fermions through the process
\beq
\phi + \bar{\phi} \rightleftharpoons f + \bar{f} \,.
\label{eq:chemical_equilibrium}
\eeq
Should direct and reverse reactions be fast enough, a chemical equilibrium is established. When the plasma temperature $T$ drops below the DM mass $m_{\phi}$, $\phi \bar{\phi}$ pairs are less and less easily produced. DM still goes on annihilating until it is so depleted that its codensity remains constant until today.
We assume that there is no asymmetry between the $\phi$ and $\bar{\phi}$ populations and denote by $n$ their densities $n_{\phi} = n_{\bar{\phi}}$. The evolution of the density $n$ with time $t$ is described by the freeze-out equation
\begin{equation}
\frac{dn}{dt} = - 3 H n \, - \, \asv n^{2} \, + \, \asv n_{\rm eq}^{2} \,,
\label{eq:freeze_out_1}
\end{equation}
where $H$ is the expansion rate of the Universe. In its right-hand side, the three terms respectively stand for the dilution resulting from the expansion of the Universe, the annihilations of $\phi \bar{\phi}$ pairs and the back reactions which regenerate scalar DM from lighter species. We have assumed detailed balance for this last process, hence the density $n_{\rm eq}$ as fixed by the thermodynamical equilibrium.
It is convenient to deal with the codensity $\tilde{n}$, i.e. the density inside a volume that expands with the expanding Universe. We define it as the ratio
\beq
\tilde{n} = \frac{n}{\theta^{3}} \,,
\eeq
where $\theta$ is a fictitious temperature which decreases as the inverse $a^{-1}$ of the scale factor $a$ of the Universe, and which is normalized in such a way that it is equal to the plasma temperature $T_{\rm F}$ at freeze-out. Notice that since the entropy of the plasma is constant in time, the factor $\theta^{3}$ is proportional to the entropy density $s = (2 \pi^{2} / 45)_{\,} h_{\rm eff\,} T^{3}$, where $h_{\rm eff}(T)$ is the effective number of entropic degrees of freedom. Our codensity $\tilde{n}$ is similar to the usual definition ${n}/{s}$ used in the literature. Relation~\ref{eq:freeze_out_1} can be recast into
\beq
\frac{d \tilde{n}}{dt} \, + \, \left\{ \asv n \right\} \tilde{n} = \asv \theta^{3} \tilde{n}_{\rm eq}^{2} \,.
\label{eq:freeze_out_2}
\eeq

This form allows for a simple interpretation of freeze-out as a mere relaxation process during which $\tilde{n}$ runs after its chemical equilibrium $\tilde{n}_{\rm eq}$ and eventually fails to reach it. The rate $\Gamma_{\rm rel}^{\rm F} \equiv \asv n$ at which $\tilde{n}$ relaxes toward $\tilde{n}_{\rm eq}$ is actually the annihilation rate. The right-hand side term of relation~\ref{eq:freeze_out_2}, i.e. the target which represents chemical equilibrium, evolves with the typical rate
\beq
\Gamma_{\rm eq}^{\rm F} \equiv \left| \frac{d}{dt} \! \ln \left\{ \asv \theta^{3} \tilde{n}_{\rm eq}^{2} \right\} \right|.
\eeq
A straightforward calculation yields
\beq
\Gamma_{\rm eq}^{\rm F} = H
\left\{ 2x - \frac{d \ln \asv}{d \ln x} - \frac{d \ln h_{\rm eff}}{d \ln T} \right\} / \left\{ 1 + \frac{1}{3} \frac{d \ln h_{\rm eff}}{d \ln T} \right\}
\sim 2 x H \,,
\label{eq:definition_Gamma_eq_F}
\eeq
where the parameter $x$ denotes the DM mass to plasma temperature ratio ${m_{\phi}}/{T}$. The Hubble expansion rate $H$ is related to the plasma temperature through
\beq
H = \frac{\left( 2 \pi \right)^{3/2}}{3 \sqrt{10}} \sqrt{g_{\rm eff}} \; \frac{T^{2}}{M_{\rm P}} \,,
\eeq
where $M_{\rm P}$ is the Planck mass and $g_{\rm eff}(T)$ is the effective number of energetic degrees of freedom of the plasma.
In deriving \ref{eq:definition_Gamma_eq_F}, we have noticed that ${d \ln \theta}/{dt} \equiv - H$ and used the identity
\beq
\frac{d \ln \theta}{d \ln T} = 1 + \frac{1}{3} \frac{d \ln h_{\rm eff}}{d \ln T} \,.
\eeq
The derivative of the annihilation cross-section with respect to the temperature depends on the one-dimensional DM dispersion velocity $\Sigma$. In the high-velocity regime where $\Sigma$ is well above $\Sigma_{0}$, the cross-section $\asv$ scales as ${1}/{\Sigma^{2}}$, hence a value for ${d \ln \asv}/{d \ln x}$ of $1$. In the Breit-Wigner enhancement regime just above $\Sigma_{0}$, this derivative reaches ${3}/{2}$ while in the low-velocity $p$-wave regime, it decreases down to $-1$.


At high temperature, the relaxation rate $\Gamma_{\rm rel}^{\rm F}$ is much larger than the evolution rate $\Gamma_{\rm eq}^{\rm F}$. The chemical equilibrium~\ref{eq:chemical_equilibrium} is established. The DM density $n$ relaxes toward its equilibrium value $n_{\rm eq}$ much faster than the latter evolves.
As time goes on, the relaxation/annihilation rate $\Gamma_{\rm rel}^{\rm F}$ decreases very rapidly. It actually drops exponentially with plasma temperature insofar as it is proportional to the DM density
\beq
n_{\rm eq} = T^{3} \left( {x}/{2 \pi} \right)^{3/2} \, \exp(-x) \,.
\label{eq:n_e_phi}
\eeq
The rate $\Gamma_{\rm eq}^{\rm F}$ decreases approximately like $T$, at a much slower pace than $\Gamma_{\rm rel}^{\rm F} \propto \exp(-{m_{\phi}}/{T})$. At some critical temperature $T_{\rm F}$, both rates are equal. Freeze-out occurs. Scalar DM decouples from chemical equilibrium insofar as its density $n$ no longer relaxes toward $n_{\rm eq}$. The freeze-out point $x_{\rm F} \equiv {m_{\phi}}/{T_{\rm F}}$ satisfies the equality
\beq
\Gamma_{\rm rel}^{\rm F}(x_{\rm F}) = \Gamma_{\rm eq}^{\rm F}(x_{\rm F}) \,.
\label{eq:freeze_out_x_F}
\eeq
This equation must be solved numerically for each set of model parameters, i.e. once the masses $m_{\phi}$ and $m_{x}$, the coupling $g_{x}$ and the mixing angle $\epsilon$ are given. Expressions~\ref{eq:asv_1}, \ref{eq:definition_a_b} and \ref{eq:definition_J_a_b} are used to calculate the average annihilation cross-section $\asv$.
Assuming that scalar DM is in thermodynamical equilibrium with the primordial plasma until it becomes non-relativistic imposes that $x_{\rm F}$ is larger than $1$. We checked that the solution of equation~\ref{eq:freeze_out_x_F} actually fulfills this condition in a large portion of parameter space, and noticeably in the domain that survives our cosmological analysis.


After freeze-out, the density $n_{\rm eq}$ vanishes rapidly and so does the right-hand side of relation~\ref{eq:freeze_out_2}. We can neglect it in deriving the final DM abundance, especially as most of the annihilation takes place well after freeze-out. The DM codensity today $\tilde{n}_{0}$ is given by the relation
\beq
\frac{1}{\tilde{n}_{0}} = \frac{1}{\tilde{n}_{\rm F}} + {\cal I}_{\rm ann}
\;\;\;\text{where}\;\;\;
{\cal I}_{\rm ann} \equiv \int_{t_{\rm F}}^{t_{0}} \! {\asv}_{\,} \theta^{3} dt \,.
\label{eq:n_tilde_0_a}
\eeq
The codensity at freeze-out is $\tilde{n}_{\rm F} \equiv \tilde{n}_{e}(T_{\rm F})$. Equation~\ref{eq:freeze_out_2} is integrated without its right-hand side from time $t_{\rm F}$, at which freeze-out occurs, until the present age $t_{0}$ of the Universe. The annihilation integral ${\cal I}_{\rm ann}$ may be conveniently recast in terms of an integral over the parameter $y = {T}/{m_{\phi}}$, where $T$ is the SM plasma temperature. This yields
\beq
{\cal I}_{\rm ann} = \frac{3 \sqrt{10}}{\left( 2 \pi \right)^{3/2}} \, {\displaystyle \int_{y_{0}}^{y_{\rm F}}} dy \,
\left\{ \frac{{\asv} M_{\rm P} m_{\phi}}{\sqrt{g_{\rm eff}(T)}} \right\}
\left\{ \frac{\theta^{3}}{T^{3}} \right\}
\left\{ 1 + \frac{1}{3} \frac{d \ln h_{\rm eff}}{d \ln T} \right\}
\;\;\;\text{where}\;\;\;
\frac{\theta^{3}}{T^{3}} \equiv \frac{h_{\rm eff}(T)}{h_{\rm eff}(T_{\rm F})} \,.
\label{eq:I_ann_1}
\eeq
The parameter $y$ runs from $y_{\rm F} = {1}/{x_{\rm F}}$ down to $y_{0} = {T_{0}}/{m_{\phi}}$, where the CMB temperature $T_{0} = 2.72548 \, {\rm K}$ has been borrowed from~\cite{Fixsen:2009ug}.
Once the DM codensity at the present epoch is known, we readily derive the number density
\beq
n_{0} = \tilde{n}_{0\,} \theta_{0}^{3} \equiv \tilde{n}_{0\,} T_{0}^{3} \left\{ \frac{h_{\rm eff}(T_{0})}{h_{\rm eff}(T_{\rm F})} \right\},
\label{eq:n_0}
\eeq
and DM mass density $\rho_{\phi}^{0} = 2_{\,}m_{\phi} n_{0} $, remembering that there are as many $\phi$ as $\bar{\phi}$ particles. Scalar DM contributes today to the Universe mass budget a fraction
\beq
\Omega_{\phi} h^{2} = \frac{\rho_{\phi}^{0}}{\rho_{\rm C}^{0}}
\;\;\;\text{where}\;\;\;
\rho_{\rm C}^{0} = \frac{3 H_{0}^{2}}{8 \pi G} \,.
\label{eq:Oh2}
\eeq
Newton's constant of gravity is $G$ while $H_{0}$ stands for a fiducial Hubble expansion rate of $100 \, {\rm km/s/Mpc}$. The actual Hubble constant is equal to $h$ in units of that benchmark value. We would like DM to be made of the scalar species $\phi$ and $\bar{\phi}$. That is why we will look in Sec.~\ref{sec:results} for configurations of parameters for which $\Omega_{\phi} h^{2}$ is equal to the cosmological measured value of $\Omega_{\rm DM} h^{2} = 0.1200$~\cite{Planck:2018vyg}.
A key ingredient in the calculation of ${\cal I}_{\rm ann}$ is the dependence of $\asv$ on the one-dimensional dispersion velocity $\Sigma$ of scalar DM, and eventually on the parameter $y$. Scalar DM is assumed to have reached inner thermalization with temperature $T_{\phi}$, hence the identity $\Sigma^{2} \equiv {T_{\phi}}/{m_{\phi}}$ in the non-relativistic limit. But the relation between DM temperature $T_{\phi}$ and plasma temperature $T$ remains to be determined. It will prove to be of paramount importance.

\subsection{Approximate expression for the relic abundance $\Omega_{\phi} h^2$}
\label{subsec:Oh2_proxy}
Before solving numerically for $\Omega_{\phi} h^{2}$, we derive approximate expressions for ${\cal I}_{\rm ann}$. These will be helpful to discuss and understand our numerical results. We will set the DM mass $m_{\phi}$, the coupling $g_{x}$ and the mixing angle $\epsilon$ constant and will concentrate on how ${\cal I}_{\rm ann}$, and eventually the DM relic abundance, vary with the mass degeneracy parameter $\Sigma_{0}^{2} \equiv 1 - {4 m_{\phi}^{2}}/{m_{x}^{2}}$.

First, we remark that the annihilation cross-section $\asv$ is not constant after freeze-out. As discussed in Sec.~\ref{sec:model}, it increases with time like ${1}/{\Sigma^{2}}$ in the high-velocity regime, or ${1}/{\Sigma^{3}}$ at the Breit-Wigner resonance, as plasma and DM temperatures decrease. It reaches a maximum when the DM dispersion velocity $\Sigma$ is equal to some critical value $\Sigma_{\rm M}$. This occurs at time $t_{\rm M}$ when the plasma temperature is $T_{\rm M}$ and the parameter $y$ is equal to $y_{\rm M} = {T_{\rm M}}/{m_{\phi}}$. Afterward, the annihilation becomes $p$-wave dominated and $\asv$ drops rapidly like $\Sigma^{2}$ to vanish. In expression~\ref{eq:I_ann_1}, the main contribution to the integral ${\cal I}_{\rm ann}$ is provided by values of $y$ between $y_{\rm M}$ and $y_{\rm F}$. From time $t_{\rm M}$ until today, DM essentially does not annihilate.
We also remark that the thermodynamical coefficients $g_{\rm eff}$, $h_{\rm eff}$ and ${d \ln h_{\rm eff}}/{d \ln T}$ vary slowly in time. They can be evaluated at time $t_{\rm M}$ and taken out of the integral over $y$.
Finally, as most of DM annihilation takes place well after freeze-out, the DM codensity $\tilde{n}_{\rm F}$ is much larger than the present codensity $\tilde{n}_{0}$ and can be removed from expression~\ref{eq:n_tilde_0_a}.
Taking into account these remarks and inserting expression~\ref{eq:asv_1} into integral~\ref{eq:I_ann_1} yields
\beq
\frac{1}{\tilde{n}_{0}} \simeq
{\cal I}_{\rm ann} \simeq {\frac{\sqrt{5/2}}{\left( 2 \pi \right)^{5/2}}}
\left\{ {g_{x}^{2} \epsilon^{2} e^{2} \tilde{Q}^{2}} \right\} \left\{ \frac{M_{\rm P}}{m_{\phi}} \right\}
\left\{ \frac{{\cal P}(T_{\rm M})}{h_{\rm eff}(T_{\rm F})} \right\} {\cal J}_{\rm ann}
\;\;\;\text{with}\;\;\;
{\cal J}_{\rm ann} \equiv
{\displaystyle \int_{y_{\rm M}}^{y_{\rm F}}} \! dy \, {\displaystyle \frac{J(a,b)}{\Sigma^{2}}} \,,
\label{eq:definition_cal_J_ann}
\eeq
where the plasma function ${\cal P}$ is defined as
\beq
{\cal P}(T) =  \frac{h_{\rm eff}(T)}{\sqrt{g_{\rm eff}(T)}}
\left\{ 1 + \frac{1}{3} \! \left. \frac{d \ln h_{\rm eff}}{d \ln T} \right|_{T} \right\}.
\eeq
The DM relic abundance may be approximated by
\beq
\Omega_{\phi} h^{2} \simeq
{\frac{{16}{\pi}^{5/2}}{\sqrt{5}}}
\left\{ \frac{{\cal F}/{{\cal J}_{\rm ann}}}{g_{x}^{2} \epsilon^{2} e^{2} \tilde{Q}^{2}} \right\}
\;\;\;\text{where}\;\;\;
{\cal F} = \left\{ \frac{m_{\phi} T_{0}^{3}}{\rho_{\rm C}^{0}} \right\}
\left\{ \frac{m_{\phi}}{M_{\rm P}} \right\}
\left\{ \frac{h_{\rm eff}(T_{0})}{{\cal P}(T_{\rm M})} \right\}.
\label{eq:Oh2_approx_a}
\eeq
It is proportional to the inverse of ${\cal J}_{\rm ann}$. Understanding how this integral varies with $\Sigma_{0}^{2}$ is paramount to our problem.
%
\begin{figure}[b!]
\centering
\includegraphics[width=0.6\columnwidth]{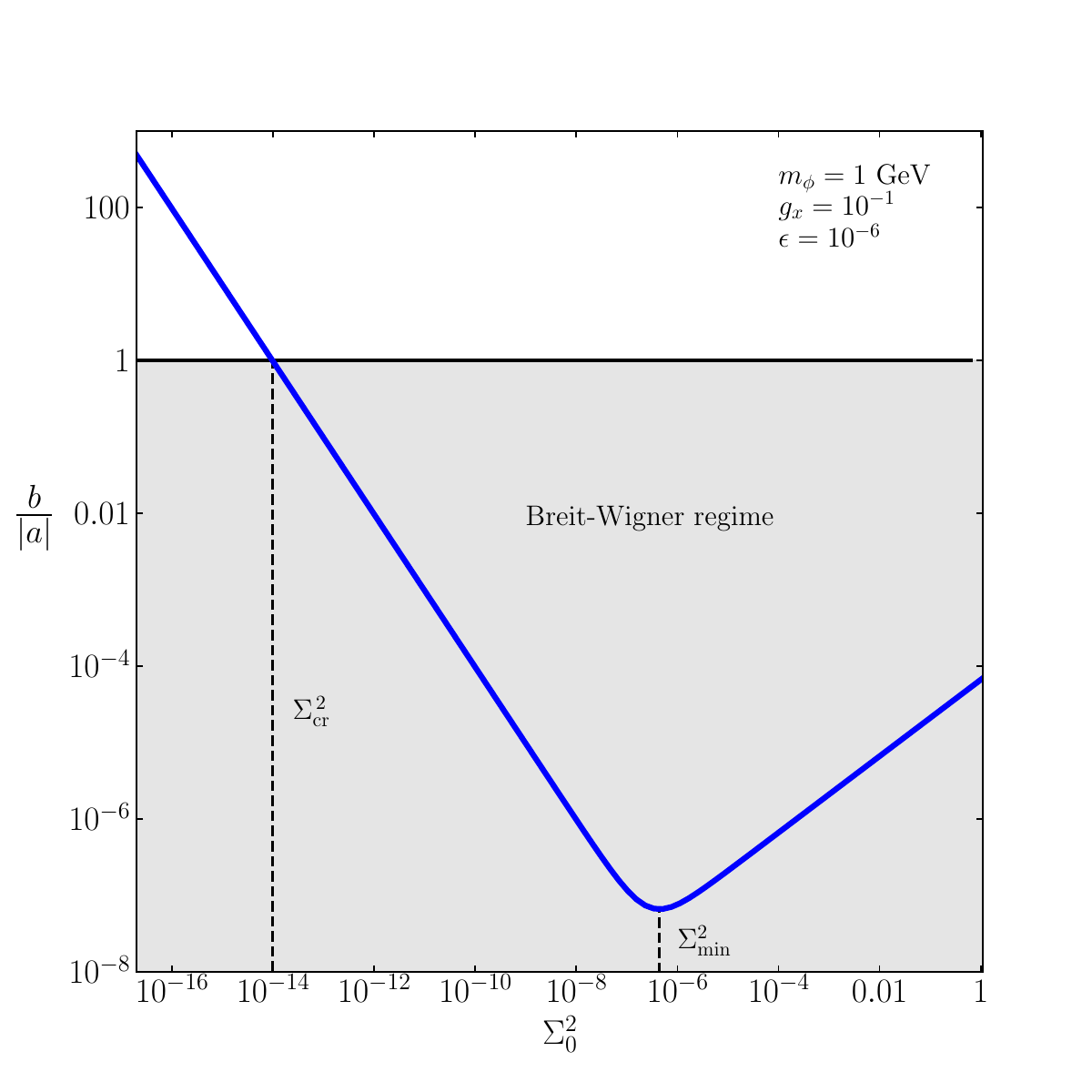}
\caption{Ratio $b/|a|$ as a function of $\Sigma_0^2$. The gray-shaded area corresponds to values of $b/|a|<1$, where the cross-section is Breit-Wigner enhanced around $\Sigma_0^2$.}
\label{fig:ab_ratio}
\end{figure}
%

The ratio ${b}/{|a|}$ is the cornerstone of our analysis. It controls how $J(a,b)$ evolves with $\Sigma^{2}$ and, in fine, with $y$. If that ratio is smaller than $1$, a Breit-Wigner enhancement appears with $\asv$ scaling like ${1}/{\Sigma^{3}}$ for dispersion velocities just above $\Sigma_{0}$. On the contrary, if it exceeds $1$, this enhancement disappears and the annihilation cross-section $\asv$ is largest for dispersion velocities of order $\Lambda_{0}$. The ratio ${b}/{|a|}$ determines also the point $y_{\rm M}$ where the integrand ${J(a,b)}/{\Sigma^{2}}$ in the integral ${\cal J}_{\rm ann}$ is maximal. It can be expressed in terms of the parameters of the model as
\beq
\frac{b}{|a|} \equiv \frac{\Lambda_{0}^{2}}{\Sigma_{0}^{2}} = {\displaystyle \frac{1}{12 \pi}}
\left\{ \frac{g_{x}^{2}}{4} \, \Sigma_{0} + \frac{\epsilon^{2} e^{2} {Q'}^{2}}{\Sigma_{0}^{2}} \right\}.
\label{eq:ratio_b_on_abs_a}
\eeq
For illustration, the ratio ${b}/{|a|}$ as a function of $\Sigma_0^2$ is presented in Fig.~\ref{fig:ab_ratio} for the same parameters as in Fig.~\ref{fig:Omegah2_nokd_wkd}. The ratio is minimal when the mass degeneracy parameter $\Sigma_{0}^{2}$ is equal to the special value
\beq
\Sigma_{\rm min}^{2} = 4 \left\{ \frac{\epsilon^{2} e^{2} {Q'}^{2}}{g_{x}^{2}} \right\}^{2/3}.
\label{eq:Sigma_min_def}
\eeq
This corresponds to a minimal ratio of
\beq
\left( {b}/{|a|} \right)_{\rm min} = \frac{1}{16 \pi}
\left\{ g_{x}^{4} \epsilon^{2} e^{2} {Q'}^{2} \right\}^{1/3}.
\label{eq:b_over_a_min}
\eeq
The mixing angle $\epsilon$ cannot be larger than $1$ by construction. The dark charge $g_{x}$ is also restricted to be smaller than~$1$ since otherwise, the theory becomes non-perturbative. According to relation~\ref{eq:definition_Q_prime_a}, the effective charge ${Q'}^{2}$ reaches a maximal value of $8$ in the unrealistic situation where the dark photon mass $m_{x}$ becomes infinite. Even in this extreme case, the minimal value of the ratio ${b}/{|a|}$ is less than $1.8 \times 10^{-2}$. As will be discussed in Sec.~\ref{subsubsec:accelerator}, collider experiments yield furthermore the constraint $\epsilon \leq  10^{-3}$ on the mixing angle for $m_{x} \leq 10 \, {\rm GeV}$ \cite{Ilten:2018crw}. This translates into an upper limit on the minimal value of ${b}/{|a|}$ of order $1.8 \times 10^{-4}$. We conclude that this minimal value is always very small compared to $1$.
We also remark that when $\Sigma_{0}^{2}$ goes to $1$, i.e. for a virtually infinite dark photon mass $m_{x}$, the ratio ${b}/{|a|}$ is still very small since
\beq
\left( {b}/{|a|} \right)_{\Sigma_{0}^{2} = 1} = {\displaystyle \frac{1}{12 \pi}}
\left\{ \frac{g_{x}^{2}}{4} + {\epsilon^{2} e^{2} {Q'}^{2}} \right\} \lesssim \frac{1}{48 \pi}\,.
\eeq
Finally, as $\Sigma_{0}^{2}$ decreases below $\Sigma_{\rm min}^{2}$, the ratio ${b}/{|a|}$ increases like ${\epsilon^{2} e^{2} {Q'}^{2}}/{12 \pi \Sigma_{0}^{2}}$ and overcomes $1$ below the critical value
\beq
\Sigma_{\rm cr}^{2} = \frac{\epsilon^{2} e^{2} {Q'}^{2}}{12 \pi} \lesssim 8 \times 10^{-9}\,.
\eeq

\subsubsection{\bf Case $\Sigma_{0}^{2} > \Sigma_{\rm cr}^{2}$ -- the Breit-Wigner regime}
We are led to conclude that, as long as the mass degeneracy parameter $\Sigma_{0}^{2}$ is larger than $\Sigma_{\rm cr}^{2}$, the ratio ${b}/{|a|}$ is always less than $1$. The thermal average $\asv$ of the DM annihilation cross-section behaves like ${1}/{\Sigma^{3}}$ for DM dispersion velocities $\Sigma$ slightly above $\Sigma_{0}$. Close to the peak, the integral $J(a,b)$ may actually be approximated by
\beq
J(a,b) \simeq J_1(a,b) = {\displaystyle \frac{\sqrt{\pi}}{b}} \, |a|^{3/2} e^{-|a|} \,,
\label{eq:approx_J1_a}
\eeq
as shown in Appendix~\ref{app:jsigmav}, and the integrand of ${\cal J}_{\rm ann}$ is proportional to
\beq
\frac{J(a,b)}{\Sigma^{2}} \propto J_1(a,b) |a| \simeq {\displaystyle \frac{\sqrt{\pi}}{b}} \, |a|^{5/2} e^{-|a|} \,.
\eeq
This integrand is maximal for $|a| = {3}/{2}$, i.e. for the DM dispersion velocity $\Sigma_{\rm M}$ equal to $\sqrt{2/3}\,({m_{x}}/{2 m_{\phi}}) \Sigma_{0}$, a value which we will approximate hereafter by $\Sigma_{0}$ since the dark photon mass $m_{x}$ is very close to $2 m_{\phi}$.
To summarize, after freeze-out, DM cools down while $\Sigma$ decreases. The cross-section $\asv$ increases like ${1}/{\Sigma^{2}}$ and, close to the peak where most of DM annihilation occurs, like ${1}/{\Sigma^{3}}$. At the lower boundary $y_{\rm M}$ of the integral ${\cal J}_{\rm ann}$, the DM dispersion velocity is $\Sigma_{\rm M} \simeq \Sigma_{0}$ and the DM temperature $T_{\phi}$ is equal to $m_{\phi} \Sigma_{0}^{2}$.

\subsubsection{\bf Case $\Sigma_{0}^{2} < \Sigma_{\rm cr}^{2}$ -- the non-resonant regime}
For a mass degeneracy parameter $\Sigma_{0}^{2}$ less than $\Sigma_{\rm cr}^{2}$, the ratio ${b}/{|a|}$ is this time larger than $1$. The integral $J(a,b)$ is equal to $1$ as long as $b$ is smaller than $1$. As shown in Appendix~\ref{app:jsigmav}, beyond that point, $J(a,b)$ can be approximated by
\beq
J(a,b) \simeq J_2(a,b) = {\displaystyle \frac{3/4}{a^{2} + b^{2}}} \simeq \frac{3/4}{b^{2}} \,.
\eeq
The transition occurs at $b = {\sqrt{3}}/{2}$, at a dispersion velocity $\Sigma_{\rm M}$ which is equal to $\left( {2}/{\sqrt{3}} \right)^{1/2}\!({m_{x}}/{2 m_{\phi}})\,\Lambda_{0}$. For simplicity, we will use from now on $\Lambda_{0}$ as the benchmark value for $\Sigma_{\rm M}$ in this case.
In the early Universe, after freeze-out, $\asv$ increases like ${1}/{\Sigma^{2}}$ until the DM dispersion velocity $\Sigma$ has decreased down to $\Lambda_{0}$. Below that point, the annihilation cross-section drops like $\Sigma^{2}$ and vanishes. The lower boundary $y_{\rm M}$ of integral ${\cal J}_{\rm ann}$ is defined in such a way that the DM dispersion velocity is $\Sigma_{\rm M} \simeq \Lambda_{0}$ and the DM temperature $T_{\phi}$ is equal to $m_{\phi} \Lambda_{0}^{2}$.

\subsection{Results without kinetic decoupling}
\label{subsec:nokd}
In Fig.~\ref{fig:Omegah2_nokd_wkd}, we consider a benchmark example of a 1~GeV DM scalar particle. The dark charge $g_{x}$ and the mixing angle $\epsilon$ have been respectively set equal to $0.1$ and $10^{-6}$. In the left panel, the green curve features the evolution of the DM relic abundance $\Omega_{\phi} h^{2}$ as a function of the mass degeneracy parameter $\Sigma_{0}^{2}$.
This curve has been obtained numerically under the assumption that DM does not undergo kinetic decoupling. We have first determined the freeze-out point $x_{\rm F}$ by solving equation~\ref{eq:freeze_out_x_F} with the help of a dichotomy. We have then calculated ${\cal I}_{\rm ann}$ by integrating expression~\ref{eq:I_ann_1}. At this stage, we have assumed that DM has always the same temperature as the SM plasma. This assumption will be challenged in the forthcoming section~\ref{subsec:withkd}. The relic codensity $\tilde{n}_{0}$ has been derived from~\ref{eq:n_tilde_0_a}, relations~\ref{eq:n_0} and \ref{eq:Oh2} yielding $\Omega_{\phi} h^{2}$.

A close inspection of the green curve allows us to clearly identify two distinct regimes. For values of $\Sigma_{0}^{2}$ smaller than approximately $10^{-13}$, $\Omega_{\phi} h^{2}$ is constant and the curve exhibits a plateau. As $\Sigma_{0}^{2}$ increases above that value, the relic abundance first decreases, reaches a minimum slightly below $10^{-6}$ and eventually increases to explode close to the boundary $\Sigma_{0}^{2}=1$. The curve exhibits a plateau for small values of $\Sigma_{0}^{2}$ and a trough for the larger values. Both features are characteristic and can be understood with the help of the approximations developed in Sec.~\ref{subsec:Oh2_proxy}. In particular, we will try to identify the values of $\Sigma_{\rm cr}^{2}$ and $\Sigma_{\rm min}^{2}$ along the green curve. As a general remark, we notice that since DM is assumed here to be always thermally coupled to the SM plasma, the temperatures of both components are equal at all times, hence the identity
\beq
y \equiv \frac{T}{m_{\phi}} = \frac{T_{\phi}}{m_{\phi}} \equiv \Sigma^{2} \,.
\eeq
In the expression~\ref{eq:definition_cal_J_ann} of ${\cal J}_{\rm ann}$, the variable $y$ can be identified with the DM dispersion velocity squared $\Sigma^{2}$. This implies that the lower boundary $y_{\rm M}$ is equal to $\Sigma_{\rm M}^{2}$.

\subsubsection{\bf The plateau}
For values of $\Sigma_{0}^{2}$ smaller than $\Sigma_{\rm cr}^{2}$, the ratio ${b}/{|a|}$ overcomes $1$ (see Fig.\ref{fig:ab_ratio}). In this regime, the annihilation cross-section $\asv$ is maximal for $\Sigma_{\rm M} \simeq \Lambda_{0}$, and the integral ${\cal J}_{\rm ann}$ is performed from $y_{\rm M} \simeq \Lambda_{0}^{2}$ to $y_{\rm F}$. In this interval, the function $J(a,b)$ is equal to $1$ and we get
\beq
{\cal J}_{\rm ann} \simeq \ln \left( {y_{\rm F}}/{\Lambda_{0}^{2}} \right) = - \ln \left( {x_{\rm F} \Lambda_{0}^{2}} \right).
\label{eq:cal_J_ann_a}
\eeq
Since the mass degeneracy parameter $\Sigma_{0}^{2}$ is smaller than $\Sigma_{\rm cr}^{2}$, which is itself smaller than $\Sigma_{\rm min}^{2}$, the reduced decay width simplifies to the constant
\beq
\Lambda_{0}^{2} = \frac{\epsilon^{2} e^{2} {Q'}^{2}}{12 \pi} \equiv \Sigma_{\rm cr}^{2} \,.
\eeq
Physically, the decay of a dark photon into a pair of DM scalars $\phi \bar{\phi}$ is kinematically suppressed when its mass is nearly degenerate with $2 m_{\phi}$. In these conditions, the dark photon decays only into fermion pairs. We notice that the reduced width $\Lambda_{0}^{2}$, the integral ${\cal J}_{\rm ann}$ and eventually the relic abundance $\Omega_{\phi} h^{2}$ no longer depend on $\Sigma_{0}^{2}$, hence the plateau. The mass degeneracy parameter $\Sigma_{0}^{2}$ has disappeared from the problem.

At this stage, we will not compare our approximation~\ref{eq:Oh2_approx_a} with the numerical value of $\Omega_{\phi} h^{2} \simeq 2.5 \times 10^{4}$ read from Fig.~\ref{fig:Omegah2_nokd_wkd}, insofar as we are mostly concerned in this article with kinetic decoupling.
%
There is also an additional complication. The plateau is expected to extend up to $\Sigma_{\rm cr}^{2}$, which is equal to $9.7 \times 10^{-15}$ in our example. This value corresponds to a DM temperature, and therefore to a plasma temperature in the absence of kinetic decoupling, of $9.7 \, {\rm {\mu}eV}$. But today the CMB temperature $T_{0}$ is equal to $235 \, {\rm {\mu}eV}$. We are in the particular situation where $y_{0}$ is larger than $y_{\rm M} \simeq \Lambda_{0}^{2} \equiv \Sigma_{\rm cr}^{2}$. This has two consequences.
The integral ${\cal J}_{\rm ann}$ must be performed from $y_{0}$ to $y_{\rm F}$, and not from $y_{\rm M}$ to $y_{\rm F}$. This yields the result $\ln \left( {T_{\rm F}}/{T_{0}} \right)$ instead of~\ref{eq:cal_J_ann_a}.
Then, the plateau extends beyond the boundary~$\Sigma_{\rm cr}^{2}$, and this as long as the Breit-Wigner enhancement of the cross-section $\asv$ does not perturb too much the integral~${\cal J}_{\rm ann}$. At least, this is so as long as $\Sigma_{0}^{2}$ is less than $y_{0}$, provided that the ratio ${b}/{|a|}$ is not too small compared to $1$, and that the function $J(a,b)$ is close to unity. In our example, we anticipate an extension of the plateau up to $y_{0} = {T_{0}}/{m_{\phi}} = 2.35 \times 10^{-13}$, more than an order of magnitude above $\Sigma_{\rm cr}^{2}$. In Fig.~\ref{fig:Omegah2_nokd_wkd}, the transition occurs around $4 \times 10^{-13}$, not too far from what is expected.

\subsubsection{\bf The trough}
For values of $\Sigma_{0}^{2}$ larger than $\Sigma_{\rm cr}^{2}$, the ratio ${b}/{|a|}$ is smaller than $1$. The annihilation cross-section undergoes a Breit-Wigner enhancement which peaks at the dispersion velocity $\Sigma_{\rm M} \simeq \Sigma_{0}$. The integral ${\cal J}_{\rm ann}$ is performed from $y_{\rm M} \simeq \Sigma_{0}^{2}$ to $y_{\rm F}$. Above the peak, $\asv$ scales like ${1}/{\Sigma^{3}}$. If DM annihilation reaches its highest intensity before the present epoch, i.e. if $y_{\rm M}$ is larger than $y_{0}$, the integral ${\cal J}_{\rm ann}$ is significantly modified with respect to the case of the plateau studied above. Actually, for values of ${b}/{|a|}$ sufficiently small with respect to $1$, the function $J(a,b)$ can be replaced, in the integrand of ${\cal J}_{\rm ann}$, by its approximation~\ref{eq:approx_J1_a}, hence
\beq
{\cal J}_{\rm ann} = {\sqrt{\pi}}
\left\{ \frac{\Sigma_{0}^{2}}{\Lambda_{0}^{2}} \right\}
\left\{ \frac{m_{x}}{2 m_{\phi}} \right\}
{\displaystyle \int_{y_{\rm M}}^{y_{\rm F}}} dy \, \frac{\Sigma_{0}}{y^{3/2}} \,.
\eeq
The ratio ${|a|}/{b}$ is by definition identical to ${\Sigma_{0}^{2}}/{\Lambda_{0}^{2}}$, while $\Sigma^{2}$ is equal to the parameter $y$ insofar as DM is kinetically coupled to the SM plasma. We have dropped the exponential and replaced it by the sharp boundary at $y_{\rm M}$. Neglecting the term with $y_{\rm F}$, and noticing that $m_{x}$ and $2 m_{\phi}$ are quasi-degenerate as long as $\Sigma_{0}^{2}$ is not too close to $1$, we get
\beq
{\cal J}_{\rm ann} \simeq {\sqrt{\pi}}
\left\{ \frac{\Sigma_{0}^{3}}{\Lambda_{0}^{2}} \right\}
\frac{2}{\sqrt{y_{\rm M}}} \simeq 2 \sqrt{\pi} \left\{ \frac{\Sigma_{0}^{2}}{\Lambda_{0}^{2}} \right\}.
\eeq
Inserting this result into relation~\ref{eq:Oh2_approx_a} yields the DM relic abundance
\beq
\left. \Omega_{\phi} h^{2} \right|_{\rm trough} \simeq
{\frac{{8}{\pi}^{2}}{\sqrt{5}}} \, {\cal F}
\left\{ \frac{{\Lambda_{0}^{2}}/{\Sigma_{0}^{2}}}{g_{x}^{2} \epsilon^{2} e^{2} \tilde{Q}^{2}} \right\}.
\label{eq:Oh2_approx_b}
\eeq
The mass degeneracy parameter $\Sigma_{0}^{2}$ appears in the argument $T_{\rm M} \simeq m_{\phi} \Sigma_{0}^{2}$ of the plasma function inside the factor ${\cal F}$, and in the ratio ${\Lambda_{0}^{2}}/{\Sigma_{0}^{2}}$. In the former case, it has little impact on $\Omega_{\phi} h^{2}$ since ${\cal P}$ is a slowly varying function of plasma temperature. The DM relic abundance depends on $\Sigma_{0}^{2}$ essentially through the ratio ${\Lambda_{0}^{2}}/{\Sigma_{0}^{2}}$.

\begin{figure}[h!]
\centering
\includegraphics[width=\columnwidth]{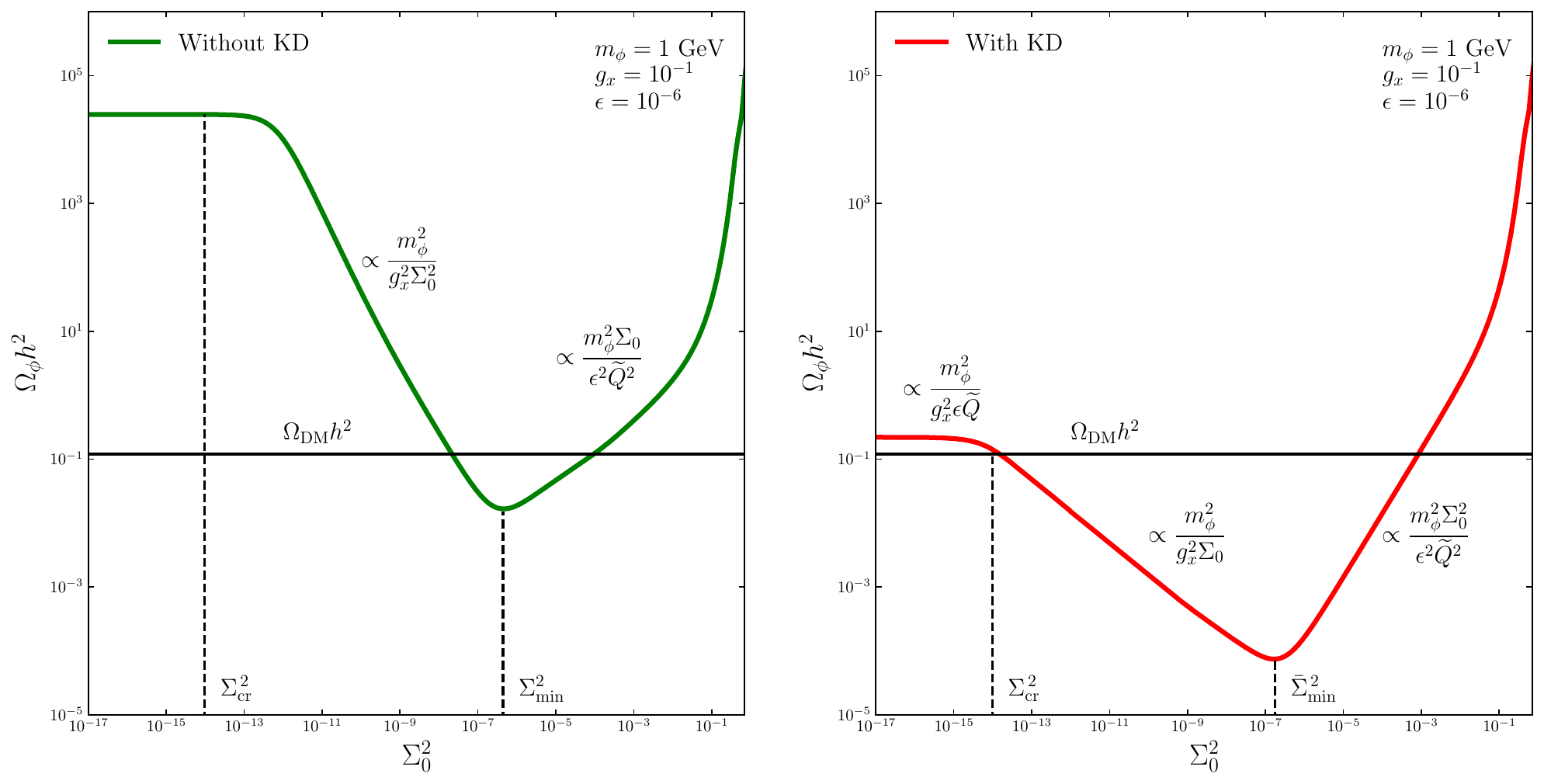}
\caption{
DM relic abundance as a function of $\Sigma_{0}^{2}$ for fixed $g_{x}$ and $\epsilon$ (respectively $10^{-1}$ and $10^{-6}$), and a DM mass of 1~GeV.
In the left panel, the case without kinetic decoupling of the DM is highlighted in green with the asymptotic scalings around $\Sigma_{\rm min}^2$.
In the right panel, the case with kinetic decoupling is highlighted in red, with the asymptotic scalings around ${\bar{\Sigma}}_{\rm min}^2$ and $\Sigma_{\rm cr}^2$. The horizontal black line indicates the Planck DM relic abundance $\Omega_{\rm DM} h^2$.
}
\label{fig:Omegah2_nokd_wkd}
\end{figure}

We can readily apply our analysis of Sec.~\ref{subsec:Oh2_proxy}. As $\Sigma_{0}^{2}$ increases from $\Sigma_{\rm cr}^{2}$ to the upper limit $1$, the ratio ${\Lambda_{0}^{2}}/{\Sigma_{0}^{2}} = b/|a|$, and hence $\Omega_{\phi} h^{2}$ to which it is proportional according to Eq.~\ref{eq:Oh2_approx_b}, decreases to a minimum reached at $\Sigma_{\rm min}^{2}$ and then increases. This is actually what we observe in Fig.~\ref{fig:Omegah2_nokd_wkd} where the green curve exhibits a trough. Its minimum corresponds to a DM dispersion velocity $\Sigma_{\rm min}^{2}$ of order $4 \times 10^{-7}$. This value can be compared to our expectation~\ref{eq:Sigma_min_def}. Plugging in it the numerical values of $g_{x}$ and $\epsilon$, and noticing that ${Q'}^{2}$ is very close to $4$ for a 1~GeV DM candidate, we derive a value of $4.4 \times 10^{-7}$, in excellent agreement with the numerical result. This gives us confidence in our approach and put our approximation on firm grounds.

According to Fig.~\ref{fig:Omegah2_nokd_wkd}, below $\Sigma_{\rm min}^{2}$, the relic abundance $\Omega_{\phi} h^{2}$ decreases with $\Sigma_{0}$ as a power law with index close to~$-2$. Above the minimum, it increases like $\Sigma_{0}$, following a power law with index~$+1$. For $\Sigma_{0}^{2}$ larger than $10^{-2}$, $\Omega_{\phi} h^{2}$ sharply increases.
To understand the trough which the green curve exhibits, we can start from the approximation~\ref{eq:Oh2_approx_b} for $\Omega_{\phi} h^{2}$ and from the expression~\ref{eq:ratio_b_on_abs_a} of the ratio ${\Lambda_{0}^{2}}/{\Sigma_{0}^{2}}$.
Below $\Sigma_{\rm min}^{2}$, i.e. for small values of $\Sigma_{0}^{2}$, that ratio simplifies to
\beq
\frac{\Lambda_{0}^{2}}{\Sigma_{0}^{2}} \simeq \frac{\epsilon^{2} e^{2} {Q'}^{2}}{12 \pi \Sigma_{0}^{2}} \,.
\label{eq:ratio_L02_on_S02}
\eeq
Noticing that the effective charges ${Q'}^{2}$ and $\tilde{Q}^{2}$ are essentially equal as long as $m_{x}$ is nearly degenerate with $2 m_{\phi}$, we infer the relic abundance
\beq
\left. \Omega_{\phi} h^{2} \right|_{\Sigma_{0}^{2} < \Sigma_{\rm min}^{2}} \simeq
{\frac{{2}{\pi}}{{3}{\sqrt{5}}}}
\left\{ \frac{\cal F}{g_{x}^{2} \Sigma_{0}^{2}} \right\} \propto \frac{m_{\phi}^{2}}{g_{x}^{2} \Sigma_{0}^{2}} \,.
\label{eq:Oh2_approx_c}
\eeq
We do find that $\Omega_{\phi} h^{2}$ scales as ${1}/{\Sigma_{0}^{2}}$, as observed in Fig.~\ref{fig:Omegah2_nokd_wkd}. We also notice that the relic abundance depends on the dark charge $g_{x}$ and no longer on the mixing angle $\epsilon$.

For values of $\Sigma_{0}^{2}$ larger than $\Sigma_{\rm min}^{2}$, the ratio ${\Lambda_{0}^{2}}/{\Sigma_{0}^{2}}$ simplifies this time to ${g_{x}^{2} \Sigma_{0}}/{48 \pi}$. This yields
\beq
\left. \Omega_{\phi} h^{2} \right|_{\Sigma_{0}^{2} > \Sigma_{\rm min}^{2}} \simeq
{\frac{\pi}{{6}\sqrt{5}}} \, {\cal F}
\left\{ \frac{\Sigma_{0}}{\epsilon^{2} e^{2} \tilde{Q}^{2}} \right\} \propto \frac{m_{\phi}^{2} \Sigma_{0}}{\epsilon^{2} \tilde{Q}^{2}} \,.
\label{eq:Oh2_approx_d}
\eeq
Our analysis allows us to understand why the DM relic abundance follows a power law in $\Sigma_{0}$ with index $+1$ above $\Sigma_{\rm min}^{2}$. We also remark that $\Omega_{\phi} h^{2}$ depends in this case on the mixing angle $\epsilon$ and not on the dark charge $g_{x}$.
When $\Sigma_{0}^{2}$ is close to $1$, our analysis and relation~\ref{eq:Oh2_approx_d} no longer apply. The dark photon mass $m_{x}$ starts to be much larger than $2 m_{\phi}$. Since the ratio ${b}/{|a|}$ is still smaller than $1$, a Breit-Wigner resonance may enhance the annihilation cross-section. But the peak dispersion velocity $\Sigma_{\rm M}$ is larger than $\Sigma_{0}$ by a factor ${m_{x}}/{2 m_{\phi}}$ which could be very large. If so, $\Sigma_{\rm M}$ exceeds the speed of light and $\asv$ is never enhanced, whatever the DM dispersion velocity. DM annihilation is, in this case, $p$-wave dominated.
The dark photon propagator in the $s$-channel yields furthermore a factor ${1}/{m_{x}^{4}}$ which suppresses DM annihilation when $m_{x}$ diverges. It is no surprise then if $\Omega_{\phi} h^{2}$ explodes when $\Sigma_{0}$ goes to $1$, as observed in Fig.~\ref{fig:Omegah2_nokd_wkd}.

As a side remark, we notice that depending on the values of $g_{x}$ and $\epsilon$, the dispersion velocity $\Sigma_{\rm min}^{2}$ may well exceed the boundary $1$. This happens actually for small $g_{x}$ and large $\epsilon$. In this case, the general behavior of $\Omega_{\phi} h^{2}$ as a function of $\Sigma_{0}^{2}$ is not qualitatively changed with respect to what has been discussed above. We still have a plateau below $\Sigma_{\rm cr}^{2}$ and a trough above, the relic abundance starting to decrease, reaching a minimum and eventually sharply increasing close to $1$. The position of the minimum is no longer given by $\Sigma_{\rm min}^{2}$, but must be defined numerically.

%
%

\subsection{Results in the presence of kinetic decoupling}
\label{subsec:withkd}
We have so far assumed that DM is in thermal contact with the SM plasma, and that their respective temperatures $T_{\phi}$ and $T$ are equal at all times relevant to our analysis. We now challenge this assumption. Thermalization of DM occurs primarily through an exchange of energy due to collisions with the SM plasma. Should DM be slightly colder than the plasma, for instance, the latter would reheat the former by injecting energy through collisions of SM fermions on DM scalars. If this occurs fast enough, $T_{\phi}$ relaxes rapidly toward $T$ and DM is thermalized.
In appendix~\ref{sec:phi_thermalization}, we present a simplified treatment of that process. Starting from Lagrangian~\ref{eq:Lagrangian_1}, we derive the rate at which energy is exchanged between DM and the plasma. DM annihilations must also be included together with collisions to determine how both DM density and temperature vary concomitantly. In appendix~\ref{sec:phi_thermalization}, we actually demonstrate that, under the specific assumptions presented at the beginning of this section, reaction~\ref{eq:chemical_equilibrium} is able alone to thermalize DM with the plasma, should collisions be turned off. Taking into account both annihilations and collisions, we establish the simplified differential equation~\ref{eq:T_phi_evol_b} which the DM temperature follows. This allows to define the rate $\Gamma_{\rm rel}^{\rm KD}$ with which $T_{\phi}$ relaxes toward $T$, and the rate $\Gamma_{\rm eq}^{\rm KD}$ with which the kinetic equilibrium itself evolves. Kinetic decoupling occurs at temperature $T_{\rm KD}$, for which both rates are equal.


In the right panel of Fig.~\ref{fig:Omegah2_nokd_wkd}, the red curve features the evolution of $\Omega_{\phi} h^{2}$ as a function of the mass degeneracy parameter $\Sigma_{0}^{2}$ when kinetic decoupling is taken into account. The other parameters have the same values as for the green curve.
The red curve has been obtained numerically by first solving Eqs.~\ref{eq:freeze_out_x_F} and \ref{eq:freeze_out_x_KD} using dichotomies. This allows to determine the freeze-out $T_{\rm F}$ and kinetic decoupling $T_{\rm KD}$ temperatures. As explained in appendix~\ref{sec:phi_thermalization}, kinetic decoupling occurs after freeze-out, hence the sequence $T_{\rm F} \geq T_{\rm KD}$.
We then calculate ${\cal I}_{\rm ann}$ by integrating expression~\ref{eq:I_ann_1} from freeze-out until now. Between freeze-out and kinetic decoupling, i.e. as long as $T$ is larger than $T_{\rm KD}$, both DM and plasma temperatures are equal. After kinetic decoupling, DM behaves as a non-relativistic gas undergoing adiabatic cooling and its temperature $T_{\phi}$ drops as ${a^{-2}}$, with $a$ the scale factor of the Universe. The plasma also cools down adiabatically, but is now decoupled from DM. The temperatures of both components are related by
\beq
T_{\phi}(T \leq T_{\rm KD}) = \left\{ \frac{h_{\rm eff}(T)}{h_{\rm eff}(T_{\rm KD})} \right\}^{2/3}
\frac{T^{2}}{T_{\rm KD}} \,.
\label{eq:tphi_tkd}
\eeq
We then determine the relic codensity $\tilde{n}_{0}$ from~\ref{eq:n_tilde_0_a} and derive the DM relic abundance $\Omega_{\phi} h^{2}$ using relations~\ref{eq:n_0} and \ref{eq:Oh2}.

The red and green curves of Fig.~\ref{fig:Omegah2_nokd_wkd} are qualitatively similar. Both exhibit a plateau for small values of $\Sigma_{0}^{2}$ and a trough for larger values. The red curve becomes flat below $5 \times 10^{-15}$. For larger values of $\Sigma_{0}^{2}$, the DM relic abundance starts to decrease and reaches a minimum for a mass degeneracy parameter of order $2 \times 10^{-7}$, slightly below $\Sigma_{\rm min}^{2}$. The red curve then follows a rising power law and eventually sharply increases close to $1$. The essential effect of including kinetic decoupling in the calculation of $\Omega_{\phi} h^{2}$ is to shift the curve downward and to get smaller values of the DM relic abundance.
This can be easily understood. When DM decouples thermally from the primordial plasma, its temperature drops faster than if thermal contact was continuously established. We have just showed that $T_{\phi}$ decreases as $a^{-2}$, while $T$ scales approximately like $a^{-1}$, where $a$ is the scale factor of the Universe. As DM cools down, the annihilation cross-section $\asv$ increases. It peaks at the DM dispersion velocity $\Sigma_{\rm M}$, where most of the annihilation takes place. When kinetic decoupling is included, this occurs at a higher plasma temperature $T_{\rm M}$, i.e. at an earlier time $t_{\rm M}$ when the DM population is denser. A stronger DM annihilation at $\asv$ peak results in a smaller relic abundance, hence the observed shift between the green and the red curves.
When kinetic decoupling is included, the relation between $\Sigma^{2}$ and $y$ is also slightly more involved. In Sec.~\ref{subsec:nokd}, we could identify $y$ and $\Sigma^{2}$ at all times. This is only possible now before kinetic decoupling. For temperatures below $T_{\rm KD}$, the new relation is
\beq
\Sigma^{2} \equiv \frac{T_{\phi}}{m_{\phi}} =  \left\{ \frac{h_{\rm eff}(T)}{h_{\rm eff}(T_{\rm KD})} \right\}^{2/3}
\frac{y^{2}}{y_{\rm KD}} \,,
\label{eq:relation_Sigma_y_wkd}
\eeq
where $y_{\rm KD}$ denotes the ratio ${T_{\rm KD}}/{m_{\phi}}$. If most of DM annihilation occurs after kinetic decoupling, i.e. if $T_{\rm M}$ is well below $T_{\rm KD}$, the integral ${\cal J}_{\rm ann}$ may be approximated by
\beq
{\cal J}_{\rm ann} \simeq y_{\rm KD} \left\{ \frac{h_{\rm eff}(T_{\rm KD})}{h_{\rm eff}(T_{\rm M})} \right\}^{2/3}
{\displaystyle \int_{y_{\rm M}}^{y_{\rm KD}}} {\frac{dy}{y^{2}}} \, J(a,b) \,.
\label{eq:int_cal_J_ann_wkd}
\eeq
The ratio ${h_{\rm eff}(T_{\rm KD})}/{h_{\rm eff}(T)}$ varies slowly in time and we have taken it at peak annihilation when the plasma temperature is $T_{\rm M}$. We also remark that the relation between $y_{\rm M}$ and the DM dispersion velocity $\Sigma_{\rm M}$ has become
\beq
y_{\rm M} = \sqrt{y_{\rm KD}}
 \left\{ \frac{h_{\rm eff}(T_{\rm KD})}{h_{\rm eff}(T_{\rm M})} \right\}^{1/3} \Sigma_{\rm M} \,.
\eeq
Equipped with these notations, we are ready to analyze the red curve.

\subsubsection{\bf The plateau}
As showed in Sec.~\ref{subsec:nokd}, the ratio ${b}/{|a|}$ overcomes $1$ when $\Sigma_{0}^{2}$ is smaller than $\Sigma_{\rm cr}^{2}$. This implies a transition at $9.7 \times 10^{-15}$, in agreement with the value of $5 \times 10^{-15}$ mentioned above. In the plateau regime, the annihilation cross-section $\asv$ is maximal for $\Sigma_{\rm M} \simeq \Lambda_{0}$ and we get
\beq
{\cal J}_{\rm ann} \simeq \frac{\sqrt{y_{\rm KD}}}{\Lambda_{0}}  \left\{ \frac{h_{\rm eff}(T_{\rm KD})}{h_{\rm eff}(T_{\rm M})} \right\}^{1/3}.
\label{eq:int_cal_J_ann_wkd_plateau}
\eeq
Using this expression into relation~\ref{eq:Oh2_approx_a} leads to the approximate DM relic abundance
\beq
\left. \Omega_{\phi} h^{2}  \right|_{\Sigma_{0}^{2} < \Sigma_{\rm cr}^{2}} \simeq
{\frac{{8}{\pi}^{2}}{\sqrt{15}}} \, {\cal F}_{\rm KD}
\left\{ \frac{\sqrt{x_{\rm KD}}}{g_{x}^{2} \epsilon^{} e^{} \tilde{Q}} \right\} \propto
\frac{m_{\phi}^{2}}{g_{x}^{2} \epsilon^{} \tilde{Q}} \,.
\label{eq:Oh2_approx_e}
\eeq
where ${\cal F}_{\rm KD}$ denotes the pre-factor
\beq
{\cal F}_{\rm KD} \equiv \left\{ \frac{m_{\phi} T_{0}^{3}}{\rho_{\rm C}^{0}} \right\}
\left\{ \frac{m_{\phi}}{M_{\rm P}} \right\}
\left\{ \frac{h_{\rm eff}(T_{0})}{{\cal P}(T_{\rm M})} \right\}
\left\{ \frac{h_{\rm eff}(T_{\rm M})}{h_{\rm eff}(T_{\rm KD})} \right\}^{1/3}.
\eeq
Along the plateau, i.e. below $\Sigma_{\rm cr}^{2}$, the dark photon mass $m_{x}$ is very close to $2 m_{\phi}$ and we can identify $\tilde{Q}$ with ${Q'}$. We have also denoted by $x_{\rm KD}$ the inverse of $y_{\rm KD}$. We notice that the mass degeneracy parameter $\Sigma_{0}^{2}$ has disappeared from the problem.
With the values of model parameters considered in Fig.~\ref{fig:Omegah2_nokd_wkd}, we find that along the plateau, freeze-out and kinetic decoupling occur respectively at $x_{\rm F} = 1.71$ and $x_{\rm KD} = 8.48$. This translates into a decoupling temperature $T_{\rm KD}$ of $0.118 \, {\rm GeV}$. The peak of DM annihilation is reached at the plasma temperature $T_{\rm M} = 58 \, {\rm eV}$ well above $T_{0}$.
Relation~\ref{eq:Oh2_approx_e} yields a DM relic abundance of $0.139$, to be compared to the numerical value of $0.220$, obtained asymptotically by setting the mass degeneracy parameter $\Sigma_{0}^{2}$ at $10^{-17}$. This result is very encouraging given the simplicity of our approximation.
We could improve the agreement by slightly increasing the critical DM dispersion velocity $\Sigma_{\rm M}$ above $\Lambda_{0}$. Shifting it from $\Lambda_{0}$ to ${3 \Lambda_{0}}/{2}$, for instance, yields a relic abundance of $0.208$, in better agreement with the numerical result.

\subsubsection{\bf The trough}
For values of $\Sigma_{0}^{2}$ larger than $\Sigma_{\rm cr}^{2}$, the ratio ${b}/{|a|}$ is smaller than $1$ and the analysis proceeds along the same line as in Sec.~\ref{subsec:nokd}. Thermal decoupling complicates the relation between the $y$ parameter and the DM dispersion velocity $\Sigma$. In the trough regime, DM annihilation is Breit-Wigner enhanced at its peak. Replacing in the integrand of expression~\ref{eq:int_cal_J_ann_wkd} the function $J(a,b)$ by its approximation~\ref{eq:approx_J1_a} leads, after some algebra, to the integral
\beq
{\cal J}_{\rm ann} \simeq \sqrt{\pi} \, y_{\rm KD}^{3/2}
\left\{ \frac{m_{x}}{2 m_{\phi}} \right\}
\left\{ \frac{\Sigma_{0}^{3}}{\Lambda_{0}^{2}} \right\}
\left\{ \frac{h_{\rm eff}(T_{\rm KD})}{h_{\rm eff}(T_{\rm M})} \right\}
{\displaystyle \int_{y_{\rm M}}^{y_{\rm KD}}} {\frac{dy}{y^{3}}}\,.
\eeq
Making use once again of the conversion~\ref{eq:relation_Sigma_y_wkd} and setting the peak dispersion velocity $\Sigma_{\rm M}$ at $\Sigma_{0}$, we get
\beq
{\cal J}_{\rm ann} \simeq \frac{\sqrt{\pi}}{2} \sqrt{y_{\rm KD}}
\left\{ \frac{\Sigma_{0}}{\Lambda_{0}^{2}} \right\}
\left\{ \frac{h_{\rm eff}(T_{\rm KD})}{h_{\rm eff}(T_{\rm M})} \right\}^{1/3}.
\label{eq:int_cal_J_ann_wkd_trough}
\eeq
As long as $\Sigma_{0}^{2}$ is not too close to $1$, we can identify $m_{x}$ with $2 m_{\phi}$. Combining relations~\ref{eq:int_cal_J_ann_wkd_trough} and \ref{eq:Oh2_approx_a} leads to the DM relic abundance
\beq
\left. \Omega_{\phi} h^{2} \right|_{\rm trough} \simeq
{\frac{{32}{\pi}^{2}}{\sqrt{5}}} \, {\cal F}_{\rm KD}
\sqrt{x_{\rm KD}} \,
\left\{ \frac{{\Lambda_{0}^{2}}/{\Sigma_{0}}}{g_{x}^{2} \epsilon^{2} e^{2} \tilde{Q}^{2}} \right\}.
\label{eq:Oh2_approx_trough_general}
\eeq
To test this approximation, we have varied $\Sigma_{0}^{2}$ from $10^{-13}$ up to $0.1$, and compared the numerical result with the value given by~\ref{eq:Oh2_approx_trough_general}. On a vast region of that interval, both results agree at the percent level and, in some cases, even at the per mille level. The agreement lessens close to the minimum located at $\Sigma_{0}^{2} \simeq 1.66 \times 10^{-7}$, where the approximation yields a relic abundance of $6.12 \times 10^{-5}$ to be compared to the numerical result of $7.45 \times 10^{-5}$.
We finally notice that above $\Sigma_{0}^{2} \simeq 5 \times 10^{-3}$, the numerical relic abundance increases more sharply than its approximation~\ref{eq:Oh2_approx_trough_general}. Reasons for this have already been presented in Sec.~\ref{subsec:nokd}.
Along the trough, $\Omega_{\phi} h^{2}$ is proportional to the ratio
\beq
\frac{\Lambda_{0}^{2}}{\Sigma_{0}} = \frac{1}{12 \pi}
\left\{ \frac{g_{x}^{2}}{4} \, \Sigma_{0}^{2} + \frac{\epsilon^{2} e^{2} {Q'}^{2}}{\Sigma_{0}} \right\},
\eeq
which is minimum at the mass degeneracy parameter
\beq
\bar{\Sigma}_{\rm min}^{2} =
\left\{ \frac{2^{} \epsilon^{2} e^{2} {Q'}^{2}}{g_{x}^{2}} \right\}^{2/3}.
\label{eq:Sigma_bar_min_def}
\eeq
So is $\Omega_{\phi} h^{2}$. If thermal contact was constantly established between DM and the SM plasma, the DM relic abundance would exhibit a minimum at $\Sigma_{\rm min}^{2}$. In the presence of kinetic decoupling, the new minimum is a factor ${4}^{2/3}$ smaller. In Fig.~\ref{fig:Omegah2_nokd_wkd}, this shift of the minimum of $\Omega_{\phi} h^{2}$ between the green and red curves can be clearly seen. Relation~\ref{eq:Sigma_bar_min_def} yields a value of $1.75 \times 10^{-7}$ for the minimum of the red curve, close to the numerical value of $1.66 \times 10^{-7}$.

Below $\bar{\Sigma}_{\rm min}^{2}$, the ratio ${\Lambda_{0}^{2}}/{\Sigma_{0}}$ can be approximated by ${\epsilon^{2} e^{2} {Q'}^{2}}/{12 \pi \Sigma_{0}}$ and expression~\ref{eq:Oh2_approx_trough_general} simplifies into
\beq
\left. \Omega_{\phi} h^{2} \right|_{\Sigma_{0}^{2} < \bar{\Sigma}_{\rm min}^{2}} \simeq
{\frac{{8}{\pi}}{{3}{\sqrt{5}}}} \, {\cal F}_{\rm KD}
\left\{ \frac{\sqrt{x_{\rm KD}}}{g_{x}^{2} \Sigma_{0}} \right\} \propto
\frac{m_{\phi}^{2}}{g_{x}^{2} \Sigma_{0}} \,,
\label{eq:Oh2_approx_trough_decreasing}
\eeq
where we have identified $\tilde{Q}^{2}$ with ${Q'}^{2}$. Above the minimum, we can replace ${\Lambda_{0}^{2}}/{\Sigma_{0}}$ by $({g_{x}^{2}}/{48 \pi})  \Sigma_{0}^{2}$ to get
\beq
\left. \Omega_{\phi} h^{2} \right|_{\Sigma_{0}^{2} > \bar{\Sigma}_{\rm min}^{2}} \simeq
{\frac{{2}{\pi}}{{3}{\sqrt{5}}}} \, {\cal F}_{\rm KD}
\sqrt{x_{\rm KD}} \,
\left\{ \frac{\Sigma_{0}^{2}}{\epsilon^{2} e^{2} \tilde{Q}^{2}} \right\} \propto
\frac{m_{\phi}^{2} \Sigma_{0}^{2}}{\epsilon^{2} \tilde{Q}^{2}} \,.
\label{eq:Oh2_approx_trough_increasing}
\eeq
In Fig.~\ref{fig:Omegah2_nokd_wkd}, the red curve decreases actually with $\Sigma_{0}$ as a power law of index $-1$ while, above the minimum, it follows a power law with index~$+2$. We also remark that $\Omega_{\phi} h^{2}$ depends in the former case on the dark charge $g_{x}$ and not on the mixing angle $\epsilon$, while it is the opposite in the latter case. This property will play a crucial role in Sec.~\ref{sec:results} and will help understand the results.
For completeness, we have derived an approximate expression for the minimal DM relic abundance. Since the minimal value of the ratio ${\Lambda_{0}^{2}}/{\Sigma_{0}}$ is given by
\beq
\left( {\Lambda_{0}^{2}}/{\Sigma_{0}} \right)_{\rm min} =
\frac{\left( 2^{} g_{x} \epsilon^{2} e^{2} {Q'}^{2} \right)^{2/3}}{16 \pi} \,,
\eeq
we infer that the DM relic abundance, in the presence of kinetic decoupling, reaches a minimal value of
\beq
\left. \Omega_{\phi} h^{2} \right|_{\rm min} \simeq
{\frac{{2}^{5/3} {\pi}}{\sqrt{5}}} \, {\cal F}_{\rm KD}
\frac{\sqrt{x_{\rm KD}}}{(g_{x}^{4} \epsilon^{2} e^{2} \tilde{Q}^{2})^{1/3}} \,,
\label{eq:Oh2_approx_trough_min}
\eeq
where we have identified once again $\tilde{Q}^{2}$ with ${Q'}^{2}$. This expression yields a minimum value of $6.13 \times 10^{-5}$ to be compared to the numerical result of $7.46 \times 10^{-5}$ for a mass degeneracy parameter $\Sigma_{0}^{2}$ of $1.75 \times 10^{-7}$. Since approximation~\ref{eq:Oh2_approx_trough_min} is based only on the behavior of the ratio ${\Lambda_{0}^{2}}/{\Sigma_{0}}$, it disregards the influence of the temperatures $T_{\rm KD}$ and $T_{\rm M}$ on $\Omega_{\phi} h^{2}$, hence the very small differences with the results quoted above. For $\Sigma_{0}^{2}$ larger than $10^{-2}$, $\Omega_{\phi} h^{2}$ sharply increases.
%

In the next section, we will impose that the DM relic abundance is equal to the observed value $\Omega_{\rm DM} h^{2}$ of $0.1200$~\cite{Planck:2018vyg}.
For this, at fixed DM mass $m_{\phi}$, dark charge $g_{x}$ and mixing angle $\epsilon$, we will look for values of the mass degeneracy parameter $\Sigma_{0}^{2}$ that fulfill this requirement. Depending on the height of the red curve with respect to the level of $\Omega_{\rm DM} h^{2}$, three configurations are possible.

\begin{enumerate}
\item
If the red curve is too high, there is no solution and the DM relic abundance $\Omega_{\phi} h^{2}$ is always larger than $\Omega_{\rm DM} h^{2}$. The Universe is overclosed by scalar DM.

\item
In the configurations of interest, two values of $\Sigma_{0}^{2}$ fulfill the condition $\Omega_{\phi} h^{2} = \Omega_{\rm DM} h^{2}$. One solution lies on the decreasing left branch of the red curve, below the critical value $\bar{\Sigma}_{\rm min}^{2}$ at which the DM relic abundance is minimal. The other solution is located above $\bar{\Sigma}_{\rm min}^{2}$, on the rising right branch of the red curve. These solutions are dubbed hereafter \textit{left} and \textit{right branch solutions}.

\item
If the red curve is too low, there is no left branch solution insofar as the plateau stands below $\Omega_{\rm DM} h^{2}$. The Universe is underclosed by scalar DM below $\bar{\Sigma}_{\rm min}^{2}$. A right branch solution still exists.
\end{enumerate}

\section{Results}
\label{sec:results}
We present our results in the form of scans on the DM model parameter space in the $g_x$ and $\epsilon$ plane for three chosen DM particle masses: 200\,MeV, 1\,GeV and 5\,GeV. We take into account kinetic decoupling as presented in Sect.~\ref{subsec:withkd}. At 200\,MeV, DM annihilation is dominated by the leptonic channels whereas at 1 and 5\,GeV, there is also a significant contribution from DM annihilating directly into quark pairs. Moreover, for these masses, we mostly avoid the stringent direct detection limits. The impact of various constraints on the parameter space is summarized in  Figs.~\ref{fig:resultsDM200MeV}, \ref{fig:resultsDM1GeV}, and \ref{fig:resultsDM5GeV}, both for the \textit{left} and \textit{right branch} cases. In  Sect.~\ref{subsec:lim_gx_epsilon}, we first discuss how the DM relic abundance shapes the allowed region in the $g_x$ and $\epsilon$ plane. Then, in Sect.~\ref{subsec:lim_xs_mw}, \ref{subsec:lim_gx_epsilon_mudistorsion}, and \ref{subsec:lim_gx_epsilon_other}, we show how other constraints, from DM annihilation in the Milky Way, in the early Universe, and from direct detection and accelerators, shrink the allowed region of the DM parameter space.

\subsection{Limits on $g_x$ and $\epsilon$ set by the relic abundance $\Omega_{\rm DM} h^2$}
\label{subsec:lim_gx_epsilon}
As discussed previously in Sect.~\ref{subsec:withkd} notably when commenting Fig.~\ref{fig:Omegah2_nokd_wkd}, requiring $\Omega_{\phi} h^2=\Omega_{\rm DM} h^2$ leads, at fixed values of the parameters $g_x$ and $\epsilon$, to two different solutions, depending on the mass degeneracy parameter $\Sigma_0^2$.
We first discuss the solutions where $\Sigma_0^2<\bar{\Sigma}_{\rm min}^2$ which we dubbed \textit{left branch solutions} in the previous section, and then consider the \textit{right branch solutions} corresponding to $\Sigma_0^2>\bar{\Sigma}_{\rm min}^2$.

\subsubsection{\bf Left branch solutions -- $\Sigma_{0}^{2} < \bar{\Sigma}_{\rm min}^{2}$}

Figure~\ref{fig:pedaplotsDM1GeV_lowsol} shows the allowed regions for the case of a 1\;GeV DM scalar after enforcing $\Omega_{\phi} h^2=\Omega_{\rm DM} h^2$ with various quantities displayed in the color bar. The trends explained in the following are similar for the other DM masses.
 The top panels of Fig.~\ref{fig:pedaplotsDM1GeV_lowsol} feature in their respective color bars the kinetic decoupling parameter $x_{\rm KD}$ (left) and the mass degeneracy parameter $\Sigma_{0}^{2}$ (right).
 In the colored band, the relic density constraint is satisfied, while in the white regions above and below either $\Omega_\phi h^2<\Omega_{\rm DM}h^2$ or $\Omega_\phi h^2 >\Omega_{\rm DM}h^2$.
First, in the region where $\Omega_\phi h^2<\Omega_{\rm DM}h^2$, the value of the plateau of $\Omega_{\phi} h^2$ at low $\Sigma_0^2$ (see right panel of Fig.~\ref{fig:Omegah2_nokd_wkd}) does not exceed $\Omega_{\rm DM}h^2$. Since the plateau height scales as $g_x^{-2} \epsilon^{-1}$, at the boundary where $\Omega_\phi h^2=\Omega_{\rm DM}h^2$, we obtain the following scaling: $\epsilon \propto g_x^{-2}$, which is the slope observed in the upper-right corner of the figures, between the white and colored regions.
Second, in the region where $\Omega_\phi h^2 >\Omega_{\rm DM}h^2$, the minimal value of $\Omega_\phi h^2$ at $\Sigma_0^2=\bar{\Sigma}_{\rm min}^2$ is always larger than $\Omega_{\rm DM}h^2$. Given the scaling of $\left. \Omega_{\phi} h^{2} \right|_{\rm min}$ shown in Eq.~\ref{eq:Oh2_approx_trough_min}, at the boundary where $\left. \Omega_{\phi} h^{2} \right|_{\rm min}=\Omega_{\rm DM}h^2$, we recover the scaling $\epsilon \propto g_x^{-2}$. This slope is observed in the lower-left corner, between the white and colored regions in the figures.

Note that the discontinuity observed around $g_x=10^{-3}$ stems from the QCD phase transition as we shall explain hereafter. As shown in appendix~\ref{sec:phi_therm_annihilation}, kinetic decoupling happens after freeze-out, and even possibly after the QCD phase transition. In the black region of the upper-left panel of Fig.~\ref{fig:pedaplotsDM1GeV_lowsol}, $x_{\rm KD}\approx 1$, which means that, since $m_\phi=1\;\rm GeV$, $T_{\rm KD}\approx 1\;\rm GeV$.  This is a temperature above the QCD phase transition. In the green region,  $x_{\rm KD}\approx 10$, thus $T_{\rm KD}\approx 0.1\;\rm GeV$ which is a temperature below the QCD phase transition.
As explained in Sec.~\ref{subsec:withkd}, the later the kinetic decoupling, the larger the relic abundance. Thus, for given model parameters, if $T_{\rm KD}$ becomes smaller than $T_{\rm QCD}$, the temperature $T_\phi$ starts dropping later, in particular, because of latent heat release, and annihilation occurs in a less dense medium resulting in a larger value for $\left. \Omega_{\phi} h^{2} \right|_{\rm min}$. Because the latter scales as $\epsilon^{-2/3}$, one has to boost the value of $\epsilon$ to satisfy $\Omega_\phi h^2 =\Omega_{\rm DM}h^2$, hence the jump in the plot observed around $g_x\approx 10^{-3}$.

Another feature we would like to explain is the sharp vertical cut in the allowed parameter space around $g_x\approx 10^{-4}$. In this region of the $(g_{x} , \epsilon)$ plane, the value of $\bar{\Sigma}_{\rm min}^{2}$, as given by Eq.~\ref{eq:Sigma_bar_min_def}, becomes larger than $1$. In the trough regime, as the mass degeneracy parameter $\Sigma_{0}^{2}$ increases, the relic abundance decreases according to relation~\ref{eq:Oh2_approx_trough_decreasing}. A minimum is reached near $\Sigma_{0}^{2} \simeq 0.01$. This corresponds here to a plasma temperature $T_{\rm M} \simeq 10 \, {\rm MeV}$. Requiring that this minimum is less than $\Omega_{\rm DM} h^{2}$ sets a lower limit on $g_{x}$. With a kinetic decoupling temperature $T_{\rm KD}$ of order 50~MeV, using relation~\ref{eq:Oh2_approx_trough_decreasing} leads to an approximate value for that bound of $g_{x} = 1.4 \times 10^{-4}$, in excellent agreement with the numerical result. We reproduced this exercise for a DM mass of $200\rm \,MeV$ and $5\,\rm GeV$, and found limiting $g_x$ values of $3.0 \times 10^{-5}$ and $5.2 \times 10^{-4}$, in agreement with the summary plots Fig.~\ref{fig:resultsDM200MeV} and Fig.~\ref{fig:resultsDM5GeV}, respectively.
The remaining panels in Fig.~\ref{fig:pedaplotsDM1GeV_lowsol} will be discussed in the following subsections.

\subsubsection{\bf Right branch solutions -- $\Sigma_{0}^{2} > \bar{\Sigma}_{\rm min}^{2}$}

The corresponding figures for the right branch solutions are shown in Fig.~\ref{fig:pedaplotsDM1GeV_highsol}. The shape of the lower boundary, below which $\Omega_\phi h^2 >\Omega_{\rm DM}h^2$ is exactly the same as in the left branch case, and is explained by the same arguments. The main difference with the left branch solution is that the upper boundary above which $\Omega_\phi h^2 <\Omega_{\rm DM}h^2$ does not exist in this case. In the right branch, it is always possible to increase the value of $\Sigma_0^2$ and thus to increase $\Omega_\phi h^2$ until we reach the required value. When $\Sigma_0^2$ approaches $1$, the dark photon mass $m_{x}$ explodes and the Breit-Wigner resonance disappears.

\subsection{Limits on $g_x$ and $\epsilon$ from the annihilation cross-section today}
\label{subsec:lim_xs_mw}
As discussed in Sect.~\ref{sec:model}, the peculiarity of our model is the non-trivial dependence of the dark matter annihilation cross-section with the dispersion velocity $\Sigma^2$, which, in the Breit-Wigner regime, peaks at $\Sigma_0^2$. The value of $\Sigma_0^2$ is set by the requirement $\Omega_\phi h^2 =\Omega_{\rm DM}h^2$. Using the scalings of Eq.~\ref{eq:Oh2_approx_trough_decreasing} and Eq.~\ref{eq:Oh2_approx_trough_increasing} (also shown in the right panel of Fig.~\ref{fig:Omegah2_nokd_wkd}) implies that $\Sigma_0^2$ scales as $g_x^{-4}$ for the left branch and as $\epsilon^{2}$ for the right branch. The resulting $\Sigma_0^2$ values are shown with a color scale, for these two cases, in the top-right panels of Fig.~\ref{fig:pedaplotsDM1GeV_lowsol} and Fig.~\ref{fig:pedaplotsDM1GeV_highsol}. As predicted by the scalings, we note the almost complete independence of $\Sigma_0^2$ values on $\epsilon$ and $g_x$ in the first and second case, respectively.

From the values of $\Sigma_0^2$, $g_x$ and $\epsilon$, we compute the corresponding DM annihilation cross-section today in the Milky Way $\langle \sigma_{\rm ann} v \rangle_{\rm MW}$ in order to compare with observational limits and to set constraints on our model.
For this, we use relation~\ref{eq:asv_1} and set the DM dispersion velocity $\Sigma$ equal to its value $\Sigma_{\rm MW}$ in the Milky Way halo.
Strictly speaking, the chosen value for $\Sigma_{\rm MW}$ should depend on the position in the Milky Way (see e.g. \cite{Lacroix:2020lhn}). However, given that the X-ray constraints that we will be using are drawn from multiple lines of sight, for simplicity, we set $\Sigma^2_{\rm MW}$ to the fiducial value $3\times10^{-7}\;c^2$. This value is obtained for an isothermal DM halo accounting for a flat rotation curve with the typical $230\,\rm km/s$ circular velocity.
We anticipate that $\langle \sigma_{\rm ann} v \rangle_{\rm MW}$ is maximal at the Breit-Wigner peak, when the Milky Way dispersion velocity $\Sigma^2_{\rm MW}$ is of order the mass degeneracy parameter $\Sigma_{0}^{2}$.
We display $\langle \sigma_{\rm ann} v \rangle_{\rm MW}$ with a color scale in the lower-left panels of Fig.~\ref{fig:pedaplotsDM1GeV_lowsol} and Fig.~\ref{fig:pedaplotsDM1GeV_highsol}, for the left and right branches, respectively. Hereafter, we discuss these two cases independently.

\subsubsection{\bf Left branch solutions -- $\Sigma_{0}^{2} < \bar{\Sigma}_{\rm min}^{2}$}

In the lower-left panel of Fig.~\ref{fig:pedaplotsDM1GeV_lowsol}, the color shows that the DM annihilation cross-section values $\langle \sigma_{\rm ann} v \rangle_{\rm MW}$ are almost independent of $\epsilon$. Going from small to large $g_x$ values, $\langle \sigma_{\rm ann} v \rangle_{\rm MW}$ increases rapidly, reaching a maximum at the Breit-Wigner peak, followed by a slow decrease toward the largest $g_x$ values. In fact, this behavior can be easily explained by the variations of $\langle \sigma_{\rm ann} v \rangle$ with $\Sigma^2$ as shown in Fig.~\ref{fig:J_figure_3}. 
Going from small to large $g_x$ values implies decreasing $\Sigma_0^2$ values as shown in the upper-right panel. When $g_{x}$ increases, the Breit-Wigner peak is shifted toward lower DM dispersion velocities. The annihilation cross-section in the Milky Way follows a $p$-wave behavior and increases until a maximum is reached when $\Sigma^2_{\rm MW}$ is of order $\Sigma_{0}^{2}$. This occurs at $g_{x}$ of order $10^{-3}$, above which $\langle \sigma_{\rm ann} v \rangle_{\rm MW}$ decreases.

The reasoning developed above is only valid at first order, since we notice deviations from a unique dependence of $\langle \sigma_{\rm ann} v \rangle_{\rm MW}$ on $g_x$, at large enough $\epsilon$ values. For example, we see that for a given value of $g_x$, e.g. $10^{-2}$, going from small to large $\epsilon$ values, $\langle \sigma_{\rm ann} v \rangle_{\rm MW}$ is constant and then increases. This leads to a chevron-like feature for $\langle \sigma_{\rm ann} v \rangle_{\rm MW}$ in the $(g_x , \epsilon)$ plane. We explain this feature by the change of behavior of $\langle \sigma_{\rm ann} v \rangle$ going from the Breit-Wigner resonance to the high-velocity regime. From Eq.~\ref{eq:asv_1}, we know that this transition typically occurs when $J(a,b)$ starts saturating to 1 for low values of $|a|$ (see also Fig.~\ref{fig:J_figure_1}). When this happens, $\langle \sigma_{\rm ann} v \rangle_{\rm MW}\propto g_x^2 \epsilon^2$, a scaling which is observed in Fig.~\ref{fig:pedaplotsDM1GeV_lowsol}. When the parameter $|a|$ is not too small, the function $J(a,b)$ is given by its approximation~\ref{eq:J1}. The DM annihilation cross-section is proportional to ${\Sigma_{0}^{2}}/{\Lambda_{0}^{2}}$. Along the left branch, we can use relation~\ref{eq:ratio_L02_on_S02} and we find that $\langle \sigma_{\rm ann} v \rangle$ scales like $g_{x}^{2} \Sigma_{0}^{3}$. The tip of the chevron, i.e. the position of this transition, can be found by equating the approximation $J_{1}(a,b)$, valid at the Breit-Wigner peak, to $1$. From Eq.~\ref{eq:J1}, this means that $\sqrt{\pi a} (|a|/b)= 1$, and we readily get that $\epsilon^2 \propto \Sigma_0^3 $. Hence, as $\Sigma_0^2 \propto g_x^{-4}$ for left branch solutions, we find that $\epsilon \propto \Sigma_0^{3/2} \propto g_x^{-3}$, which corresponds to the line one can draw through the tips of the chevrons in the lower-left panel.

\subsubsection{\bf Right branch solutions -- $\Sigma_{0}^{2} > \bar{\Sigma}_{\rm min}^{2}$}

On the lower-left panel of Fig.~\ref{fig:pedaplotsDM1GeV_highsol}, 
we see that $\langle \sigma_{\rm ann} v \rangle_{\rm MW}$ values are almost independent of $g_x$. Going from large to small $\epsilon$ values, this cross-section exhibits a rapid increase, reaches a maximum, and then decreases with $\epsilon$. In the same way, as for the left branch case, this behavior is explained by the variations of $\langle \sigma_{\rm ann} v \rangle$ with $\Sigma^2$ as shown in Fig.~\ref{fig:J_figure_3}. Going from large to small $\epsilon$ values implies decreasing $\Sigma_0^2$ values (upper-right panel of Fig.~\ref{fig:pedaplotsDM1GeV_highsol}), and shifting the position of the Breit-Wigner peak of $\langle \sigma_{\rm ann} v \rangle$ toward smaller $\Sigma^2$. Since $\langle \sigma_{\rm ann} v \rangle_{\rm MW}$ is evaluated at fixed DM dispersion velocity $\Sigma^2_{\rm MW}$, the annihilation cross-section in the Milky Way follows a $p$-wave behavior as $\epsilon$ decreases. It increases until a maximum is reached when $\Sigma^2_{\rm MW}$ is of order $\Sigma_{0}^{2}$. This occurs at $\epsilon$ of order $10^{-8}$. Below that value, $\langle \sigma_{\rm ann} v \rangle_{\rm MW}$ decreases.

As for the left branch, this picture is only valid at first order, important corrections occurring for $\epsilon > 10^{-7}$. For example, at $g_x=10^{-2}$, going from $\epsilon=10^{-7}$ to  $10^{-2}$, $\langle \sigma_{\rm ann} v \rangle_{\rm MW}$ decreases and then increases again. The latter behavior can be explained by going into some more detail. From Eq.~\ref{eq:asv_1}, we know that this behavior occurs in the $p$-wave regime when $J(a,b)$ is well approximated by $J_2(a,b)$ (see Eq.~\ref{eq:definition_J2} and Fig.~\ref{fig:J_figure_1} of the appendix). In this approximation, one can easily show that $\langle \sigma_{\rm ann} v \rangle_{\rm MW}\propto \epsilon^2 / \Sigma_0^4$ and, given the scaling of the right branch $\Sigma_0^2 \propto \epsilon^2$ (see Eq.~\ref{eq:Oh2_approx_trough_increasing}), we recover that $\langle \sigma_{\rm ann} v \rangle_{\rm MW}\propto \epsilon^{-2}$. However, as one can notice from the right panel of Fig.~\ref{fig:Omegah2_nokd_wkd}, the scaling drawn from Eq.~\ref{eq:Oh2_approx_trough_increasing} fails to reproduce the sharp rise of $\Omega_\phi h^2$ as $\Sigma_0^2$ tends to 1. Instead, we can assume that $\Sigma_0^{2 \beta} \propto \epsilon^2$ with $\beta$ being a number larger than 1. In that case $\langle \sigma_{\rm ann} v \rangle_{\rm MW}\propto \epsilon^{2-4/\beta}$, which means that, when $\beta$ gets larger than 2, $\langle \sigma_{\rm ann} v \rangle_{\rm MW}$ starts increasing with growing $\epsilon$ values. This explains the violet-blue spot of the lower-left panel of Fig.~\ref{fig:pedaplotsDM1GeV_highsol}.

\textit{X-rays constraints}

\begin{figure}[h!]
\centering
\includegraphics[width=0.49\columnwidth]{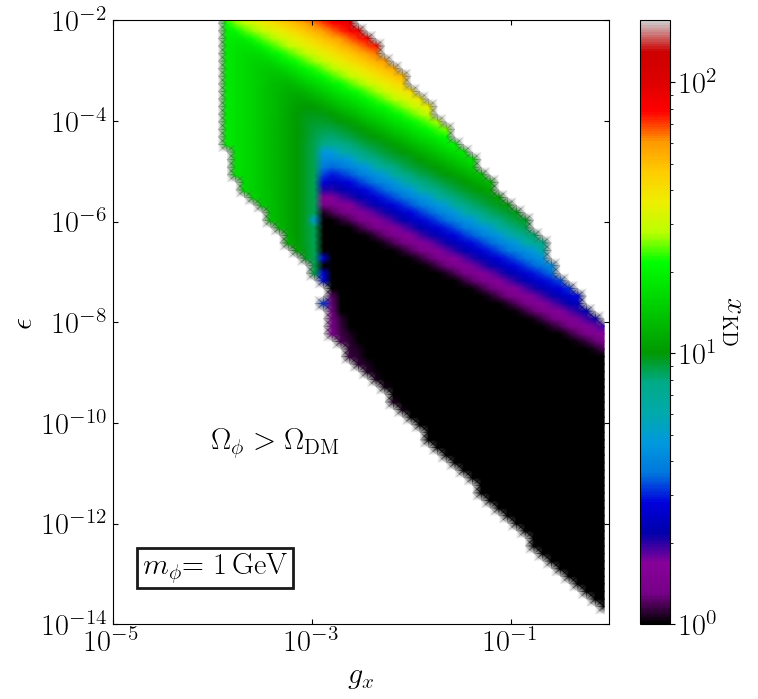}
\includegraphics[width=0.49\columnwidth]{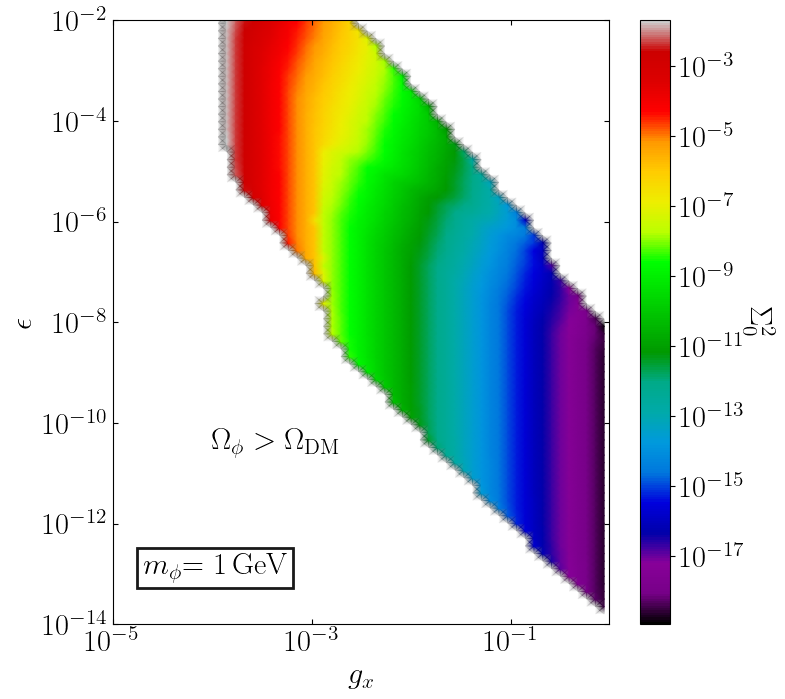}
\includegraphics[width=0.49\columnwidth]{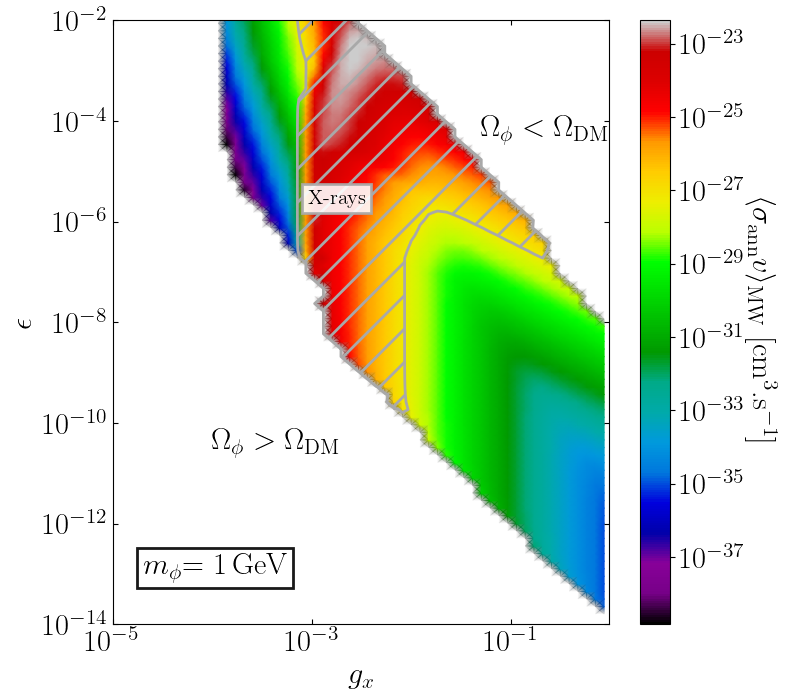}
\includegraphics[width=0.49\columnwidth]{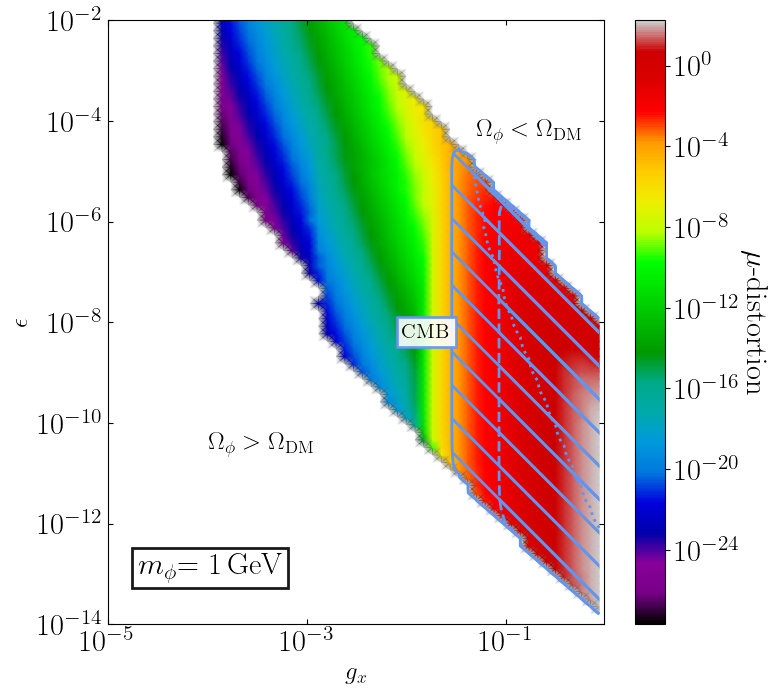}
\caption{
Allowed parameter space in the $(g_x , \epsilon)$ plane, for \textit{left branch solutions} ($\Sigma_{0}^{2} < \bar{\Sigma}_{\rm min}^{2}$), and for a DM mass of $1\;\rm GeV$. From the upper-left to the lower-right panel, the color code shows $x_{KD}$, $\Sigma_0^2$, the corresponding DM annihilation cross-section in the Milky Way $\langle \sigma_{\rm ann} v \rangle_{\rm MW}$ and the CMB $\mu$-distortion. In the bottom-left panel, the hatched region features the exclusion constraints by XMM X-ray measurements. In the bottom-right panel, the forbidden hatched region is drawn from CMB $\mu$-distortion as observed by FIRAS (COBE). It encompasses the domains excluded by CMB $y$-distortions (dashed line) and anisotropies (dotted line).%
The full interpretation of these plots is provided in the text.
}
\label{fig:pedaplotsDM1GeV_lowsol}
\end{figure}

\begin{figure}[h!]
\centering
\includegraphics[width=0.49\columnwidth]{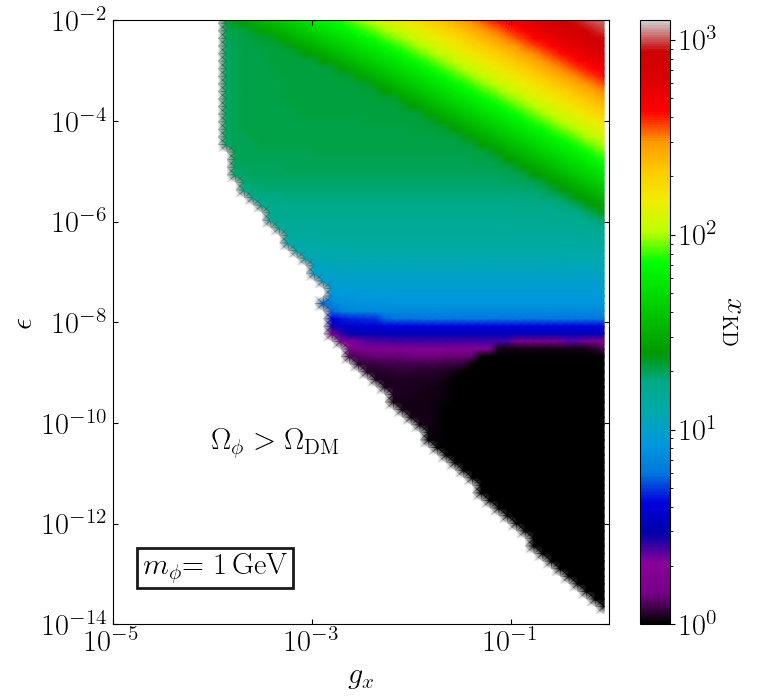}
\includegraphics[width=0.49\columnwidth]{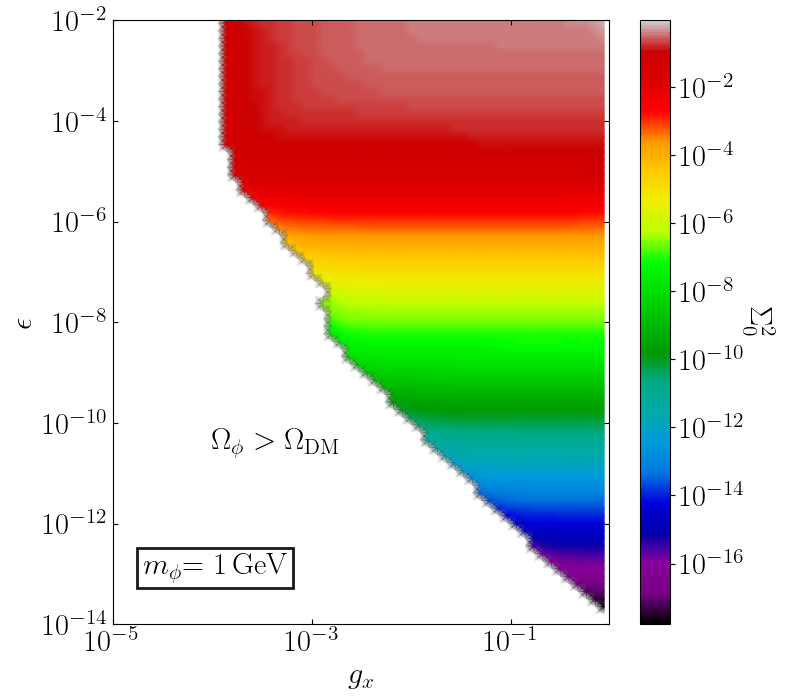}
\includegraphics[width=0.49\columnwidth]{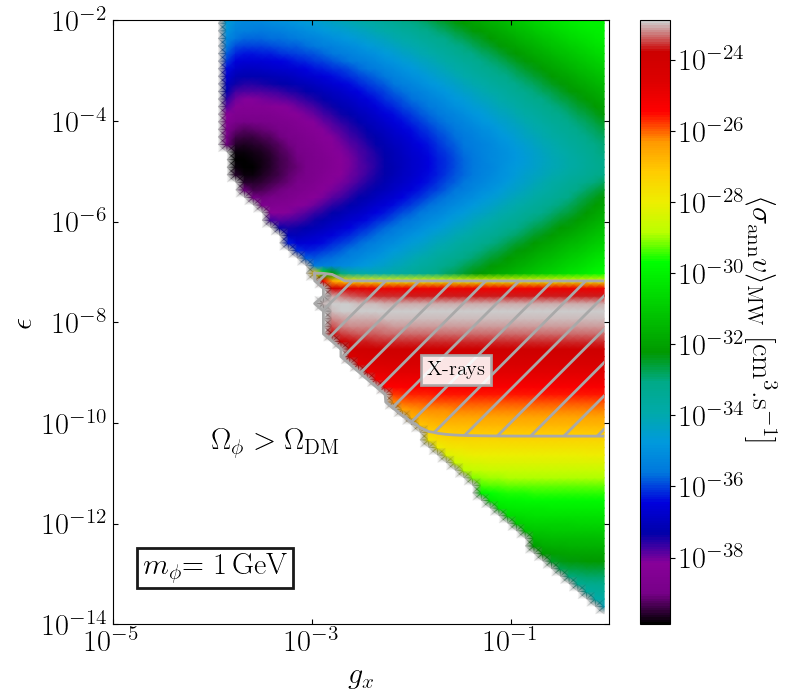}
\includegraphics[width=0.49\columnwidth]{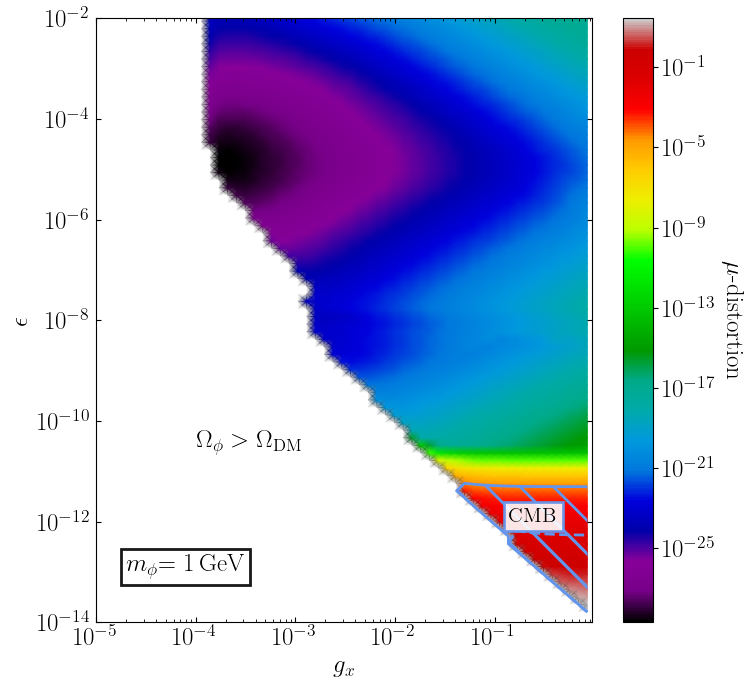}
\caption{
Allowed parameter space in the $(g_x , \epsilon)$ plane, for \textit{right branch solutions} ($\Sigma_{0}^{2} > \bar{\Sigma}_{\rm min}^{2}$) and for a DM mass of $1\;\rm GeV$. From the upper-left to the lower-right panel, the color code shows $x_{KD}$, $\Sigma_0^2$, the corresponding DM annihilation cross-section in the Milky Way $\langle \sigma_{\rm ann} v \rangle_{\rm MW}$ and the CMB $\mu$-distortion, respectively. In the bottom panels, the hatched regions draw the exclusion constraints by XMM X-ray measurements and FIRAS (COBE) on the $\mu$-distortion (solid line) and on the $y$-distortion (dashed line). The full interpretation of these plots is provided in the text.
}
\label{fig:pedaplotsDM1GeV_highsol}
\end{figure}

Light stable fermions (typically $e^+e^-$ or $\mu^+\mu^-$) from DM annihilating today, may energize the low-energy photons of the interstellar radiation field of the Galaxy and generate a sizable X-ray emission, via the Inverse Compton effect. In the lower-left panels of Fig.~\ref{fig:pedaplotsDM1GeV_lowsol} and Fig.~\ref{fig:pedaplotsDM1GeV_highsol}, we draw with a hatched region, the X-ray constraints of XMM Newton taken from~\cite{Cirelli:2023tnx}, that we find to be the strongest of the literature. The constraints shown include only DM annihilating into $e^+e^-$ pairs. This annihilation channel provides the strongest limit on $\langle \sigma_{\rm ann} v \rangle_{\rm MW}$. The analysis by~\cite{Cirelli:2023tnx} provides the following upper bound on $\langle \sigma_{\rm ann} v \rangle_{\rm MW}$ for DM annihilating into $e^+e^-$: $7.8 \times 10^{-29}$, $1.5 \times 10^{-28}$, and $2.3 \times 10^{-27} \; {\rm cm^3 \, s^{-1}}$ for $200\,\rm MeV$, $1\,\rm GeV$ and $5\,\rm GeV$ DM mass, respectively.
In our case, we apply these limits on the product $B_{e^+e^-} \times \langle \sigma_{\rm ann} v \rangle_{\rm MW}$, with the branching ratio $B_{e^+e^-} \equiv \tilde{Q}^2_{e^+e^-}/\tilde{Q}^2$ (see Eq.~\ref{eq:definition_tilde_Q_b}, where $\tilde{Q}^2_{e^+e^-}$ only includes $e^+e^-$ in the sum). The other constraints discussed in~\cite{Cirelli:2023tnx}, which are based on $\mu^+\mu^-$ and $\pi^+\pi^-$, are not competitive and have not been implemented.

%
%

%
%

\subsection{Limits on $g_x$ and $\epsilon$ from CMB spectral and angular distortions}
\label{subsec:lim_gx_epsilon_mudistorsion}
DM annihilation in the early Universe injects energy in the primordial plasma and may generate distortions in the radiation spectrum. These distortions get frozen after recombination and leave indelible deviations of the CMB from a pure black body spectrum (see e.g. \cite{Chluba:2018cww} for an introduction). The characteristics of the distortions can be addressed very generally by solving the so-called Kompaneets equation~\cite{1956ZhETF..31..876K} which describes the Comptonization of photons by free thermal electrons. However, depending on the epoch at which the energy is released, approximations can be accurately used and avoid solving this non-linear equation. If the injection occurs for redshifts larger than $z_{DC}=1.98\times 10^6$~\cite{Chluba:2015bqa}, the energy is rapidly redistributed to the photons via Compton scattering, while the number of photons is adjusted by photon non-conserving processes: double Compton (DC), which sets $z_{DC}$, and thermal Bremsstrahlung. In this case, the CMB spectrum just undergoes a temperature shift. If the energy injection occurs after $z_{DC}$, but before $z_C=5.8\times 10^4$~\cite{Chluba:2015bqa}, the number of photons remains unchanged while Comptonization, which sets $z_C$, ensures an efficient redistribution of energy among photons. The energy per photon is increased and leads to a non-vanishing chemical potential $\mu$ in the Bose-Einstein spectrum, referred to as $\mu$-distortion. Finally, if the energy injection happens for redshifts smaller than $z_C$, a typical Compton $y$-distortion arises when scatterings become inefficient in exchanging energy. This classification is useful to easily compare with present constraints from FIRAS~\cite{Fixsen_1996,Bianchini:2022dqh} which set the upper limits $|\mu| < 4.7 \times 10^{-5}$ and $|y|< 1.5\times 10^{-5}$, but in general, there is a large variety of possible distortions with a smooth transition from $y$ to $\mu$ distortions that could give complementary information.

In practice we compute the $\mu$ and $y$-distortions using the prescription presented in~\cite{Chluba:2015bqa} and \cite{Chluba2016}, such as:
\begin{align}
&\mu=\int d\mu \approx 1.4\int_{0}^{\infty} {\cal J}_\mu(z)\frac{d \rho_\gamma}{\rho_\gamma} \\
&y=\int d\mu \approx \frac{1}{4}\int_{0}^{\infty} {\cal J}_y(z)\frac{d \rho_\gamma}{\rho_\gamma}\;,
\label{eq:mudist}
\end{align}
where we have introduced the window functions ${\cal J}_\mu(z)$ and ${\cal J}_y(z)$ defined by:
\begin{align}
 &{\cal J}_\mu(z)= \left(1 - \exp \left[ - \left( \frac{1 + z}{z_{C}} \right)^{1.88} \right] \right)\times \exp\left[- \left(\frac{z}{z_{DC}}\right)^{5/2} \right]\;,\\
 & {\cal J}_y(z)= \left(1 + \left(\frac{1 + z }{6.0\times 10^4}\right)^{2.58} \right)^{-1}\;.
\end{align}
We compute the energy released in the photon bath $d \rho_\gamma = f_{\rm em} \times \asv n_{\phi}^2 dt \times 2 m_\phi$ corresponding to the energy injected by DM annihilation during $dt$. The pre-factor $f_{em}$ accounts for the fraction of the energy injected in the form of electromagnetic energy and depends on the dark matter mass. This factor has already been computed for this model in Ref.~\cite{Bernreuther:2020koj} (see their Eq.~3.4 and Fig.~2), and we extract the following values: 0.52, 0.28, 0.27, for $m_\phi = 200\rm \,MeV$, $1\rm\,GeV$ and $5\rm\,GeV$. However, in this paper, the authors were interested in the impact of the ionization on the CMB and so included ionization efficiencies. Hence, using their values for $f_{em}$, we slightly underestimate the $\mu$ and $y$ values we compute. The constraints we will draw from FIRAS~\cite{Fixsen_1996,Bianchini:2022dqh} upper bounds on $\mu$ an $y$, thus give conservative limits in the $(g_x , \epsilon)$ parameter space.

The results for the $\mu$-distortion are shown with a color scale on the lower-right panel of Fig.~\ref{fig:pedaplotsDM1GeV_lowsol} and Fig.~\ref{fig:pedaplotsDM1GeV_highsol}, for the left and right branch, respectively. As for $\langle \sigma_{\rm ann} v \rangle_{\rm MW}$, the $\mu$ values are primarily dependent on $g_x$ (resp. $\epsilon$) in the left (resp. right) branch case. The evolution of $\mu$ with these parameters follows the same trend as for $\langle \sigma_{\rm ann} v \rangle_{\rm MW}$. This is not a coincidence since, by definition, $\mu$ traces the DM annihilation history in the redshift window $[z_{DC},z_C]$. In particular, we remark first that the peak of $\mu$ values occurs at larger $g_x$ (resp. smaller $\epsilon$) compared to the peak in $\langle \sigma_{\rm ann} v \rangle_{\rm MW}$ in the left (resp. right) branch. This is explained by the fact that the DM dispersion velocity $\Sigma^2$ at the epoch $[z_{DC},z_C]$ is smaller than the virialized one in the DM galactic halo today. Second, the peak of the $\mu$ values is broader (the red-colored region is wider) than the one seen on $\langle \sigma_{\rm ann} v \rangle_{\rm MW}$. This is because the DM annihilation is integrated over the full redshift window $[z_{DC},z_C]$, including a broad range of $\Sigma^2$ values with respect to the plot of $\langle \sigma_{\rm ann} v \rangle_{\rm MW}$ which displays a picture of the annihilation for the specific $\Sigma^2_{\rm MW}$ value. The hatched-region delineated by the solid line corresponds to the region of the parameter space excluded by the upper bound $|\mu| < 4.7 \times 10^{-5}$ from FIRAS~\cite{Fixsen_1996,Bianchini:2022dqh}. The results for the $y$-distortion look very similar in shape to the ones for the $\mu$-distortion, and we only show with a dashed line the region of the parameter space excluded by the upper bound $|y|< 1.5\times 10^{-5}$ from FIRAS~\cite{Fixsen_1996}. We note that the constraints from the $y$-distortion are always weaker than the ones from the $\mu$-distortion. In the summary plots of Fig.~\ref{fig:resultsDM200MeV}, \ref{fig:resultsDM1GeV} and \ref{fig:resultsDM5GeV}, we call CMB constraints the limits from $\mu$-distortion.

For completeness, we also report the constraints coming from CMB anisotropies. Following Ref.~\cite{Bernreuther:2020koj}, we have computed the effective annihilation parameter defined in their Eq.~(3.1):
\beq
p_{\rm ann} = \frac{R^{2}}{2} \, f_{\rm em} \, \frac{\asv_{\rm CMB}}{m_{\phi}} \,,
\eeq
where $R\equiv {n_{\phi}(\rm CMB)}/{n_{\phi}(\text{today})}$ being the ratio between the dark matter density at the CMB epoch and today. As the injection of energy in the primordial plasma from DM annihilation has maximal effect for $z \approx 600$, we have computed $p_{\rm ann}$ at that redshift. We have applied the limits on $p_{\rm ann}$ from \cite{Planck:2018vyg}. Thus, for the \textit{left branch solutions} ($\Sigma_{0}^{2} < \bar{\Sigma}_{\rm min}^{2}$), the corresponding exclusion region stands to the right of the dotted line in the lower-right panel of Fig.~\ref{fig:pedaplotsDM1GeV_lowsol} for a 1~GeV DM scalar. We find the same trend for the 200~MeV and 5~GeV cases, with much weaker constraints for the latter case. This is expected insofar as the peak of DM annihilation takes place earlier for heavier masses. For all the masses considered, we always find the $p_{\rm ann}$ constraint to be sub-leading with respect to the $\mu$-distortions one.
For \textit{right branch solutions} ($\Sigma_{0}^{2} > \bar{\Sigma}_{\rm min}^{2}$), we find no constraints from $p_{\rm ann}$ since the peak of DM annihilation occurs well before recombination.

\subsection{Limits on $g_x$ and $\epsilon$ from direct detection and collider experiments}
\label{subsec:lim_gx_epsilon_other}
Other significant limits on the $(g_x , \epsilon)$ parameter space come from the DM direct detection experiments as well as from the accelerator searches.

\subsubsection{\bf Direct detection limits }
\label{subsubsec:direct_detection}
Dark photon interacting with $\phi$ and the SM particles can be constrained by the direct detection limits because DM can scatter off nucleons or electrons in the detector material through $A'^\mu$  exchange in $\phi$ SM $\to \phi$ SM processes. The spin-independent elastic scattering cross-section is given by~\cite{Essig:2015cda,Feng:2017drg}
\beq
\sigma^{\rm SI}_{\phi T}=\frac{1}{\pi}\frac{q_T^2\,e^2\,\epsilon^2\,g_x^2}{(m_x^2+q^2)^2} \, \mu_T^2 \,.
\label{dd_xs}
\eeq
where $\mu_T$ is the reduced mass of DM with the target species $T$ = electron/nucleon and $q_T$ is the ${\rm U'}(1)$ charge of the target particle, which is $1/2$ for a proton, $-1/2$ for a neutron and $1$ for an electron. $q$ is the momentum transfer.
For our parameter range of interest, $m_x \approx 2 m_\phi$, which implies the zero-momentum transfer limit for the case of a heavy mediator, i.e. $m_x \gg q$.  Eq. \ref{dd_xs} can be approximated as:
\beq
\sigma^{\rm SI}_{\phi T}\approx\frac{1}{\pi}\frac{q_T^2\,e^2\,\epsilon^2\,g_x^2}{m_x^4} \, \mu_T^2 \,.
\label{dd_xsApprox}
\eeq

The limits from electron recoil experiments are relevant for lighter DM whereas for larger $m_\phi$( $ie$, $m_\phi \gtrsim 100$ MeV), the limits from DM-nucleon scattering kick in. Although detection sensitivity rapidly decreases at low recoil energies, there have recently been substantial efforts in low-mass detection techniques, with experiments like PandaX-4T and XENON IT using liquid Xenon (LXe) detectors. DM scattering off the target nuclei produces scintillation photons (S1) and ionized electrons (S2) (produced through Migdal effect). Both S1 and S2 signals are used for low-mass DM detection, although the S2-only signal is more efficient toward lower energy. For nuclear recoil experiments, around $\sim 100$ MeV DM mass, the most efficient limits come from the S2-only bounds. In this category, PandaX-4T limits improve the previous limits from CRESST-III and XENON-IT(M) by several orders~\cite{PandaX:2023xgl,Billard:2021uyg}. For heavier DM of a few GeV mass, the most stringent limits come from PandaX-4T (S1-S2), which improves the old XENON-IT(S2) limits by $\sim$ a factor of 2~\cite{PandaX:2023xgl,Billard:2021uyg}.
In the case of electron recoil, for DM mass $m_\phi\lesssim 10$ MeV, the most stringent limits come from SENSEI~\cite{SENSEI:2020dpa}, while for larger DM mass these limits are surpassed by DarkSide-50~\cite{Franco:2023sjx} and eventually XENON-IT(S2) limits take over around $m_\phi \sim 30$ MeV~\cite{XENON:2019gfn}. Recent results from PandaX-4T improve XENON-1T limits by almost an order for DM mass $\gtrsim 50$ MeV~\cite{PandaX:2022xqx}.

In Figs.~\ref{fig:resultsDM200MeV}-\ref{fig:resultsDM5GeV}, the exclusion limits by the existing direct detection experiments are illustrated in the $(g_x , \epsilon)$ plane. As obvious from Eq.~\ref{dd_xsApprox}, the scattering cross-sections are effectively independent of $\Sigma_0^2$ (it only introduces a negligible correction when taking $2 m_\phi\simeq m_x$), which implies that the exclusion regions in $(g_x , \epsilon)$ plane do not change for the left and right branch solutions.  
For $m_\phi=200$ MeV, the orange shaded region in Fig.~\ref{fig:resultsDM200MeV} corresponds to the region excluded by PandaX-4T (S2 only+Migdal) limits, which is the most constraining so far with the allowed upper limit of spin-independent scattering cross-section reaching down to $\sigma^{\rm SI}_N \simeq 3.25\times 10^{-38}\,{\rm cm^2}$~\cite{PandaX:2023xgl}. The electron recoil limits can be obtained from PandaX-4T (constant $W$ model)~\cite{PandaX:2022xqx}, where the exclusion region corresponds to $\sigma_e^{\rm SI} \gtrsim 2.1\times 10^{-41}\,{\rm cm^2}$. These limits are much less constraining in comparison with the nuclear recoil bounds and therefore are not shown in the figure. In Fig~\ref{fig:resultsDM1GeV} and \ref{fig:resultsDM5GeV}, the orange colored exclusion regions are obtained from  PandaX-4T (S2 only+Migdal) and PandaX-4T (S1-S2)~\cite{PandaX:2023xgl} respectively. We quote the following numbers for the respective allowed upper limits on $\sigma_N^{\rm SI}$ : $\sigma_N^{\rm SI} \simeq 1.61\times 10^{-39}\,{\rm cm^2}$ for $m_\phi=1$ GeV and $\sigma_N^{\rm SI} \simeq 1.22\times 10^{-44}\,{\rm cm^2}$ for $m_\phi=5$ GeV. We find that the electron recoil limits are practically irrelevant for these masses and do not constrain the range of $g_x$ and $\epsilon$ considered in the figures.

In Fig.~\ref{fig:resultsDM200MeV}\,-\,\ref{fig:resultsDM5GeV}, the projected limits from the near-future direct detection experiments are shown with orange dashed lines. For 200 MeV and 1 GeV DM, we show projections from DARKSPHERE~\cite{NEWS-G:2023qwh}, an experiment proposed by the NEWS-G collaboration. DARKSPHERE uses a spherical proportional counter, optimised for detecting nuclear recoils with sub-keV energy. We quote the projected exclusions of $\sigma_N^{\rm SI} \gtrsim 2.47\times 10^{-42}\,{\rm cm^2}$ for $m_\phi=200$ MeV and $\sigma_N^{\rm SI} \gtrsim 1.7\times 10^{-43}\,{\rm cm^2}$ for $m_\phi=1$ GeV. For 5 GeV DM, the best projections are obtained from the SBC experiment~\cite{Giampa:20211P,Akerib:2022ort} which uses liquid Argon (LAr) spiked with liquid Xenon (LXe) in a bubble chamber setup. We have used the projected sensitivity of $\sigma_N^{\rm SI} \simeq 7.28\times 10^{-46}\,{\rm cm^2}$ for $m_\phi=5$ GeV in the figures.

\subsubsection{\bf Accelerator limits }
\label{subsubsec:accelerator}
Dark photons can also be probed in various accelerator searches. Depending on the mass scale, the production and the decay of the dark photon in these experiments dictate the sensitivity of detection. While sub-GeV dark photons are best probed in the electron and the proton beam-dumps and other fixed-target experiments, they can also be tested in $e^+e^-$ colliders, up to a few GeV mass~\cite{Graham:2021ggy, Ilten:2018crw}. For dark photons of mass $\gtrsim$ 10 GeV, there are constraints from several LHC searches~\cite{LHCb:2019vmc,LHCb:2017trq}. The signal for all these probes can come from both visible and invisible decay of dark photons. For visible dark photon searches, the typical signal is a pair of leptons whereas the invisible searches are sensitive to the missing energy signal, due to the dominant decay of the dark photon into DM.

In our model, the bounds can be obtained from both visible and invisible dark photon accelerator searches. The partial decay widths of the dark photon into a pair of DM and  SM particles respectively can be obtained from Eq.~\ref{eq:xdecay}, which implies that ${\rm BR}(A' \to {\rm visible}) =(\epsilon^2 e^2 Q^{\prime^2})/(g_x^2\Sigma_0^3/4+\epsilon^2 e^2 Q^{\prime^2})$ and ${\rm BR}(A' \to {\rm invisible}) =(g_x^2\Sigma_0^3/4)/(g_x^2\Sigma_0^3/4+\epsilon^2 e^2 Q^{\prime^2})$. When both decay modes are kinematically accessible, one should take into account the respective branching ratios while computing the constraints. The constraints will therefore depend on $g_x,\Sigma_0$ and $\epsilon$. The constraints on dark photons found in the literature only depend on $\epsilon$ since it is generally assumed that the branching ratio for either visible or invisible decay modes is 1. The $\epsilon$ dependence comes only from the dark photon production which is determined by its interactions with SM particles. For the dark photon mass range of our interest, the following search strategies yield significant limits:

\begin{enumerate}
\item {\bf visible searches :} The best limits for dark photons of mass $\sim 100$ MeV are obtained from BaBar~\cite{BaBar:2014zli} in $e^+e^-$ collisions. The production channel for the dark photon is $e^+e^-\to \gamma\, A'$, and the signal is observed when $A'^\mu$ promptly decays into visible final states: $A'\to e^+e^-$ and $\mu^+\mu^-$. BaBar provides the most stringent limits on visible dark photon decays for 100 MeV $\lesssim m_x\lesssim$ 200 MeV and 1 GeV $\lesssim m_x \lesssim$ 10 GeV. For the mass window $2\,m_\mu\lesssim m_x\lesssim \,0.5~{\rm GeV}$, the prompt and the displaced vertex searches ($A'\to \mu^+\mu^-$) at LHCb produce the best limits~\cite{LHCb:2019vmc}. Here $A'^\mu$ is produced through meson decays : $\pi^0\to A'\gamma$ or $\eta \to A' \gamma$. $m_x \gtrsim 10$ GeV is best constrained by the di-muon searches at LHCb and CMS~\cite{CMS:2022qej}. A heavier dark photon, which is relevant for these searches can be produced in Drell-Yann process $qq^\prime\to A'$ at the LHC. 

\item {\bf invisible searches :} Light dark photons decaying into a pair of DM particles can be probed in the invisible search experiments. For $m_x$ in our range of interest, there are only a few experiments that produce the relevant constraints. For both Babar and LEP, the process under consideration is the production of an ordinary photon accompanied by a dark photon. The decay channel $A'\to \phi\phi^\dag$  gives a signature of monophoton and missing energy.  For $25\ {\rm MeV} \lesssim m_x \lesssim 8\ {\rm GeV}$, BaBar limits are the most stringent~\cite{BaBar:2017tiz}.  For $m_x \gtrsim 8\  {\rm GeV}$, LEP limits prevail~\cite{Graham:2021ggy, DELPHI:2003dlq, DELPHI:2008uka}.
\end{enumerate}

\begin{figure}[h!]
\centering
\includegraphics[width=0.45\columnwidth]{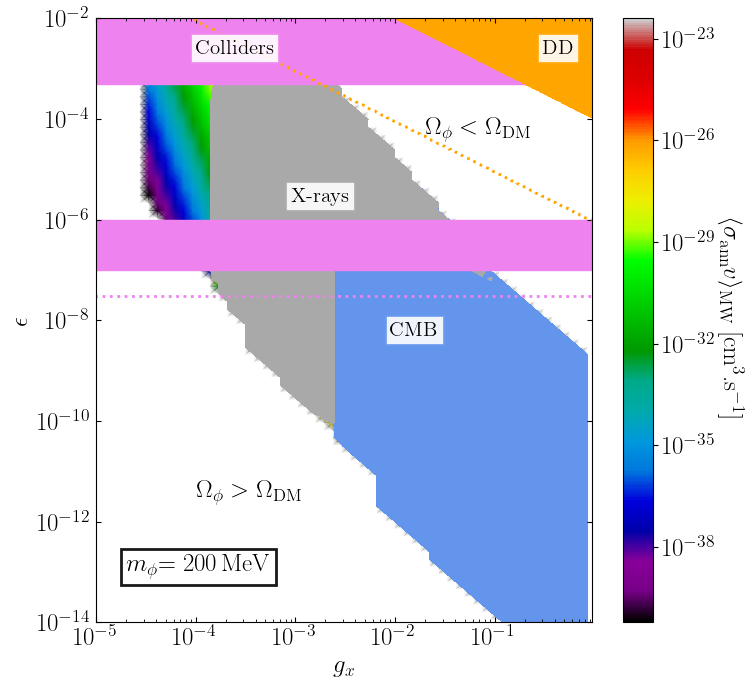}
\includegraphics[width=0.45\columnwidth]{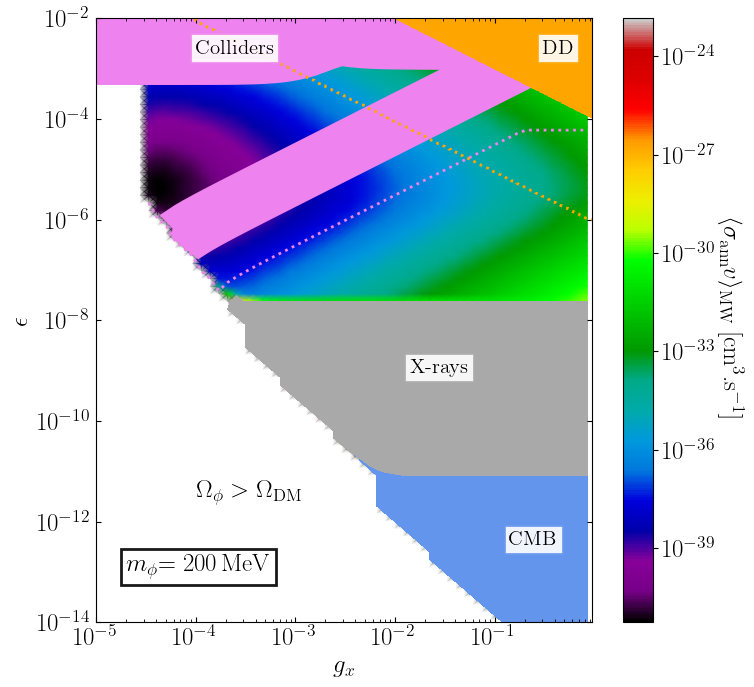}
\caption{
Summary plot for $m_\phi=200\;\rm MeV$: constraints on the parameter space $(g_x , \epsilon)$. The left and right panels correspond to the left ($\Sigma_{0}^{2} < \bar{\Sigma}_{\rm min}^{2}$) and right ($\Sigma_{0}^{2} > \bar{\Sigma}_{\rm min}^{2}$) branch solutions, respectively. The color scale indicates the resulting $\asvMW$. The colored patches show excluded regions from CMB $\mu$-distortion constraints (blue), X-rays measured by XMM (gray), colliders (pink) and direct detection (orange). Projected exclusion limits are depicted for future direct detection experiments (orange dashed lines) and future accelerator searches (pink dashed lines). More details are provided in the Sect.~\ref{sec:results}. 
}
\label{fig:resultsDM200MeV}
\end{figure}

\begin{figure}[h!]
\centering
\includegraphics[width=0.45\columnwidth]{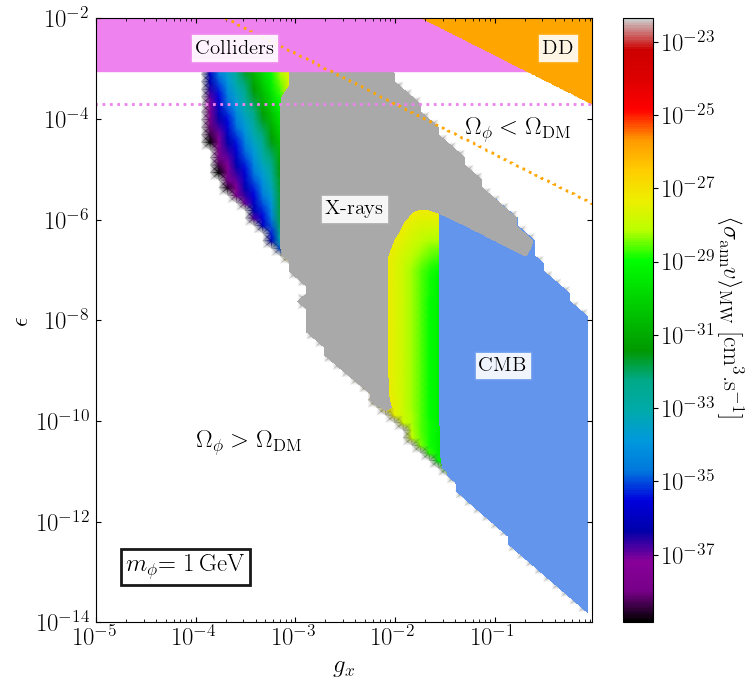}
\includegraphics[width=0.45\columnwidth]{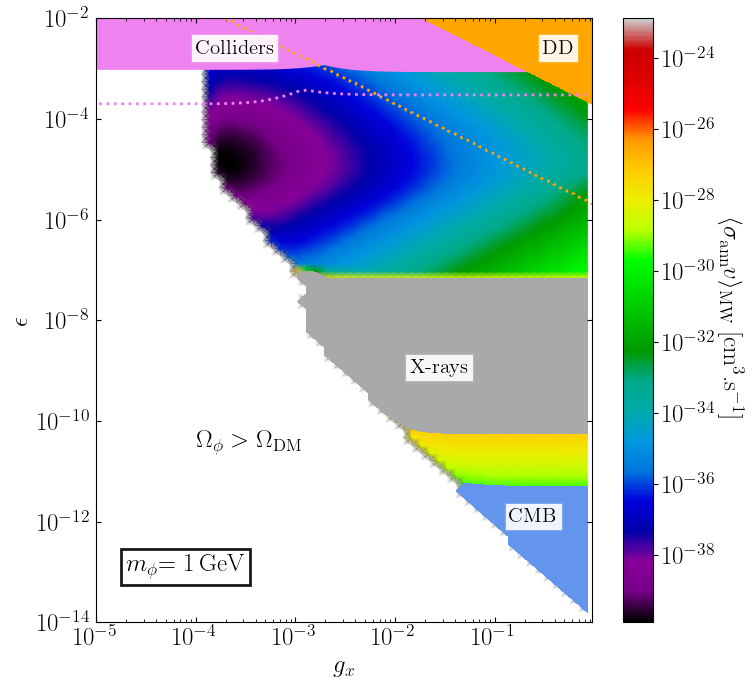}
\caption{
Same legend as Fig.~\ref{fig:resultsDM200MeV}, for $m_\phi=1\;\rm GeV$.
}
\label{fig:resultsDM1GeV}
\end{figure}

\begin{figure}[h!]
\centering
\includegraphics[width=0.45\columnwidth]{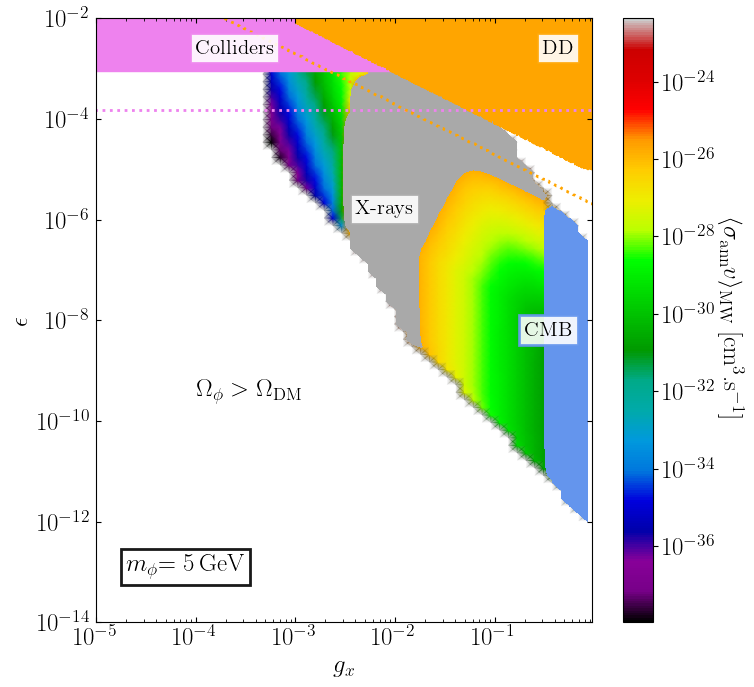}
\includegraphics[width=0.45\columnwidth]{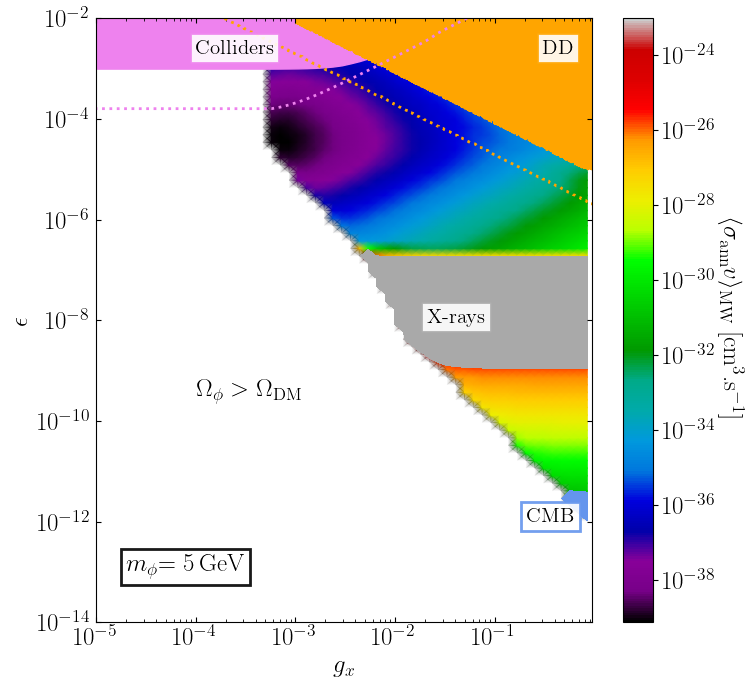}
\caption{
Same legend as Fig.~\ref{fig:resultsDM200MeV}, for $m_\phi=5\;\rm GeV$. 
}
\label{fig:resultsDM5GeV}
\end{figure}

In Figs.~\ref{fig:resultsDM200MeV}-\ref{fig:resultsDM5GeV}, the exclusion regions corresponding to the accelerator limits are shown in magenta shade. We find that the dark photon decays almost entirely into SM particles in the parameter space of interest for the left branch solutions, therefore we use only the limits from visible decays. For the right branch solutions, the dark photon can either decay into SM particles or mostly invisibly, the latter occurs typically in the region at large $g_x$. Therefore, we take the limits from both visible and invisible decay and rescale them in proportion to their respective branching ratios.
For $m_\phi=200\, {\rm MeV}$, (which implies $m_x=400\, {\rm MeV}$), the LHCb di-muon searches constrain $\epsilon_{{\rm LHCb}}\sim 5\times 10^{-4}$. In addition, the displaced vertex searches from the proton beam-dump such as $\nu-{\rm CAL}$~\cite{Ilten:2018crw,Bauer:2018onh,Graham:2021ggy} and CHARM~
\cite{Gninenko:2012eq} constrain smaller values of $
\epsilon$, namely, $10^{-7}\lesssim \epsilon \lesssim 9 \times 10^{-7}$. In these searches, the dark photon is produced from the meson decays such as $\eta\,(\eta^\prime)\to \gamma\, A'$ and is subsequently decayed within the detector into a pair of displaced muons.
As mentioned above, the exclusion bands in Fig.~\ref{fig:resultsDM200MeV} ({\it left}) correspond to the visible decay limits alone. In Fig.~\ref{fig:resultsDM200MeV} ({\it right}),  the magenta-shaded upper band corresponds to the region excluded by LHCb (for visible decay) and BaBar (for invisible decays). When the branching ratios are $\sim 100\%$, the limits on $\epsilon$ correspond to $\epsilon_{\rm LHCb}\sim 5\times 10^{-4}$ and $\epsilon_{\rm BaBar}\sim 1\times 10^{-3}$. As obvious from the expressions for the branching ratios above, we find that toward smaller $g_x$ and larger $\epsilon$, the visible decays dominate and as $g_x$ increases, invisible decays gradually take over. Thus the magenta band for smaller $g_x$ corresponds to the purely visible decay limits from LHCb and large $g_x$ to the purely invisible decay bounds from BaBar. For intermediate $g_x$, where we find substantial contribution for both decay modes, the limits are rescaled with the respective branching ratios as the number of signal events is $\propto \epsilon^2\times {\rm BR}$. However, for the lower band, the only relevant limits are for visible decays. Therefore, for the right branch solution, the bound gradually weakens for larger $g_x$ where we find a dominant invisible decay contribution. Fig.~\ref{fig:resultsDM1GeV} corresponds to $m_x\sim 2\,{\rm GeV}$, which is best constrained by BaBar for both visible and invisible searches. The allowed upper limits of $\epsilon$ from both types of searches are comparable, i.e. $\epsilon \sim 10^{-3}$ for invisible limits and $\sim 9\times 10^{-4}$ for visible limits. Therefore, the magenta exclusion band is almost independent of $g_x$. For Fig.~\ref{fig:resultsDM5GeV}, where the dark photon has a mass of 10 GeV, the limit from BaBar corresponds to $\epsilon\sim 9\times 10^{-4}$ for the dark photon decaying visibly while LEP constrains the invisible decays for $\epsilon\sim 4\times 10^{-2}$. Thus the limits for intermediate $g_x$ are rescaled for the right branch solution and become weaker at large values of $g_x$, see Fig.~\ref{fig:resultsDM5GeV}~({\it right}) plot.

In Fig.~\ref{fig:resultsDM200MeV}\,-\,\ref{fig:resultsDM5GeV}, we also show the projected sensitivities of the upcoming accelerator searches with pink dashed lines. For $m_x=400\, {\rm MeV}$, (which corresponds to $m_\phi=200\, {\rm MeV}$ here), DUNE is expected to provide the most stringent limits among the near-future visible search experiments~\cite{Gori:2022vri}. In DUNE, a light dark photon is produced in the decays of neutral mesons ($\pi^0$ and $\eta$). We quote the projected sensitivity of $\epsilon \sim 3\times 10^{-8}$ when the dark photon decays 100\% into visible particles. On the other hand, LDMX~\cite{LDMX:2018cma}, a proposed electron beam dump experiment, is expected to constrain the invisible searches at $\epsilon \sim 6\times 10^{-5}$ when $x$ mostly decays invisibly. For a GeV scale dark matter, the best visible search projections are obtained from Belle-II prompt searches into dileptons~\cite{Gori:2022vri}. In the figures, we have used the projected upper limits of $\epsilon \sim 2\times 10^{-4}$ for $m_x=2\, {\rm GeV}$ (corresponding to $m_\phi=1\, {\rm GeV}$) and $\epsilon \sim 1.5\times 10^{-4}$ for $m_x=10\, {\rm GeV}$ (for $m_\phi=5\, {\rm GeV}$) from Belle-II. Projections for the invisible searches are taken as $\epsilon \sim 3\times 10^{-4}$ for $m_x=2\, {\rm GeV}$  (for $m_\phi=1\, {\rm GeV}$) from Belle-II~\cite{ipa2018-darkphoton}  when dark photon decays 100\% into invisibles. While realising the future limits, we rescale the numbers according to the branching ratio of dark photon decays, following the same strategy as with the current constraints described before.

%
%
%

%
%

\subsubsection{\bf Summary plots}
\label{subsubsec:sumplots}

The impact of the various constraints on the parameter space of the dark photon model as displayed in Figs.~\ref{fig:resultsDM200MeV}\,-\,\ref{fig:resultsDM5GeV} can be summarized as follows.

\paragraph{Left branch solutions -- $\Sigma_{0}^{2} < \bar{\Sigma}_{\rm min}^{2}$}

For the left branch solutions, large $g_x$ values are constrained both by CMB and X-rays while accelerator searches for dark photons constrain the region at large kinetic mixing.
When the dark photon is about 400 MeV, a constraint from decays of light mesons into dark photons applies as well, the allowed parameter space corresponds to a narrow region with $\epsilon$ roughly between $10^{-6}< \epsilon < 5 \times10^{-4}$ and $g_x<  10^{-4}$. This allowed region shifts toward higher values of $g_x$ as the mass of DM increases.
In the cases where $m_{\phi}$ is equal to 1 and 5~GeV, there is an additional region that survives both X-ray and CMB constraints in the ranges $8 \times 10^{-3} < g_{x} <  3 \times 10^{-2}$ (1~GeV) and $2 \times 10^{-2} < g_{x} <  0.3$ (5~GeV). Finally note that the direct detection constraint does not further restrict the parameter space.

\paragraph{Right branch solutions -- $\Sigma_{0}^{2} > \bar{\Sigma}_{\rm min}^{2}$}

 For the right branch solutions which correspond to a larger $\Sigma_0^2$, that is a larger mass gap $\Delta=m_x-2\, m_\phi$, 
it is the region at low values of $\epsilon$ that is constrained by both CMB and X-rays while, as in the previous case,  larger values of $\epsilon$ are constrained by accelerator searches for dark photons. Direct detection constraints are relevant in the region of parameter space where $\epsilon$ and $g_x$ are large, especially when  $m_\phi=5\ {\rm GeV}$.

In the future, direct detection experiments dedicated to detecting low-energy threshold nuclear recoils provide promising limits to constrain DM interaction with SM particles more efficiently. Experiments such as DARKSPHERE~\cite{NEWS-G:2023qwh} could reach a sensitivity of $\sigma_N^{\rm SI} \simeq 10^{-43}\,{\rm cm^2}$~\cite{Akerib:2022ort} for DM mass around 1 GeV and SBC 1 ton~\cite{Giampa:20211P,Akerib:2022ort} could probe $\sigma_N^{\rm SI} \simeq 7\times 10^{-46}\,{\rm cm^2}$ for a 5 GeV DM. This corresponds to more than four (one) orders of magnitude improvement for a DM mass of 1 (5) GeV, allowing direct detection to significantly probe the allowed region.

Again, for a DM mass $m_{\phi}$ of 1 and 5~GeV, a horizontal band opens up in parameter space, which escapes CMB and X-ray constraints, in the range $8 \times 10^{-12} < \epsilon < 8 \times 10^{-11}$ (1~GeV) and $3 \times 10^{-12} < \epsilon < 10^{-9}$ (5~GeV). It is expected that future CMB missions such as PIXIE or PRISM will partially probe these regions.

In the future, various searches for dark photons through their visible or invisible decay modes will be able to probe the region with prompt decays for large values of $\epsilon$ and displaced decays for small values of $\epsilon$. Near-future prompt and displaced visible decay search experiments~\cite{Gori:2022vri} like LHCb, Belle-II, DarkQuest and DUNE are proposed to exclude the kinetic mixing parameter down to $\epsilon \sim 3\times 10^{-8}$ for a 200-500 MeV dark photon. Moreover, the sensitivity reach for the invisible decay searches from BaBar is expected to improve by 2 orders of magnitude with LDMX~\cite{Filippi:2020kii} in this mass range.
For dark photons mass around the GeV,  Belle-II could probe a few $10^{-4}$ with dilepton (visible) ~\cite{Graham:2021ggy} and invisible searches~\cite{Filippi:2020kii}.

\section{Conclusion}
\label{sec:conclusion}
In this work, we have explored a model where the DM candidate is a GeV-scale scalar species $\phi$ charged under a new local gauge group ${\rm U'}(1)$. The gauge boson associated to this gauge group, dubbed throughout the article `dark photon', acts as a vector boson portal between the dark and the SM sectors of the theory. DM annihilation into light SM fermions proceeds through the exchange in the $s$-channel of this dark photon.
In these conditions, DM annihilation has a $p$-wave behavior at low energy. However, it can be significantly enhanced if two conditions are met. First, the dark photon mass $m_{x}$ must be slightly larger than twice the DM mass $m_{\phi}$ and, second, the decay width of the dark photon must be smaller than twice the mass gap $\Delta$ between $m_{x}$ and $2 m_{\phi}$. If both conditions are met, a Breit-Wigner resonance appears at a DM dispersion velocity directly related to this mass gap, i.e. when the DM temperature is equal to $\Delta$. The smaller the mass gap, the smaller the DM dispersion velocity at peak annihilation.

This has profound implications for the cosmological behavior of our DM candidate. In most models in the literature, the bulk of DM annihilation takes place during the freeze-out process. Shortly after freeze-out, DM annihilation stops and the cosmological abundance of DM reaches rapidly its relic level.
In our model, on the contrary, the peak of DM annihilation may occur well after freeze-out if the mass gap is very small. Moreover, the relic abundance is always shaped at that particular moment. Determining the evolution of the DM temperature during the expansion of the Universe turns out to be crucial, insofar as DM annihilation mostly occurs when that temperature is of order the mass gap $\Delta$.
That is why kinetic decoupling between DM and the primordial plasma must be dealt with. As we showed, failure to do so results in a DM relic abundance being overestimated by several orders of magnitude.

At fixed DM mass $m_{\phi}$, dark charge $g_{x}$ and mixing angle $\epsilon$, we found in general two values for the dark photon mass $m_{x}$ for which the DM relic abundance is equal to the cosmological measurement~\cite{Planck:2018vyg}.
The smallest value, for which $\Sigma_{0}^{2} < \bar{\Sigma}_{\rm min}^{2}$, is classified as the \textit{left branch solution} while the largest value, for which $\Sigma_{0}^{2} > \bar{\Sigma}_{\rm min}^{2}$, is classified as the \textit{right branch solution}.
For both possibilities, we explored the $( g_{x} , \epsilon )$ plane for a DM mass $m_{\phi}$ set equal to $200 \, {\rm MeV}$, $1 \, {\rm GeV}$ and $5 \, {\rm GeV}$.
For small values of $g_{x}$ and $\epsilon$, the scalar DM candidate overshoots the cosmological observed value. For \textit{left branch solutions} only, large values of $g_{x}$ and $\epsilon$ are also excluded since in this case scalar DM undershoots the measurement.
We then applied several constraints to the surviving regions. In the domains where $\epsilon$ is small, the peak of DM annihilation occurs when the CMB energy spectrum is most sensitive to energy injection. We found $\mu$ and $y$-distortions exceeding by far the bounds set by observations.
For smaller values of $g_{x}$ (\textit{left branch}) and larger values of $\epsilon$ (\textit{right branch}), the regions are excluded by X-ray observations.
These astrophysical constraints are complemented by the limits set by colliders and direct detection experiments for large values of $\epsilon$.

In all the cases which we investigated, we found regions having successfully passed all the tests.
In the \textit{left branch} case, this region lies at the upper-left corner of the band where $g_{x}$ is small and $\epsilon$ is large.
In the \textit{right branch} case, the surviving domain corresponds roughly to values of $\epsilon$ in the range between $10^{-7}$ and $10^{-3}$. In these allowed regions, the mass gap $\Delta$ is not too small and the peak of DM annihilation occurs shortly after freeze-out.

Although energy injection at late cosmological times is severely constrained, we found that new domains open up in parameter space, all the more so when DM is heavy. Vertical (horizontal) bands appear in the \textit{left} (\textit{right}) \textit{branch} plots of Figs.~\ref{fig:resultsDM1GeV} and \ref{fig:resultsDM5GeV}. We expect the future CMB missions PIXIE (NASA) or PRISM (ESA) to partially close these possibilities since the future instruments will reach a sensitivity of $10^{-8}$ on both $y$ and $\mu$-distortions~\cite{Kogut:2019vqh}. We anticipate that these regions will nevertheless survive.

In the future, the region at large values of $g_x$ and $\epsilon$ could be further probed with dedicated low-energy threshold direct detection experiments such as DARKSPHERE~\cite{NEWS-G:2023qwh} and SBC~\cite{Akerib:2022ort,Giampa:20211P} and improve the current limits by a few orders in our mass range of interest. Moreover, several planned searches for dark photons through their visible and invisible decays at accelerators such as LHCb, Belle-II, DUNE, DarkQuest or LDMX ~\cite{Gori:2022vri,Graham:2021ggy} are proposed to improve the upper bound on $\epsilon$ by a few orders of magnitude for a few hundreds of MeV and by an order of magnitude for a few GeV dark photon. We have discussed this in details in the text.

Our results should finally be taken with a grain of salt. Our conclusions are based on the key assumption that DM is always in thermal equilibrium with itself. In the regions having passed all our tests, kinetic decoupling occurs well after DM has become non-relativistic. Actually, before kinetic decoupling happens, DM reaches a state of inner thermal equilibrium through its close contact with the primordial plasma, exactly like photons do with electrons just before recombination. But, after kinetic decoupling, our assumption that DM is still in thermal contact with itself needs to be scrutinized. Such an investigation was beyond the scope of this exploratory analysis. In a follow-up study, we plan to investigate this point by solving, for instance, the transport Boltzmann equation for DM species and evolve in time the particle momentum distribution~\cite{Binder:2021bmg}.
A more ambitious goal would be to embed the Lagrangian~\ref{eq:Lagrangian_1} in a more general set-up and to study the cosmological consequences of the overall theory. This would allow to determine whether or not the DM scalar $\phi$ is actually in thermodynamical equilibrium with the primordial plasma before it becomes non-relativistic.

\acknowledgments
The work of GB and SC was funded in part by the Indo-French Centre for the Promotion of Advanced
Research (Project title: Beyond Standard Model Physics with Neutrino and Dark Matter
at Energy, Intensity and Cosmic Frontiers, Grant no: 6304-2). S. C. also acknowledges support from the UKRI Future Leader Fellowship “DARKMAP” (Grant No:
MR/T042575/1) and the hospitality of LAPTh Annecy where part of this work was done. 

\appendix
\section{ Discussion of $J(a,b)$ and $\asv$}
\label{app:jsigmav}
We highlight some properties and approximations of $J(a,b)$ below
\begin{itemize}
    \item The integral $J$ has been normalized so that $J(0,0) = 1$
    \item  {\boldmath $a>0$ :} With reference to Eq.~\ref{eq:definition_a_b}, this corresponds to $m_{x} < 2 m_{\phi}$. 
\begin{enumerate}
\item{
If $a$ and $b$ are both smaller than or of order $1$, i.e. if the dispersion velocity $\Sigma$ is larger than ${\rm Max}\,(\Sigma_{0} , \Lambda_{0})$, the integral $J$ is roughly equal to $1$ and the average cross-section $\asv$ scales as $\Sigma^{-2}$. It decreases as $\Sigma$ increases. This behavior is reminiscent of a Sommerfeld enhanced annihilation.}
\item{
In the opposite situation where $\Sigma$ is less than ${\rm Max}\,(\Sigma_{0} , \Lambda_{0})$, at least one of the parameters $a$ and $b$, if not both, is larger than $1$ and the denominator of the integrands in expressions~\ref{eq:definition_J_a_b} is roughly approximated as $a^{2} + b^{2}$. This yields a value of $J$ equal to
\beq
J_{2}(a,b) = {\displaystyle \frac{3/4}{a^{2} + b^{2}}} \,.
\label{eq:definition_J2}
\eeq
The averaged cross-section $\asv$ scales this time as $\Sigma^{2}$. The annihilation is $p$-wave suppressed at low velocities.}
\end{enumerate}

\item {\boldmath $a<0$ :} In this case, the dark photon is heavier than a $\phi$-$\bar{\phi}$ pair at rest. The annihilation can be enormously enhanced if the velocities of the incoming DM scalars can fill the gap $\Delta$ between $2 m_{\phi}$ and $m_{x}$ and if the decay width of the dark photon that mediates the process is very narrow. The annihilation proceeds then through a Breit-Wigner resonance. The smaller is $\Lambda_{0}$ with respect to $\Sigma_{0}$, the narrower and higher the resonance. The peak is actually reached when the one-dimensional dispersion velocity $\Sigma$ of the DM scalars is of order $\Sigma_{0}$.

We remark with reference to Eq.~\ref{eq:definition_a_b} that when $\Sigma$ increases, both $|a|$ and $b$ decrease while the ratio $|a|/b$ remains fixed at ${\Sigma_{0}^{2}}/{\Lambda_{0}^{2}}$.
The behavior of the integral $J$ is now more involved than when $a$ was positive. The left panel of Fig.~\ref{fig:J_figure_1} features the evolution of $J$ as a function of $|a|$ for a fixed ratio ${b}/{|a|}$ of $10^{-4}$. Three regimes are clearly visible :
%
\begin{figure}[t!]
\centering
\includegraphics[width=0.45\textwidth]{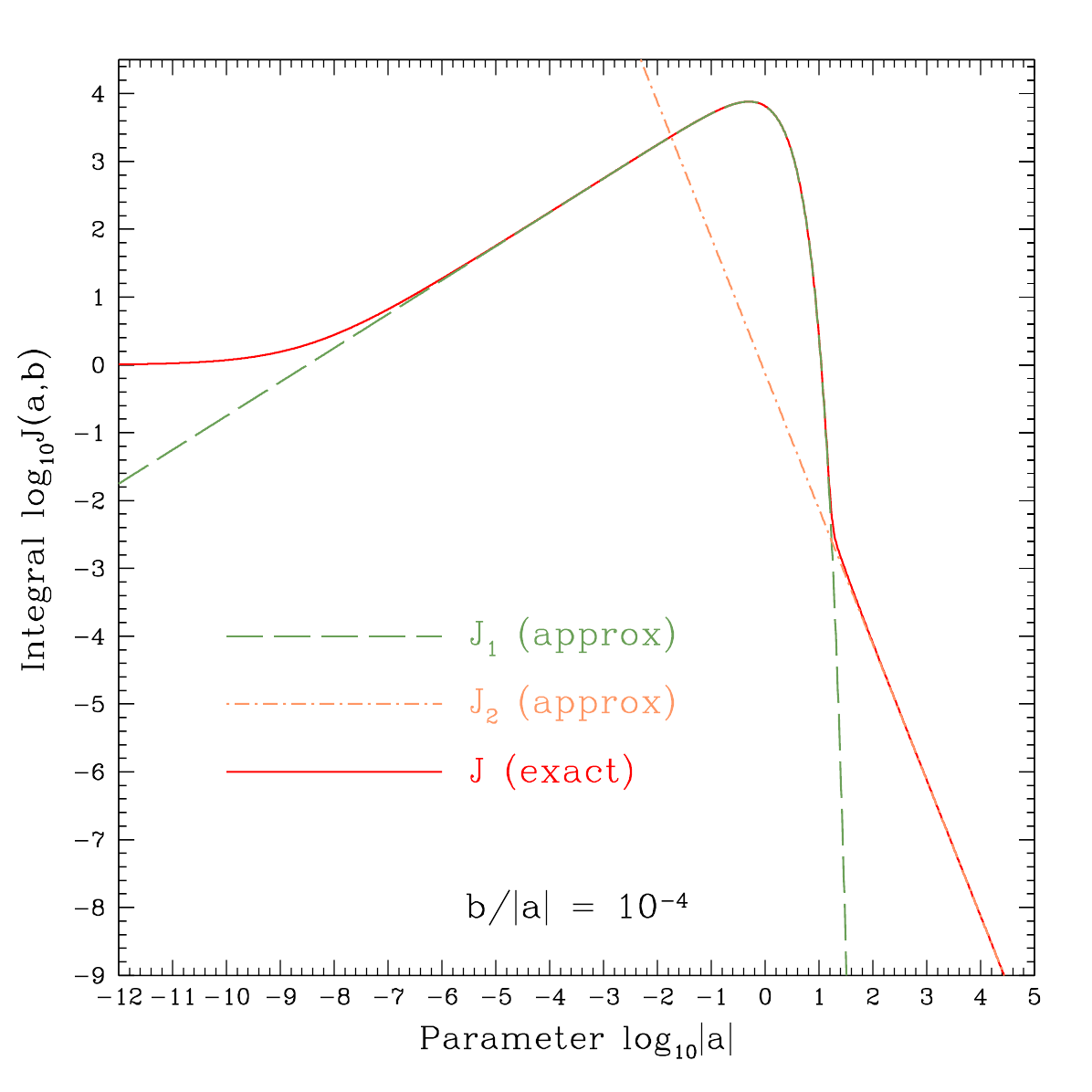}
\includegraphics[width=0.45\textwidth]{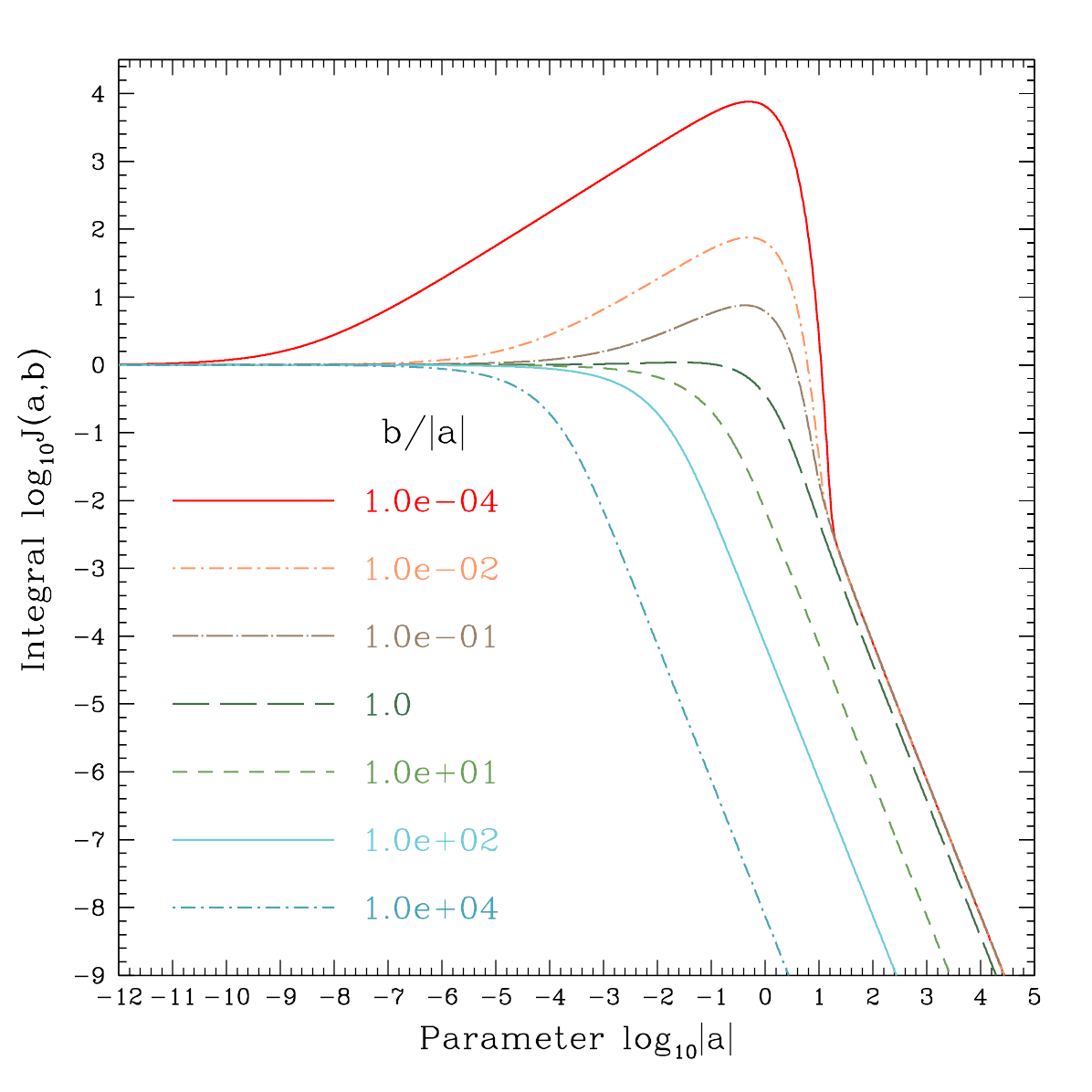}
\caption{
The integral $J(a,b)$ is plotted as a function of $|a|$ ($a<0$) for a fixed ratio ${b}/{|a|}$ of $10^{-4}$ (left) and several values of the ratio ${b}/{|a|}$ (right).
In the left panel, three behaviors are clearly visible - (i) When the parameter $|a|$ goes to zero, the integral $J$ tends to $1$. (ii) The Breit-Wigner regime sits in the intermediate region where $J$ steadily increases like $\sqrt{|a|}$ until $|a|$ reaches ${1}/{2}$. Beyond that point, $J$ drops exponentially until the third regime is reached. (iii) At large values of $|a|$, the integral $J$ decreases actually as ${1}/{a^{2}}$.
In the right panel, the curves for which the ratio is small, exhibit a Breit-Wigner enhancement, while for larger values of the ratio, we observe the smooth transition from the constant value of 1 to the asymptotic form.
}
\label{fig:J_figure_1}
\end{figure}
%

\begin{enumerate}
\item{
When the parameter $|a|$ goes to $0$, the integral converges toward $1$ as it should, since $J(0,0) = 1$.}
\item{
For small values of the ratio ${b}/{|a|}$, and as long as $|a|$ is not too large with respect to $1$, the Breit-Wigner resonance sets in. The region of integration which contributes most to $J$ is centered around $t = |a|$. It corresponds to the annihilation peak whose full width at half-maximum is $2b \ll |a|$. We can write the integral as
\beq
J(a,b) =
\frac{1}{\sqrt{\pi}\,} {\displaystyle \int_{0}^{+ \infty}} \!
{\displaystyle \frac{t^{3/2\,} e^{-t}}{(t - |a|)^{2} + b^{2}}} \; dt \simeq
{\displaystyle \frac{|a|^{3/2} e^{-|a|}}{\sqrt{\pi}}}
\left\{
{\displaystyle \int_{0}^{+ \infty}}\!\!\!{\displaystyle \frac{dt}{(t - |a|)^{2} + b^{2}}} \simeq
{\displaystyle \int_{-\infty}^{+ \infty}}\!{\displaystyle \frac{du}{u^{2} + b^{2}}} = {\displaystyle \frac{\pi}{b}}
\right\} \,,
\eeq
where the variable $u = t - |a|$. The integral $J$ is well approximated by the function
\beq
J_{1}(a,b) = {\displaystyle \frac{\sqrt{\pi}}{b}} \, |a|^{3/2} e^{-|a|} \,.
\label{eq:J1}
\eeq
It reaches its maximum at $|a| = {1}/{2}$. Beyond that point, $J$ decreases exponentially like $e^{-|a|}$.
}
\item{
For large values of $|a|$, a new regime sets in where $J$ is well approximated by the function $J_{2}(a,b)$ defined in~\ref{eq:definition_J2}. In general, this approximation is a good description of $J$ for values of $\sqrt{a^{2} + b^{2}}$ much larger than $1$, irrespective of the sign of $a$.
}
\end{enumerate}
\end{itemize}

The evolution of $J$ as a function of $|a|$ is plotted in the right panel of Fig.~\ref{fig:J_figure_1} for several values of the ratio ${b}/{|a|}$. When this ratio is small, i.e. when the Breit-Wigner resonance through which the annihilation proceeds is narrow, this  translates into $\Lambda_{0}$ being much smaller than $\Sigma_{0}$ and the integral $J$ has the same behavior as in Fig.~\ref{fig:J_figure_1}. It features a bump whose highest point is reached at $|a| = {1}/{2}$. On the contrary, when $\Lambda_{0}$ is larger than $\Sigma_{0}$, i.e. for large values of the ratio ${b}/{|a|}$, the contribution from $J_{1}$ becomes subdominant and the integral $J$ does not exhibit any enhancement. It evolves smoothly from the constant value of $1$ to its asymptotic form $J_{2}$. The transition between these regimes takes place for $|a| \sim {|a|}/{b}$. We recover the same behavior as if $a$ were positive.

\section{ Scalar Dark Matter thermalization}
\label{sec:phi_thermalization}
In this section, we present a simplified analysis of the thermalization of scalar DM with the SM plasma.
A complete treatment would require the knowledge of the overall theory, which is beyond the scope of our exploratory work. We will concentrate here on the interactions between scalar DM and SM fermions that are encoded in Lagrangian~\ref{eq:Lagrangian_1}, the starting point of our analysis.

We will also assume that scalar DM is thermalized with itself, and that a DM temperature $T_{\phi}$ can be defined at all times. This is certainly true whenever collisions between DM scalars and SM fermions are rapid enough to ensure efficient energy exchange between these populations and to establish thermalization. We will assume that DM reaches a state of inner thermal equilibrium after kinetic decoupling from the SM plasma has occured, allowing the DM temperature $T_{\phi}$ to be defined also in this situation. This may appear as an oversimplification. Going beyond it would make the problem orders of magnitude more complicated. We would have to solve directly the Boltzmann equation and study the evolution in time of the DM distribution function in momentum space~\cite{Binder:2021bmg}. We will defer such an investigation to a future work.
Assuming that DM reaches inner thermal equilibrium with temperature $T_{\phi}$ leads already to a particularly rich and complex phenomenology, which could be the starting point for further investigations.

The question of DM thermalizing with the primordial plasma is paramount insofar as the annihilation cross-section~\ref{eq:asv_1} crucially depends on the DM dispersion velocity $\Sigma$ and DM temperature $T_{\phi} \equiv m_{\phi} \Sigma^{2}$. As encrypted in Lagrangian~\ref{eq:Lagrangian_1}, scalar DM exchanges energy with the SM plasma through annihilations into, recreations from and collisions upon SM light fermions. Our aim is to model these processes to establish an equation that drives the evolution of the DM temperature $T_{\phi}$ with respect to the SM plasma temperature $T$. Notice that the DM heat capacity is small compared to that of the SM plasma. As it becomes non-relativistic, DM annihilates and its density actually drops, hence a negligible contribution to the overall heat capacity of the primordial plasma. The temperature of the latter still evolves at constant entropy, decreasing approximately like $a^{-1}$, where $a$ is the scale factor of the Universe. The SM plasma is not affected by its thermal contact with DM, or the breaking of it. This is not so for the DM temperature which relaxes rapidly toward the SM temperature at early times. At some point, called thermal or kinetic decoupling, this relaxation slows down, and $T_{\phi}$ cannot follow $T$ anymore. The DM temperature decreases afterward more rapidly than the plasma temperature.

In Sec.~\ref{sec:phi_therm_annihilation}, we show that DM can be thermalized through its annihilation into and recreation from SM fermions. We show that kinetic decoupling occurs always after freeze-out.
In Sec.~\ref{sec:phi_therm_collisions}, we model the energy exchanged between DM and the SM plasma through collisions.
We eventually establish the master equation for $T_{\phi}$ in Sec.~\ref{sec:kinetic_decoupling} and explain how we find the kinetic decoupling point.

\subsection{Thermalization through annihilations}
\label{sec:phi_therm_annihilation}
We investigate here whether or not the thermalization of $\phi$ and $\bar{\phi}$ particles with the SM plasma could be ensured solely through their annihilation into, and production from, standard model fermions
\beq
\phi + \bar{\phi} \rightleftharpoons f + \bar{f} \,.
\label{eq:chemical_equilibrium_appendix}
\eeq
We want to determine if DM species are thermalized, and thermal contact is established with the SM plasma, should this reaction be fast enough.
We assume that DM particles are thermalized with each other so that a DM temperature $T_{\phi}$ can be defined. We would like to determine how fast $T_{\phi}$ relaxes toward the plasma temperature $T$.

For this, let us consider a volume ${\cal V}$ of space. It is filled with DM species $\phi$ and $\bar{\phi}$ whose densities $n_{\phi}$ and $n_{\bar{\phi}}$ are equal insofar as no asymmetry is assumed. Let us define $n \equiv n_{\phi} \equiv n_{\bar{\phi}}$. The number of DM particles inside the volume ${\cal V}$ is
\beq
N = (n_{\phi} + n_{\bar{\phi}}) {\cal V} = 2 n {\cal V} \,.
\label{eq:N_phi}
\eeq
The DM energy stored inside that volume is
\beq
U = N \left\{ m_{\phi} + \frac{3}{2} T_{\phi} \right\} .
\label{eq:U_phi}
\eeq
We assume DM to be non-relativistic as we are dealing with the freeze-out process at temperatures below $m_{\phi}$. The pressure of the DM gas is $P = (n_{\phi} + n_{\bar{\phi}}) T_{\phi} = 2 n T_{\phi}$.
During the lapse of time dt, the number $N$ of DM species inside volume ${\cal V}$ varies according to the chemical reaction~\ref{eq:chemical_equilibrium_appendix}
\beq
\frac{dN}{dt} =
- 2_{\,} \asv_{T_{\phi}} n^{2\,} {\cal V} \, + \, 2_{\,} \asv_{T} n_{\rm eq}^{2\,} {\cal V} \,.
\label{eq:N_dot_a}
\eeq
Two DM species disappear or are created per reaction. The creation term can be derived by noticing that it should cancel the annihilation term at thermodynamical equilibrium. The density $n_{\rm eq}$ corresponds to a population at temperature $T$ with vanishing chemical potential. In the non-relativistic regime under consideration, it is given by Eq.~\ref{eq:n_e_phi}. On the other hand, relation~\ref{eq:N_phi} leads to
\beq
\frac{dN}{dt} = 2_{\,} {\frac{dn}{dt}}_{\,} {\cal V} \, + \, 2 n \left\{ \frac{d{\cal V}}{dt} = 3 H {\cal V} \right\},
\label{eq:N_dot_b}
\eeq
where $H = {\dot{a}}/{a}$ is the expansion rate and $a$ is the scale factor of the Universe at time $t$. Combining both equalities~\ref{eq:N_dot_a} and \ref{eq:N_dot_b} yields the well-known equation
\beq
\frac{dn}{dt} = - 3 H n \, - \, \asv_{T_{\phi}} n^{2} \, + \, \asv_{T} n_{\rm eq}^{2} \,.
\eeq
Alternatively, we can recast relation~\ref{eq:N_dot_a} into
\beq
\frac{dN}{dt} \, + \, \left\{ \asv_{T_{\phi}} n \right\} N = \left\{ \asv_{T} n_{\rm eq} \right\} N_{\rm eq} \,.
\label{eq:N_evol_a}
\eeq
As long as the annihilation rate
\beq
\Gamma_{\rm \! ann} \equiv \asv_{T_{\phi}} n
\label{eq:Gamma_ann}
\eeq
is larger than the rate with which the right hand side term evolves, a dynamical equilibrium is reached and ${\dot N}$ can be disregared in equation~\ref{eq:N_evol_a}. If both temperatures $T_{\phi}$ and $T$ are furthermore equal, the DM density $n$ is given by its chemical equilibrium value~\ref{eq:n_e_phi}.

During the lapse of time $dt$, the DM internal energy~\ref{eq:U_phi} varies by an amount
\beq
dU = - P d{\cal V} \, - \,
2_{\,} \asv_{T_{\phi}} n^{2\,} {\cal V}_{\,} dt \left\{ m_{\phi} + \frac{3}{2} T_{\phi} \right\} + \,
2_{\,} \asv_{T} n_{\rm eq}^{2\,} {\cal V}_{\,} dt \left\{ m_{\phi} + \frac{3}{2} T \right\} .
\label{eq:dU_phi_a}
\eeq
We have also applied detailed balance, noticing that each time a pair of DM scalars disappears, an average energy twice equal to $m_{\phi} + {3}/{2\,}T_{\phi}$ is removed from the DM population. The amount of energy given to DM each time a pair of scalars $\phi \, \bar{\phi}$ is created is twice equal to $m_{\phi} + {3}/{2\,}T$, with this time $T$ instead of $T_{\phi}$. Actually when thermodynamical equilibrium is reached between DM and the SM plasma, the only change in the DM internal energy comes from the pressure work $- P d{\cal V}$. The variation of the DM internal energy~\ref{eq:U_phi} can also be written as
\beq
dU = dN \left\{ m_{\phi} + \frac{3}{2} T_{\phi} \right\} \, + \, \frac{3}{2} N dT_{\phi} \,.
\label{eq:dU_phi_b}
\eeq
Using expressions~\ref{eq:N_evol_a}, \ref{eq:dU_phi_a} and \ref{eq:dU_phi_b}, we derive the equation fulfilled by the DM temperature
\beq
\frac{dT_{\phi}}{dt} = - 2 H T_{\phi} \, - \left\{ \asv_{T} \frac{n_{\rm eq}^{2}}{n} \right\}_{\!} \left( T_{\phi} - T \right) .
\eeq
At early times, before freeze-out starts, we can asume DM to be in thermodynamical equilibrium with the rest of the plasma so that $n = n_{\rm eq}$ while $T_{\phi} = T$. The annihilation rate $\Gamma_{\rm \! ann}$ is given by $\asv_{T} n_{\rm eq}$ and is equal to the rate $\Gamma_{\rm rel}^{\rm F}$ defined in Sec~\ref{subsec:omegah2}. The DM temperature evolves as
\beq
\frac{dT_{\phi}}{dt} = - 2 H T_{\phi} \, - \, \Gamma_{\rm \! ann} \left( T_{\phi} - T \right) .
\eeq
This equation can be recast into
\beq
\frac{dT_{\phi}}{dt} \, + \, \left\{ \Gamma_{\rm \! ann} + 2H \right\} T_{\phi} =  \Gamma_{\rm \! ann} T \,,
\label{eq:T_phi_evol_a}
\eeq
with the same structure as~\ref{eq:N_evol_a} which boils down to
\beq
\frac{dN}{dt} \, + \, \Gamma_{\rm \! ann \, } N = \Gamma_{\rm \! ann} N_{\rm eq} \,.
\label{eq:N_evol_b}
\eeq
As both Eqs.~\ref{eq:T_phi_evol_a} and \ref{eq:N_evol_b} are similar, we conclude that freeze-out (aka chemical decoupling) is concomitant with thermal decoupling (aka kinetic decoupling). We can even guess that freeze-out occurs slightly before kinetic decoupling.
Actually, the codensity $N$ relaxes toward its dynamical equilibrium value $N_{\rm eq}$ with rate $\Gamma_{\rm \! ann}$, while the DM temperature $T_{\phi}$ relaxes toward the plasma temperature $T$ with the slightly larger rate $\Gamma_{\rm \! ann} + 2H$. We also notice that the right hand side term $\Gamma_{\rm \! ann} N_{\rm eq}$ of the codensity equation~\ref{eq:N_evol_b} drops much faster than $ \Gamma_{\rm \! ann} T$, its temperature counterpart in equation~\ref{eq:T_phi_evol_a}. The codensity $N_{\rm eq}$ has an exponential dependence $\exp(- {m_{\phi}}/{T})$ and decreases much faster than $T$.

To conclude, we have proved that DM annihilation into, and production from, SM light fermions is able alone to thermalize DM with the plasma. Kinetic decoupling in that case occurs slightly after freeze-out. Reaction~\ref{eq:chemical_equilibrium_appendix} results from the interaction Lagrangian~\ref{eq:Lagrangian_1} which also implies the existence of collisions between DM and light fermions discussed in the next section.

\subsection{Thermalization through collisions with SM species}
\label{sec:phi_therm_collisions}
There are of course collisions between light SM fermions and DM. These contribute to the thermalization of DM since energy is exchanged between both populations. Let us develop a simplified calculation of the energy transferred from the fermions $f$ to the DM species. The latter are non-relativistic since we are interested here in the period of kinetic decoupling which occurs after freeze-out, for a plasma temperature below $m_{\phi}$. We can safely treat the DM scalars as if they were at rest in the cosmological frame and compute their recoil energy as they are impacted by incident fermions. We focus on the collision
\beq
f  + \phi \longrightarrow f + \phi \,.
\label{eq:collision_f_phi}
\eeq
A dark photon is exchanged in the $t$-channel between the fermionic line and the scalar line. The fermion $f$ is ultra-relativistic as we are interested this time in plasma temperatures $T$ above the mass $m_{f}$. In the opposite situation, pair annihilation drives the population of fermions $f$ to extinction, with number density $n_{f}$ dropping like $\exp(-{m_{f}}/{T})$, and energy transfer stops.
The incident fermion in reaction~\ref{eq:collision_f_phi} has energy $\epsilon_{f}$ and is scattered through the angle $\theta$ with respect to its initial direction. The scalar $\phi$, initially at rest, recoils with kinetic energy $E_{R}$. A straightforward but exact calculation yields
\beq
E_{R} = \left\{ \frac{u}{1 + u} \right\} \epsilon_{f}
\;\;\;\text{where}\;\;\;
u = \frac{\epsilon_{f}}{m_{\phi}} (1 - \cos\theta) \,.
\label{eq:E_R_a}
\eeq
We recover the well-known relation of the Compton effect. In the regime where the plasma temperature $T$ is significantly smaller than the DM mass $m_{\phi}$, so is the average fermion energy. The parameter $u$ is small with respect to $1$ and the recoil energy boils down to
\beq
E_{R} \simeq \frac{\epsilon_{f}^{2}}{m_{\phi}} \, (1 - \cos\theta) \,.
\label{eq:E_R_b}
\eeq
The collisions of the fermions $f$ and $\bar{f}$ on the scalars $\phi$ and $\bar{\phi}$ result in the increase of the DM internal energy contained in volume ${\cal V}$ with rate
\beq
\frac{dU_{\rm col}}{dt} = N
{\displaystyle \int} \! dn_{f} \, v_{f} {\displaystyle \int_{4 \pi}} \!\! d\Omega \; \frac{d\sigcol}{d\Omega} \, E_{R}(\Omega) \,,
\label{eq:dU_collision_a}
\eeq
where $E_{R}(\Omega)$ is given by~\ref{eq:E_R_b}. As fermions are ultra-relativistic, their velocity $v_{f}$ is equal to the speed of light. Their number density obeys Fermi-Dirac statistics
\beq
dn_{f} = \frac{4 \pi \epsilon_{f}^{2} d\epsilon_{f}}{8 \pi^{3}}
\left\{ f(\epsilon_{f}) = \frac{g_{f}}{\exp({\epsilon_{f}}/{T}) + 1} \right\} ,
\eeq
where $g_{f}$ denotes the spin degeneracy factor.
The Feynman diagram associated to reaction~\ref{eq:collision_f_phi} yields the differential collision cross-section
\beq
v_{f} \, \frac{d\sigcol}{d\Omega} = \frac{1}{8 \pi^{2}}
{\displaystyle \frac{g_{x}^{2} \epsilon^{2} e^{2} Q_{\!f}^{2}}{m_{x}^{4}}} \;
\epsilon_{f}^{2} \left( 1 + \cos\theta \right) .
\eeq
Plugging the two last relations into the rate of collisional energy transfer~\ref{eq:dU_collision_a} yields
\beq
\frac{dU_{\rm col}}{dt} = N \, {\displaystyle \frac{g_{f}}{8 \pi^{3}}} {\displaystyle \frac{g_{x}^{2} \epsilon^{2} e^{2} Q_{\!f}^{2}}{m_{x}^{4} m_{\phi}}}
\left\{ {\displaystyle \int_{0}^{\infty}} \! \frac{\epsilon_{f}^{6} \, d\epsilon_{f}}{\exp({\epsilon_{f}}/{T}) + 1} =
\Gamma(7) \zeta(7) \! \left(1 - \frac{1}{2^{6}} \right) \! T^{7} \right\}
\left\{ {\displaystyle \int_{0}^{\pi}} \! d(-\cos\theta) \left( 1 - \cos^{2}\theta \right) = \frac{4}{3} \right\}.
\label{eq:dU_collision_b}
\eeq
We can apply detailed balance to describe now the energy flowing from scalar DM to fermions as a result of collisions. This boils down to replacing $T^{7}$ by $T_{\phi}^{7}$ in the previous expression. Going a step further by linearizing the net gain of DM internal energy per unit time with respect to the temperature difference $T_{\phi} - T$, we get
\beq
\frac{dU_{\rm col}}{dt} = \frac{3}{2} N \, {\cal C}_{\rm col} T^{6} \left( T - T_{\phi} \right) \,.
\eeq
The coefficient ${\cal C}_{\rm col}$ is given by
\beq
{\cal C}_{\rm col} = {\cal A}_{\rm \, col \,} {Q}_{\rm eff \,}^{2} {\displaystyle \frac{g_{x}^{2} \epsilon^{2} e^{2}}{m_{x}^{4} m_{\phi}}},
\eeq
where
\beq
{\cal A}_{\rm \, col} = \frac{2205}{4 \pi^{3}} \, \zeta(7) \simeq 17.9271
\;\;\;\text{while}\;\;\;
{Q}_{\rm eff}^{2} \equiv {\displaystyle \sum_{m_{f} \leq T}} g_{f} Q_{\!f}^{2} \,.
\eeq
We have summed over all fermionic populations which are ultra-relativistic when the plasma temperature is $T$, hence the effective charge ${Q}_{\rm eff}^{2}$.

Notice that we have so far concentrated on fermions but we might be in the situation where charged pions also collide on DM scalars. This may happen at the end of the quark-hadron phase transition, although pions are not strictly ultra-relativistic at that time. We can slightly modify the definition of ${Q}_{\rm eff}^{2}$ to include these charged scalar species. Their collision cross-section is enhanced by a factor of $3$ with respect to fermions and the statistical factor must also be modified. We describe now ${Q}_{\rm eff}^{2}$ as the smooth function of the plasma temperature $T$
\beq
{Q}_{\rm eff}^{2} \equiv {\displaystyle \sum_{f}} \; g_{f} Q_{\!f}^{2} \exp(- {m_{f}}/{T}) \,+ \,
\frac{192}{63} \, g_{\pi\,} Q_{\pi}^{2} \exp(- {m_{\pi}}/{T}) \,,
\eeq
where $g_{\pi} = 2$ and $Q_{\pi}^{2} = 1$.

\subsection{Kinetic decoupling}
\label{sec:kinetic_decoupling}
Taking into account both annihilations and collisions, the evolution of the scalar DM temperature becomes
\beq
\frac{dT_{\phi}}{dt} = - 2 H T_{\phi} \, -
\left\{ \asv_{T} \frac{n_{\rm eq}^{2}}{n} + {\cal C}_{\rm col} T^{6} \right\}_{\!} \left( T_{\phi} - T \right).
\eeq
For numerical resolution purposes, this equation can be recast into a form similar to relation~\ref{eq:freeze_out_2} 
\beq
\frac{dT_{\phi}}{dt} \, + \,
\left\{ \Gamma_{\rm \! ann} +  \Gamma_{\rm \! col} + 2H \right\} T_{\phi} =
\left\{ \Gamma_{\rm \! ann} + \Gamma_{\rm \! col} \right\} T \,,
\label{eq:T_phi_evol_b}
\eeq
where
\beq
\Gamma_{\rm \! ann} = \asv_{T} \frac{n_{\rm eq}^{2}}{n} \simeq \asv_{T} n_{\rm eq} = \Gamma_{\rm rel}^{\rm F}
\;\;\;\text{while}\;\;\;
\Gamma_{\rm \! col} = {\cal C}_{\rm col} T^{6} = {\cal A}_{\rm \, col \,} {Q}_{\rm eff \,}^{2} {\displaystyle \frac{g_{x}^{2} \epsilon^{2} e^{2}}{m_{x}^{4} m_{\phi}}} \, T^{6} \,.
\eeq
The DM temperature $T_{\phi}$ relaxes toward the plasma temperature $T$ with rate
\beq
\Gamma_{\rm rel}^{\rm KD} =  \Gamma_{\rm \! ann} +  \Gamma_{\rm \! col} + 2H \,,
\eeq
while the rate of evolution of kinetic equilibrium can be defined as in~\ref{eq:definition_Gamma_eq_F} through the identity
\beq
\Gamma_{\rm eq}^{\rm KD} \equiv \left| \frac{d}{dt} \! \ln \left\{ \left( \Gamma_{\rm \! ann} + \Gamma_{\rm \! col} \right) T \right\} \right|.
\eeq
At early times, $\Gamma_{\rm rel}^{\rm KD}$ exceeds $\Gamma_{\rm eq}^{\rm KD}$ by orders of magnitude and DM is thermally connected to the SM plasma. But as time goes on, $\Gamma_{\rm \! ann}$ drops exponentially like $\exp(-{m_{\phi}}/{T})$ while $\Gamma_{\rm \! col}$ drops like $T^{6}$. The relaxation rate $\Gamma_{\rm rel}^{\rm KD}$ decreases faster than the rate $\Gamma_{\rm eq}^{\rm KD}$ at which the kinetic equilibrium evolves. Kinetic decoupling occurs when both rates are equal
\beq
\Gamma_{\rm rel}^{\rm KD}(x_{\rm KD}) = \Gamma_{\rm eq}^{\rm KD}(x_{\rm KD}) \,,
\label{eq:freeze_out_x_KD}
\eeq
where the kinetic-decoupling point $x_{\rm KD}$ is defined as the ratio ${m_{\phi}}/{T_{\rm KD}}$. Afterward, the DM temperature follows the equation
\beq
\frac{dT_{\phi}}{dt} \, + \,  2 H T_{\phi} = 0 \,,
\eeq
and $T_{\phi}$ decreases like $a^{-2}$, where $a$ is the scale factor of the Universe. DM behaves then like a non-relativistic gas undergoing adiabatic expansion.

%
\end{document}